\documentclass[draftclsnofoot,onecolumn,11pt]{IEEEtran}
\IEEEoverridecommandlockouts

\usepackage[utf8]{inputenc} 
\usepackage[T1]{fontenc}
\usepackage{url}
\usepackage{ifthen} 
\newboolean{short_version}
\setboolean{short_version}{false} 
\usepackage[cmex10]{amsmath} 

\usepackage[url,hyperrefblack,notheorems,IEEEtran]{research17} 
\usepackage{balance} 
\usepackage{amssymb,amsfonts,amsbsy}
\usepackage{enumitem}
\usepackage{mathtools}
\usepackage[lined,boxed,commentsnumbered,linesnumbered,ruled]{algorithm2e}
\usepackage{comment}
\setlist[description]{leftmargin=\parindent,labelindent=\parindent}

\newtheorem{theorem}{\mytheoremname}
\newtheorem{lemma}[theorem]{\mylemmaname}
\newtheorem{corollary}[theorem]{\mycorollaryname}

\newtheorem{proposition}[theorem]{\mypropositionname}

\newtheorem{definition}[theorem]{\mydefinitionname}
\newtheorem{remark}[theorem]{\myremarkname}
\newtheorem{example}[theorem]{\myexamplename}
\newcommand{\Hwt}[1]{\wH\left(#1\right)} 
\newcommand{\Lwt}[1]{w_{\tn{Lee}}\left(#1\right)} 
\newcommand{\Ewt}[1]{w_{\tn{E}}\left(#1\right)} 
\renewcommand{\ker}[1]{\operatorname{ker}\left(#1\right)} 
\newcommand{\ConstrA}[1]{\Lambda_\textnormal{A}(#1)} 
\newcommand{\eConstrA}[1]{\Lambda_\textnormal{A}(#1)} 

\newcommand{\GammaA}[1]{\Gamma_\textnormal{A}\left(#1\right)} 
\newcommand{\eGammaA}[1]{\Gamma_\textnormal{A}(#1)} 
\newcommand{\ConstrAfour}[1]{\Lambda_{\textnormal{A}_4}\left(#1\right)} 
\newcommand{\eConstrAfour}[1]{\Lambda_{\textnormal{A}_4}(#1)}

\newcommand{\ConstrC}[2]{\Lambda_{\textnormal{C}}(#1,#2)} 
\newcommand{\PConstrC}[2]{\Gamma_{\textnormal{C}}(#1,#2)} 
\renewcommand{\Rationals}{\mathbb{Q}} 
\renewcommand{\Reals}{\mathbb{R}} 
\newcommand{\we}[1]{W_{#1}} 
\newcommand{\jwe}[2]{\textnormal{jwe}_{#1,#2}} 
\newcommand{\swe}[1]{\textnormal{swe}_{#1}} 
\newcommand{\na}[2]{n_{#1}(#2)} 

\newcommand*{\Vrule}{\rule[-1ex]{0.5pt}{2.5ex}}

\newcommand*{\Scale}[2][4]{\scalebox{#1}{\ensuremath{#2}}} 
\usepackage{multirow}
\usepackage{tikz}
\usetikzlibrary{calc,shapes, patterns,decorations.text, decorations.pathreplacing}
\usepackage{ctable} 
\usepackage{afterpage} 
\usepackage{lscape} 
\usepackage{pdflscape} 
\usepackage{rotating} 
\usepackage{pbox} 
\usepackage{longtable}
\usepackage{nicematrix} 
\makeatletter
\def\LT@makecaption#1#2#3{%
  \LT@mcol\LT@cols c{\hbox to\z@{\hss\parbox[t]\LTcapwidth{%
        \footnotesize\bgroup\par\centering\@IEEEtabletopskipstrut{\normalfont\footnotesize #2}\\{\normalfont\footnotesize\scshape #3}\par\addvspace{0.5\baselineskip}\egroup\endgraf%
        \@IEEEtablecaptionsepspace}%
      \hss}}}
\makeatother

\allowdisplaybreaks

\usepackage[colorinlistoftodos, textsize=footnotesize]{todonotes}
\definecolor{darkgreen}{rgb}{0, 0.5, 0}

\begin{document}

\title{Secrecy Gain of Formally Unimodular Lattices\\ from Codes over the Integers Modulo 4
\thanks{This work was supported in part by the Norwegian Research Council through the qsIoT grant, project number 274889.}
}


\author{%
  \IEEEauthorblockN{Maiara F.~Bollauf, Hsuan-Yin Lin, and {\O}yvind Ytrehus}
\thanks{All authors are affiliated to Simula UiB, N--5006 Bergen, Norway. Their respective e-mail addresses are maiara@simula.no, lin@simula.no, and oyvindy@simula.no.}
\thanks{This paper was published partially in the proceedings of the IEEE International Symposium on Information Theory (ISIT'22), Espoo, Finland, 2022~\cite{BollaufLinYtrehus22_2} and in the proceedings of the Information Theory Workshop (ITW'23), Saint-Malo, France, 2023~\cite{BollaufLinYtrehus23_2}.}}%

\maketitle

\begin{abstract}

  Recently, a design criterion depending on a lattice's volume and theta series, called the \emph{secrecy gain}, was proposed to quantify the secrecy-goodness of the applied lattice code for the Gaussian wiretap channel. To address the secrecy gain of Construction $\textnormal{A}_4$ lattices from \emph{formally self-dual} $\Integers_4$-linear codes, i.e., codes for which the \emph{symmetrized weight enumerator (swe)} coincides with the swe of its dual, we present new constructions of $\Integers_4$-linear codes which are formally self-dual with respect to the swe. For even lengths, formally self-dual $\Integers_4$-linear codes are constructed from nested binary codes and double circulant matrices. For odd lengths, a novel construction called \emph{odd extension} from double circulant codes is proposed. Moreover, the concepts of Type I/II formally self-dual codes/unimodular lattices are introduced. Next, we derive the theta series of the \emph{formally unimodular lattices} obtained by Construction $\textnormal{A}_4$ from formally self-dual $\Integers_4$-linear codes and describe a universal approach to determine their secrecy gains. The secrecy gain of Construction $\textnormal{A}_4$ formally unimodular lattices obtained from formally self-dual $\Integers_4$-linear codes is investigated, both for even and odd dimensions. Numerical evidence shows that for some parameters, Construction $\textnormal{A}_4$ lattices can achieve a higher secrecy gain than the best-known formally unimodular lattices from the literature. Results concerning the flatness factor, another security criterion widely considered in the Gaussian wiretap channel, are also discussed.
  
\end{abstract}

\begin{IEEEkeywords}
  Lattices, Codes over $\Integers_4$, Construction C, Construction $\textnormal{A}_4$, secrecy gain, Gaussian wiretap channel.
\end{IEEEkeywords}

\section{Introduction}
\label{sec:introduction}

\IEEEPARstart{P}{\emph{hysical layer security (PLS)}} has recently received a great deal of attention in 5G and beyond 5G (B5G) wireless communications~\cite{PoorSchaefer16_1, CostaOggierCampelloBelfioreViterbo17_1, WuKhisti-etal18_1, Bloch-etal21_1, MaengYapiciGuvencBhuyanDai22_1}. In contrast to cryptographic algorithms, PLS approaches utilize the properties of the physical layer of the transmitting parties and provide \emph{information-theoretically unbreakable security} for safeguarding confidential data. PLS originates from Aaron D.~Wyner's landmark paper~\cite{Wyner75_1} in 1975, which showed that based on the \emph{communication channel} characteristics, one can achieve communication that is reliable and at the same time secure against an adversarial eavesdropper. Moreover, PLS has drawn significant attention in connection with industrial standards for the next generation of wireless communications~\cite{Mucchi-etal21_1, Chorti-etal22_1, Ruzomberka-etal23_1sub}.

In the \emph{wiretap channel (WTC)} introduced in~\cite{Wyner75_1}, a single transmitter (Alice) tries to communicate with a receiver (Bob) while keeping the transmitted messages secure from an unauthorized eavesdropper (Eve). The secure and confidential achievable rate between Alice and Bob for WTC is defined as the \emph{secrecy rate}. There is a recent focus on designing practical wiretap codes that achieve a high secrecy rate based on lattices over Gaussian WTCs~\cite{LingLuzziBelfioreStehle14_1, OggierSoleBelfiore16_1, OggierBelfiore18_1}. In~\cite{BelfioreOggier10_1, OggierSoleBelfiore16_1, OggierBelfiore18_1}, the authors introduce the \emph{secrecy function}, which is evaluated in terms of the lattice's volume and \emph{theta series}, and can be interpreted as the security coding gain of a specifically designed lattice for Eve compared to an uncoded lattice. The maximum of the secrecy function, namely the \emph{(strong) secrecy gain}, has been shown to be an essential design criterion for wiretap lattice codes. Please see Section~\ref{sec:secrecy-function_lattice} for an explicit description.

Another design criterion for wiretap lattice codes, called the \emph{flatness factor}, was proposed by Ling \emph{et al.}~\cite{LingLuzziBelfioreStehle14_1}. The flatness factor quantifies how much secret information can leak to Eve in terms of mutual information, while the secrecy gain characterizes the success probability for Eve to correctly guess the transmitted messages. Both the secrecy gain and the flatness factor require the minimization of the theta series of the lattice designed to confuse Eve at a given point to guarantee secrecy-goodness~\cite{LingLuzziBelfioreStehle14_1, LinLingBelfiore14_1}.  Even though this work is mainly focused on the secrecy gain, we also provide a brief analysis of the flatness factor.

Secrecy gains of the so-called \emph{unimodular} lattices have been studied for well over a decade~\cite{BelfioreOggier10_1}. In this pioneering work, Belfiore and Sol{\'{e}} discovered that there exists a symmetry point in the secrecy functions. Further, they conjectured that for unimodular lattices, the secrecy gain is achieved at the symmetry point of its secrecy function. The conjecture has been further investigated and verified for unimodular (or \emph{isodual}) lattices in dimensions less than $80$~\cite{OggierSoleBelfiore16_1, Ernvall-Hytonen12_1, LinOggier13_1, Pinchak13_1}. The study of secrecy gain was recently also extended to the \emph{$\ell$-modular lattices}~\cite{OggierSoleBelfiore16_1, LinOggierSole15_1, OggierBelfiore18_1}, where it is believed that the higher the parameter $\ell$ is, the better secrecy gain we can achieve. Most recently, a new family of lattices, called \emph{formally unimodular lattices}, or lattices with the same theta series as their dual, was introduced~\cite{BollaufLinYtrehus22_1, BollaufLinYtrehus23_3}.\footnote{A \emph{formally-self dual} code has the same weight enumerator as its dual.} It was shown that formally unimodular lattices have the same symmetry point as unimodular and isodual lattices, and the Construction A lattices obtained from the formally self-dual codes can achieve a higher secrecy gain than the unimodular lattices. Moreover, for formally unimodular lattices obtained by Construction A from even formally self-dual codes,\footnote{In an \emph{even} code, all codewords have even weights. Otherwise, the code is \emph{odd}.} a sufficient condition to verify Belfiore and Sol{\'{e}}'s conjecture on the secrecy gain was also provided.

This paper especially focuses on the original analysis of the secrecy gain for formally unimodular lattices obtained by Construction $\textnormal{A}_4$ from  linear codes over the ring $\Integers_4\eqdef\{0,1,2,3\}$ (also called \emph{quaternary codes})~\cite{ConwaySloane93_1, HammonsKumarCalderbankSloaneSole94_1, BonnecazeSoleCalderbank95_1, Wan97_1}. The main contributions are listed as follows.
\begin{enumerate}[nosep,label=\roman*)]
\item For codes over $\Integers_4$, a code is said to be \emph{formally self-dual} if it has the same \emph{symmetrized weight enumerator (swe)} as its dual. We show that if $\code{C}$ is formally self-dual, then its corresponding Construction $\textnormal{A}_4$ lattice is formally unimodular. Moreover, we introduce the concepts of Type I/II formally self-dual codes/unimodular lattices.

\item To study the secrecy gain of Construction $\textnormal{A}_4$ lattices from formally self-dual $\Integers_4$-linear codes, we present new code constructions of formally self-dual $\Integers_4$-linear codes with respect to swe. Little is known about the code construction in the literature~\cite{GulliverHarada01_1, BetsumiyaHarada03_1, YooLeeKim17_1}. For even lengths, formally self-dual $\Integers_4$-linear codes are constructed from nested binary codes and double circulant matrices. A sufficient condition for two nested binary codes to construct a formally self-dual code $\Integers_4$-linear is provided (see Theorem~\ref{thm:FSD-Z4codes_A1plus2A2}).

\item It is important to emphasize that, unlike binary codes, codes over $\Integers_4$ admit self-dual (and formally self-dual) codes of \emph{odd length}. By combining the generator matrix's standard form of a $\Integers_4$-linear code and the double circulant construction, a novel construction called \emph{odd extension} based on double circulant matrices is proposed for odd-length formally self-dual codes (see Section~\ref{sec:double-circulant-its-odd-extension}).
  
\item The theta series of $2$-level Construction C and Construction $\textnormal{A}_4$ lattices are discussed in Section~\ref{sec:theta-series_2-level-ConstrC-ConstructionAfour-lattices}, and the expressions of the corresponding theta series are derived in terms of Jacobi theta functions (see Theorem~\ref{thm:theta-series_2-level-constructionC}).
  
\item Based on the theta series expression derived from Section~\ref{sec:theta-series_2-level-ConstrC-ConstructionAfour-lattices}, we provide a novel and universal approach to determine the secrecy gain for Construction $\textnormal{A}_4$ lattices obtained from formally self-dual codes over $\Integers_4$ (see Theorem~\ref{thm:inv_secrecy-function_SymmetrizedWeightEnumerator}). Moreover, we provide a sufficient condition to verify the Belfiore and Sol{\'{e}} conjecture on the (strong) secrecy gain for Construction $\textnormal{A}_4$ formally unimodular lattices obtained from Type I formally self-dual codes over $\Integers_4$ (see Theorem~\ref{thm:strong-secrecy-gain_TypeI-FSDcodes_Z4}). An upper bound on the secrecy gain of Type I formally unimodular lattices is also re-derived on a comparative basis (see Lemma~\ref{lem:upper-bound_SG_TypeI-FUL}).

  \ifthenelse{\boolean{short_version}}{}
  \item The flatness factor of formally unimodular lattices is briefly studied, focused on its relations to the secrecy gain. We have performed exhaustive code searches to find the \emph{best} even-length double circulant codes (DCCs) for lengths up to $20$. Numerical results are presented to indicate an interesting observation: if the best code is determined regarding secrecy gain, then such code is also the best in terms of the flatness factor. This observation demonstrates the usefulness of the secrecy gain design criterion for wiretap lattices codes (see Section~\ref{sec:flatness-factor_FU-lattices}). 
  
\item Finally, numerical results on the secrecy gain of Construction $\textnormal{A}_4$ lattices obtained from formally self-dual $\Integers_4$-linear codes are presented. The best even-length DCCs and odd extension codes with respect to secrecy gain found by exhaustive code searches are summarized in Table~\ref{tab:table_secrecy-gains_FU-lattices_z4_summary}. 
  For even dimensions $2\leq n\leq 20$, we note that the Construction $\textnormal{A}_4$ lattices obtained from DCCs generally achieve better secrecy gain than the ones obtained from nested binary codes. For odd dimensions, we also demonstrate the high secrecy gain of the newly presented odd extension codes.
\end{enumerate}

To the best of our knowledge, most of the efforts to solve Belfiore and Sol{\'{e}}'s conjecture on the secrecy gain of formally unimodular lattices have only been based on lattices obtained by Construction A from binary codes. The investigation of Construction $\textnormal{A}_4$ lattices obtained from formally self-dual codes over $\Integers_4$ has not been addressed in the previous literature. Furthermore, it is worth mentioning that previous contributions were mainly focused on even-dimensional lattices, while in this work, we extensively study \emph{odd-dimensional} formally unimodular lattices obtained from formally self-dual codes over $\Integers_4$.

\section{Definitions and Preliminaries}
\label{sec:definitions-preliminaries}

\subsection{Notation}
\label{sec:notation}

We denote by $\Naturals$, $\Integers$, $\Rationals$, and $\Reals$ the set of naturals, integers, rationals, and reals, respectively. $[i:j]\eqdef\{i,i+1,\ldots,j\}$ for $i,j\in\Integers$, $i\leq j$. Vectors are \emph{row} vectors and are represented by boldfaced, e.g., $\vect{x}$. The all-zero vector is denoted by $\vect{0}$. Matrices and sets are represented by capital sans serif letters and calligraphic uppercase letters, respectively, e.g., $\mat{X}$ and $\set{X}$. $\mat{X}_{k\times n}$ represents a matrix of size $k\times n$, and a square matrix of size $n$ is denoted by $\mat{X}_n$. We omit the subscript of a matrix if it is clearly understood from the context. An identity matrix and an all-zero matrix are denoted by $\mat{I}$ and $\mat{O}$, respectively. We denote by, respectively, $\Hwt{\vect{x}}$, $\Lwt{\vect{x}}$, and $\Ewt{\vect{x}}$ the \emph{Hamming} weight, the \emph{Lee} weight, and the \emph{Euclidean} weight of a vector $\vect{x}\in\Integers_m^n$ or $\Integers^n$, where $\Integers_m=\{0,\dots,m-1\}$ is the ring of integers modulo $m$. In this work, $m$ can be $2$ or $4$. We use the code parameters $[n,M]$, $[n,M,d_{\textnormal{Lee}}]$, or $[n,M,d_{\textnormal{Lee}},d_\textnormal{E}]$ to denote a linear code $\code{C}$ of length $n$, $M$ codewords, minimum \emph{Lee distance} $d_{\textnormal{Lee}}\eqdef\min_{\vect{x},\vect{y}\in\code{C}}\Lwt{\vect{x}-\vect{y}}$, and minimum squared Euclidean distance $d_{\textnormal{E}}\eqdef\min_{\vect{x},\vect{y}\in\code{C}}\Ewt{\vect{x}-\vect{y}}$.\footnote{For general readers, see~\cite[Ch.~2.1]{NebeRainsSloane06_1} for the detailed definitions of weights.} A generator matrix of a code $\code{C}$ is represented by $\mat{G}^{\code{C}}$, while $\code{C}^{\mat{G}}$ represents the corresponding linear code generated by $\mat{G}$. $\phi_m\colon\Integers^n_m \rightarrow \Integers^n$ is defined as the natural embedding, i.e., the elements of $\Integers_m$ are mapped to the respective integer by $\phi_m$ element-wisely. $\trans{(\cdot)}$ represents the transpose of its argument and $\inner{\vect{x}}{\vect{y}}$ denotes the inner product and the $\vect{x}\circ\vect{y}$ represents the element-wise (Hadamard/Schur) product between two vectors over $\Integers_m$, respectively.

\subsection{Basics on Codes and Lattices}
\label{sec:basics_codes-lattices}

We recall some definitions of codes over $\Integers_m$, $m\in\{2,4\}$, and lattices.

Let $\code{A}$ be an $[n,M]$ binary code. Its \emph{weight enumerator} is 
\begin{IEEEeqnarray*}{c}
  W_{\code{A}}(x,y)=\sum_{\vect{a}\in\code{A}} x^{n-\Hwt{\vect{a}}}y^{\Hwt{\vect{a}}}.
\end{IEEEeqnarray*}

Let $\code{A}_1,\code{A}_2$ be two binary linear codes. For $\vect{a}_1=(a_{1,1},\ldots,a_{1,n}) \in \code{A}_1, \vect{a}_2=(a_{2,1},\ldots,a_{2,n}) \in \code{A}_2$, we define 
\begin{IEEEeqnarray}{rCl}
  \IEEEyesnumber
  \IEEEyessubnumber*
  d_{0,0}(\vect{a}_1,\vect{a}_2)& \eqdef & \card{\{j\in [1:n]\colon (a_{1,j},a_{2,j}) = (0,0)\}},  \label{eq:def_d00}
  \\ [1mm]
  d_{0,1}(\vect{a}_1,\vect{a}_2) & \eqdef & \card{\{j\in [1:n]\colon (a_{1,j}, a_{2,j})=(0,1)\}},  \label{eq:def_d01}
  \\[1mm]
  d_{1,0}(\vect{a}_1,\vect{a}_2) & \eqdef & \card{\{j\in [1:n]\colon (a_{1,j},a_{2,j}) = (1,0)\}},  \label{eq:def_d10}
  \\[1mm]
  d_{1,1}(\vect{a}_1,\vect{a}_2) & \eqdef &\card{\{j\in [1:n]\colon (a_{1,j}, a_{2,j})=(1,1)\}} \label{eq:def_d11}.
\end{IEEEeqnarray}
Observe that $d_{0,0}(\vect{a}_1,\vect{a}_2) + d_{0,1}(\vect{a}_1,\vect{a}_2) + d_{1,0}(\vect{a}_1,\vect{a}_2)+ d_{1,1}(\vect{a}_1,\vect{a}_2) = n$. 
The \emph{joint weight enumerator} of $\code{A}_1$ and $\code{A}_2$ is given by
\begin{IEEEeqnarray}{c}
  \jwe{\code{A}_1}{\code{A}_2}(a,b,c,d)\eqdef\sum_{\vect{a}_1 \in \code{A}_1} \sum_{\vect{a}_2 \in \code{A}_2} a^{d_{0,0}} b^{d_{0,1}} c^{d_{1,0}} d^{d_{1,1}},
  \label{eq:def_jwe}
\end{IEEEeqnarray}
where we use the shorthand $d_{i,j}$ for $d_{i,j}(\vect{a}_1,\vect{a}_2)$ defined in~\eqref{eq:def_d00}-\eqref{eq:def_d11}. Detailed properties and MacWilliams identities of the joint weight enumerator can be found in~\cite[Ch.~5, pp.~147--149]{MacWilliamsSloane77_1}.

A \emph{${\Integers_4}$-linear code} $\code{C}$ of length $n$ is an additive subgroup of $\Integers_4^n$. If $\code{C}$ is a $\Integers_4$-linear code of length $n$, then $\code{C}^\perp\eqdef\{\vect{x}\in\Integers_4^n\colon\inner{\vect{x}}{\vect{y}} = 0,\,\forall\,\vect{y}\in\code{C}\}$ is the \emph{dual code} of $\code{C}$. 
    
From~\cite[Prop.~1.1]{Wan97_1}, it is well-known that any $\Integers_4$-linear code is \emph{permutation equivalent} to a code $\code{C}$ with a generator matrix $\mat{G}^{\code{C}}$ in \emph{standard form}
\begin{IEEEeqnarray}{c}
\label{eq:generator_cs}
  \mat{G}^{\code{C}}=\begin{pNiceMatrix}
    \mat{I}_{k_1} & \mat{A} & \mat{B} \\
    \mat{O}_{k_2 \times k_1} & 2\mat{I}_{k_2} & 2\mat{C}
    \label{eq:Generator_Matrix}
  \end{pNiceMatrix},
\end{IEEEeqnarray}
where $\mat{A}$ and $\mat{C}$ are binary matrices, and $\mat{B}$ is defined over $\Integers_4$. Such a code $\code{C}$ is said to be a code of \emph{type $4^{k_1}2^{k_2}$} and $\code{C}$ contains $2^{2k_1+k_2}$ codewords. Also, the dual code $\code{C}^\perp$ of $\code{C}$ has generator matrix
\begin{IEEEeqnarray}{C}
  \mat{G}^{\code{C}^\perp}=
  \begin{pNiceMatrix}
    -\trans{\mat{B}} - \trans{\mat{C}}\trans{\mat{A}} & \trans{\mat{C}} & \mat{I}_{n-k_1-k_2} \\
    2\trans{\mat{A}} & 2\mat{I}_{k_2} & \mat{O}_{k_2 \times (n-k_1-k_2)}
    \label{eq:Generator_Matrix_Dual}
  \end{pNiceMatrix},
\end{IEEEeqnarray}
and type $4^{n-k_1-k_2}2^{k_2}$ and $\code{C}^\perp$ contains $2^{2n-2k_1-k_2}$.

The \emph{symmetrized weight enumerator (swe)} of a $\Integers_4$-linear code $\code{C}$ is defined as
\begin{IEEEeqnarray*}{c}
  \swe{\code{C}}(a,b,c)=\sum_{\vect{c}\in\code{C}} a^{n_0(\vect{c})}b^{n_1(\vect{c})+n_3(\vect{c})}c^{n_2(\vect{c})},
\end{IEEEeqnarray*}
where $n_i(\vect{c})\eqdef\card{\{j\in [1:n]\colon c_j=i\}}$, $i\in\Integers_4$.\footnote{The exponent of $b$ combines weights $1$ and $3$ according to the Lee distance definition.} The corresponding MacWilliams identity for $\Integers_4$-linear codes is given by~\cite[Th.~2.3]{Wan97_1}
\begin{IEEEeqnarray}{c}
    \swe{\code{C}}(a,b,c)
   = \frac{1}{\card{\dual{\code{C}}}}\swe{\dual{\code{C}}}(a+2b+c,a-c,a-2b+c). 
  \label{eq:swe-MacWilliams-identity_Z4}
\end{IEEEeqnarray}

Following the notion of swe, we have the following families of codes over $\Integers_4$.
\begin{definition}[Self-dual, isodual, formally self-dual codes]
  \begin{itemize}
  \item If $\code{C}=\dual{\code{C}}$, $\code{C}$ is a \emph{self-dual} code.
  \item If there is a permutation of coordinates and a (possible) change of signs carried out by a mapping $\pi$, such that $\code{C}=\pi(\dual{\code{C}})$, $\code{C}$ is called \emph{isodual}.
  \item If $\code{C}$ and $\dual{\code{C}}$ have the same swe, i.e., $\swe{\code{C}}(a,b,c)=\swe{\dual{\code{C}}}(a,b,c)$, $\code{C}$ is a \emph{formally self-dual} code.
  \end{itemize}
\end{definition}

From~\eqref{eq:swe-MacWilliams-identity_Z4}, we can conclude that a code in any of these classes has its swe satisfying
\begin{IEEEeqnarray}{c}
  \swe{\code{C}}(a,b,c)
  =\frac{1}{\card{{\code{C}}}}\swe{\code{C}}(a+2b+c,a-c,a-2b+c).
  \IEEEeqnarraynumspace\label{eq:swe-MacWilliams-identity_FSD-codes_Z4}
\end{IEEEeqnarray}

A (full rank) \emph{lattice} $\Lambda\subset\Reals^n$ is a discrete additive subgroup of $\Reals^{n}$, and it can be seen as
$\Lambda=\{\vect{\lambda}=\vect{u}\mat{L}_{n\times n}\colon\vect{u}\in\Integers^n\}$,
where the $n$ rows of $\mat{L}$ form a lattice basis in $\Reals^n$. 
The \emph{volume} of $\Lambda$ is $\vol{\Lambda} = \ecard{\det(\mat{L})}$. \ifthenelse{\boolean{short_version}}{} A \emph{fundamental region} $\set{R}(\Lambda)$ of a lattice $\Lambda$ is a bounded set such that $\bigcup_{\vect{\lambda} \in \Lambda} (\set{R}(\Lambda) + \vect{\lambda}) = \Reals^n$ and $(\set{R}(\Lambda) + \vect{\lambda}) \cap (\set{R}(\Lambda) + \vect{\lambda}')=\emptyset$ for any $\vect{\lambda} \neq \vect{\lambda}' \in \Lambda$.

If a lattice $\Lambda$ has generator matrix $\mat{L}$, then the lattice $\Lambda^\star\subset\Reals^n$ generated by $\trans{\bigl(\inv{\mat{L}}\bigr)}$ is called the \emph{dual lattice} of $\Lambda$. For lattices, the analogue of the weight enumerator of a code is the \emph{theta series}.
\begin{definition}[Theta series]
  \label{def:theta-series}
  Let $\Lambda$ be a lattice, its \emph{theta series} is given by
  \begin{IEEEeqnarray*}{c}
    \Theta_\Lambda(z) = \sum_{{\bm \lambda} \in \Lambda} q^{\norm{\vect{\lambda}}^2},
  \end{IEEEeqnarray*}
  where $q\eqdef\ope^{i\pi z}$ and $\Im{z} > 0$. 
\end{definition}

Analogously, the spirit of the MacWilliams identity can be captured by the \emph{Jacobi's formula}~\cite[eq.~(19), Ch.~4]{ConwaySloane99_1}
\begin{IEEEeqnarray}{c}
  \Theta_{\Lambda}(z)=\vol{\Lambda^\star}\Bigl(\frac{i}{z}\Bigr)^{\frac{n}{2}}\Theta_{\Lambda^\star}\Bigl(-\frac{1}{z}\Bigr).
  \label{eq:Jacobi-formula}
\end{IEEEeqnarray}

In some particular cases, the theta series of a lattice can be expressed in terms of the \emph{Jacobi theta functions} defined as follows.
\begin{IEEEeqnarray*}{rCl}
  \vartheta_2(z)& \eqdef &\sum_{m\in\Integers} q^{\bigl(m+\frac{1}{2}\bigr)^2}=\Theta_{\mathbb{Z} + \frac{1}{2}}(z),
  \nonumber\\
  \vartheta_3(z)& \eqdef &\sum_{m \in \mathbb{Z}} q^{m^2}=\Theta_{\mathbb{Z}}(z)
  = 1 + 2q + 2q^{4} +  2q^{9} + 2q^{16} + 2q^{25} + \cdots
  \IEEEyesnumber\label{eq:Jacobi-theta3-function}\\
  \vartheta_4(z)& \eqdef &\sum_{m\in \mathbb{Z}} (-q)^{m^2}.
\end{IEEEeqnarray*}

A lattice is said to be \emph{integral} if the inner product of any two lattice vectors is an integer. An integral lattice such that $\Lambda = \Lambda^\star$ is called a \emph{unimodular} lattice. A lattice $\Lambda$ is called \emph{isodual} if it can be obtained from its dual $\Lambda^\star$ by (possibly) a rotation or reflection. In~\cite{BollaufLinYtrehus22_1}, a new and broader family was presented, namely, the \emph{formally unimodular lattices}, that consists of lattices having the same theta series as their duals, i.e., $\Theta_{\Lambda}(z)=\Theta_{\Lambda^\star}(z)$. We remark here that isodual and formally unimodular lattices are not necessarily integral.

Lattices can be constructed from binary linear codes through the so-called Constructions A and C~\cite{ConwaySloane99_1}.
\begin{definition}[Construction A]
  \label{def:def_ConstrA}
  Let $\code{A}$ be a binary $[n,M]$ code, then $\ConstrA{\code{A}}\eqdef\frac{1}{\sqrt{2}}(\phi_2(\code{A}) + 2\Integers^n)$ is a lattice.
\end{definition}

To introduce a multilevel construction, we need the auxiliary definition notion of a chain of codes being closed under the element-wise (also known in the literature as Schur or Hadamard) product.

\begin{definition}[Chain closed under element-wise product] The chain $\code{A}_1 \subseteq \code{A}_2$ is called \emph{closed under the element-wise product} if for all $\vect{a}_1,\vect{a}'_1\in \code{A}_1$, we have $\vect{a}_1\circ \vect{a}'_1 = (a_{1,1}a_{1,1}', \dots, a_{1,n}a_{1,n}')\in\code{A}_{2}$.
\end{definition}

\begin{definition}[$2$-level Construction C]
  \label{def:def_ConstrC}
  Let $\code{A}_1,\code{A}_2$ be two binary linear codes and $\code{A}_1 \subseteq \code{A}_2$. If the chain $\code{A}_1 \subseteq \code{A}_2$ is closed under the element-wise product, then 
  \begin{IEEEeqnarray}{c}
    \label{eq:construction_c}
    \ConstrC{\code{A}_1}{\code{A}_2}\eqdef\phi_2(\code{A}_1) + 2\phi_2(\code{A}_2) + 4\Integers^n
    \label{eq:ConstrC}
  \end{IEEEeqnarray}
  generates a lattice~\cite{KositwattanarerkOggier14_1}.
\end{definition}

For general choices of $\code{A}_1$ and  $\code{A}_2$, \eqref{eq:ConstrC} is a nonlattice packing, and we will denote by $\PConstrC{\code{A}_1}{\code{A}_2}$. If $\code{A}_1$ is the zero code, and $\code{A}_2$ is linear, then $\ConstrC{\code{A}_1}{\code{A}_2}=2\sqrt{2}\ConstrA{\code{A}_2}$. If $\code{A}_2$ is the universe code $\mathbb{F}_2^n$ and $\code{A}_1$ is linear, then $\ConstrC{\code{A}_1}{\code{A}_2}=\sqrt{2}\ConstrA{\code{A}_1}$. Under the assumptions of Definition~\ref{def:def_ConstrC}, Constructions C and D coincide, but we have chosen to work with the above characterization since some of our further results could be generalized to nonlattice cases (see Theorem~\ref{thm:theta-series_2-level-constructionC}, for example).

    
A packing $\Gamma\subset\Reals^n$ is \emph{geometrically uniform} if for any two elements $\vect{x},\vect{x}'\in\Gamma$, there exists an isometry $T_{\vect{x},\vect{x}'}$ such that $\vect{x}'=T_{\vect{x},\vect{x}'}(\vect{x})$ and $T_{\vect{x},\vect{x}'}(\Gamma)\eqdef\{T_{\vect{x},\vect{x}'}(\vect{x}'')\colon\vect{x}''\in\Gamma\}=\Gamma$. It was demonstrated that $\PConstrC{\code{A}_1}{\code{A}_2}$ is geometrically uniform~\cite{Forney91_1,BollaufZamir16_1}, for linear codes $\code{A}_1$ and $\code{A}_2$.

There is an analogue of Construction A for codes over $\Integers_4$, which is called \emph{Construction $\textnormal{A}_4$}~\cite[Ch.~12.5.3]{HuffmanPless03_1}.
\begin{definition}[Construction $\textnormal{A}_4$]
  \label{def:def_ConstrAfour}
  If $\code{C}$ is a $\Integers_4$-linear code, then $\ConstrAfour{\code{C}}=\frac{1}{2}(\phi_4(\code{C})+4\mathbb{Z}^n)$ is a lattice.
\end{definition}

It is known that $\ConstrAfour{\code{C}}$ is a unimodular lattice if and only if the $\Integers_4$-linear code $\code{C}$ is self-dual~\cite[Prop.~12.2]{Wan97_1}. For notational convenience, from now on, the mapping $\phi_m$ is omitted. 

The following example illustrates Construction $\textnormal{A}_4$ of the octacode.
\begin{example}
  \label{ex:E8_octacode}
  The self-dual $\Integers_4$-linear code, known as the octacode $\code{O}_8$, is generated by $\mat{G} =(\mat{I}_{4}\,\,\,\mat{B}_{4})$, where 
  \begin{IEEEeqnarray*}{c}
    \mat{B}_{4} =
    \begin{pmatrix}
      3 & 1 & 2 & 1 \\
      1 & 2 & 3 & 1 \\
      3 & 3 & 3 & 2 \\
      2 & 3 & 1 & 1
    \end{pmatrix}.  
  \end{IEEEeqnarray*}
  It is of type $4^4$ and its swe~\cite[Ex.~12.5.13]{HuffmanPless03_1} is given by
  \begin{IEEEeqnarray}{c}
  \label{eq:swe_octacode}
    \swe{{\code{O}_8}}(a,b,c)
    = a^8+ 16b^8+ c^8+ 14a^4c^4+ 112a^3 b^4c + 112a b^4 c^3.
    \label{eq:swe_octacode}
  \end{IEEEeqnarray}
  A unimodular lattice can be constructed by performing $\widebar{\lattice{E}}_8=\Lambda_{\textnormal{A}_4}(\code{O}_8) = \frac{1}{2}\left(\code{O}_8+4\mathbb{Z}^8\right)$. It is equivalent to the well-known Gosset lattice $\lattice{E}_8$~\cite[Ex.~12.5.13]{HuffmanPless03_1}. Note that the theta series of the $\lattice{E}_8$ lattice in terms of the Jacobi theta functions is
  \begin{IEEEeqnarray}{rCl}
    \Theta_{\lattice{E}_8}(z)& = &\frac{1}{2}\bigl[\vartheta^8_2(z) + \vartheta^8_3(z)+\vartheta^8_4(z) \bigr]=\frac{1}{2}\bigl[(\vartheta^4_2(z))^2 + \vartheta^8_3(z)+\vartheta^8_4(z)\bigr] \nonumber \label{eq:theta-series-1_GossetE8}
    \\
    & \stackrel{\eqref{eq:useful-identities-2}}{=} &\frac{1}{2}\bigl[(\vartheta^4_3(z)-\vartheta^4_4(z))^2+\vartheta^8_3(z)+\vartheta^8_4(z)\bigr]=\vartheta_3(z)^8-\vartheta^4_3(z)\vartheta^4_4(z)+\vartheta^8_4(z)\IEEEeqnarraynumspace \nonumber \label{eq:theta-series-2_GossetE8}
    \\
    & = &1+240q^2+2160q^4+6720q^6+\cdots.
    \label{eq:theta-series-expension_GossetE8}
  \end{IEEEeqnarray}
  
  The $\lattice{E}_8$ lattice can also be constructed via the binary Construction A, using the $[8,4,4]$ extended Hamming code.\hfill\exampleend
\end{example}

\subsection{Type I Codes and Lattices}
\label{sec:TypeI-lattices}

In the coding theory and lattice literature, one can define a \emph{Type I} self-dual code and a Type I unimodular lattice. We briefly summarize the concepts below.
\begin{remark}[{\cite[Remark 2.3.1]{NebeRainsSloane06_1}}]
  \begin{itemize}
  \item Let $m\in\{2,4\}$. If a code over $\Integers_{m}$ is self-dual, all the Euclidean weights of codewords are divisible by $m$~\cite[Ths.~1.4.5 and 12.1.5]{HuffmanPless03_1}. A self-dual code is Type II if the Euclidean weight of every codeword is a multiple of $2m$. Otherwise, it is called a Type I self-dual code.
  \item A unimodular lattice must be integral. Hence, the inner product of any two lattice vectors is either even or odd. A unimodular lattice is Type II if the inner product of any two lattice points is even. Otherwise, it is called a Type I unimodular lattice~\cite[Sec.~2.4, Ch.~2]{ConwaySloane99_1}.
  \end{itemize}
\end{remark}

In this paper, since we work on the generalization of formally self-dual codes/unimodular lattices, we also adapt the Type I concepts for formally self-dual codes/unimodular lattices.

\begin{definition}
  \label{def:TypeI-FSD-codes}
  Let $\code{C}$ be a code over $\Integers_{m}$, where $m\in\{2,4\}$. A formally self-dual code is said to be of Type I if all its codewords have Euclidean weight divisible by $m$ and of Type II if all its codewords have Euclidean weights divisible by $2m$. Analogously, a formally unimodular lattice is of Type I if the inner product of any two lattice vectors is either even or odd and is of Type II if the inner product of any two lattice vectors is even.
\end{definition}

\section{Constructions of Formally Self-Dual $\Integers_4$-Linear Codes}
\label{sec:construction_FSD-Z4-codes}

In this section, we present constructions of formally self-dual $\Integers_4$-linear codes from binary codes and also from double circulant matrices, as well as an original construction of odd-length codes, denoted by the \emph{odd extension}.

\subsection{Formally Self-Dual $\Integers_4$-Linear Codes from Nested Binary Codes}
\label{sec:FSD-Z4-codes_nested-binary-codes}

In this subsection, we present a novel construction of formally self-dual codes over $\Integers_4$ via two nested binary linear codes $\code{A}_1\subseteq\code{A}_2$.

\begin{proposition}[{\cite[Lemma 2.1]{BonnecazeSoleCalderbank95_1}}]
  \label{prop:Z4-linear_ConsC}
  Consider two binary linear codes $\code{A}_1\subseteq\code{A}_2$, and let $\code{C}=\code{A}_1+2\code{A}_2\eqdef\{\vect{a}_1+2\vect{a}_2\colon\vect{a}_1\in\code{A}_1,\vect{a}_2\in\code{A}_2\}$. Then, the code $\code{C}$ over $\Integers_4$ is linear if and only if $\code{A}_1 \subseteq \code{A}_2$ is closed under the element-wise product.
\end{proposition}
On one hand, the condition that the chain $\code{A}_1 \subseteq \code{A}_2$ is closed under element-wise product guarantees that  $\ConstrC{\code{A}_1}{\code{A}_2}$ as in~\eqref{eq:construction_c} is a lattice. 
On the other hand, $\code{C}$ being $\Integers_4$-linear assures that $\ConstrAfour{\code{C}}$ is a lattice. Therefore, Proposition~\ref{prop:Z4-linear_ConsC} standardizes the $2$-level Construction C and Construction $\textnormal{A}_4$, together with their respective conditions to be a lattice.

The dual of a $\Integers_4$-linear code $\code{C}=\code{A}_1+2\code{A}_2$ can be described as follows.
\begin{lemma}
  \label{lem:dual_C1-2C2}
  Let $\code{C} = \code{A}_1 + 2\code{A}_2$ be a $\Integers_4$-linear code. Then, $\code{C}^\perp = \code{A}_2^\perp + 2\code{A}_1^\perp$.
\end{lemma}
\begin{IEEEproof}
  First, notice that $\code{A}_2^\perp \subseteq \code{A}_1^\perp$. Since $\code{C} = \code{A}_1 + 2\code{A}_2$ is linear over $\Integers_4$, $\code{A}_1 \subseteq \code{A}_2$ is closed under element-wise product. We next show that $\code{A}_2^\perp + 2\code{A}_1^\perp \subseteq \code{C}^\perp$. Consider an element $\vect{a}'_2 + 2\vect{a}'_1 \in \code{A}_2^\perp + 2\code{A}_1^\perp$ and $\vect{a}_1 + 2\vect{a}_2 \in \code{C}$. Then, 
  \begin{IEEEeqnarray}{c}
    \inner{\vect{a}'_2 + 2\vect{a}'_1}{\vect{a}_1 + 2\vect{a}_2}=\inner{\vect{a}'_2}{\vect{a}_1}+2\inner{\vect{a}'_2}{\vect{a}_2}+ 2\inner{\vect{a}'_1}{\vect{a}_1}+ 4\inner{\vect{a}'_1}{\vect{a}_2}\equiv 0 \pmod 4.
  \end{IEEEeqnarray}
  Therefore, $\vect{a}'_2 + 2\vect{a}'_1 \in \code{C}^\perp$. Now, based on the arguments presented in \cite[pp.~33--34]{ConwaySloane93_1}, we observe that $\card{\code{C}} \card{\code{C}^\perp} = \card{\code{A}_1}\card{\code{A}_2}\card{\code{A}_1^\perp}\card{\code{A}_2^\perp} = 2^{k_1}2^{k_2}2^{n-k_1}2^{n-k_2} = 2^{2n} = 4^n$, which is the dimension of $\Integers_4^n$ and here, $k_1$ is the dimension of $\code{A}_1$ and $k_2$ is the dimension of $\code{A}_2$. The proof is then complete.
\end{IEEEproof}

Lemma~\ref{lem:dual_C1-2C2} implies that if $\code{A}_2 = \code{A}_1^\perp$, then $\code{C}=\code{C}^\perp$ and $\code{C}$ is self-dual.

The result below gives a condition to construct formally self-dual $\Integers_4$-linear codes.    
\begin{theorem}
  \label{thm:FSD-Z4codes_A1plus2A2}
  Let $\code{C} = \code{A}_1 + 2\code{A}_2$ be a $\Integers_4$-linear code. If $W_{\code{A}_1}(x,y) = W_{\code{A}_2^\perp}(x,y)$ and $W_{\code{A}_2}(x,y) = W_{\code{A}_1^\perp}(x,y)$, then $\code{C}$ is formally self-dual. 
\end{theorem}
\begin{IEEEproof}
  We start the proof by using the following useful identities~\cite[Ch.~5, pp.~148]{MacWilliamsSloane77_1}:
  \begin{IEEEeqnarray}{rCl}
    \we{\code{A}_1}(x,y)\we{\code{A}_2}(z,t)& = &\jwe{\code{A}_1}{\code{A}_2}(xz,xt,yz,yt),
    \label{eq:wes-jwe_C1-C2}
    \\
    \jwe{\code{A}_1}{\code{A}_2}(a,b,c,d)& = &\jwe{\code{A}_2}{\code{A}_1}(a,c,b,d).
    \label{eq:swap_jwe_C1-C2}
    \\
    \jwe{\dual{\code{A}}_1}{\dual{\code{A}}_2}(a,b,c,d)  &  = &  \frac{1}{|\code{A}_1|~|\code{A}_2|}\jwe{\code{A}_1}{\code{A}_2}\big( a+b+c+d,a-b+c-d,\nonumber\\
    && \hspace*{3.25cm}\>  a+b-c-d,a-b-c+d \bigr).
    \label{eq:mac-williams-jwe}
  \end{IEEEeqnarray}
  Observe that
  \begin{IEEEeqnarray}{rCl}
    \IEEEeqnarraymulticol{3}{l}{%
      \jwe{\code{A}_1}{\code{A}_2}(xz,xt,yz,yt)}\nonumber\\*\quad%
    & = &\we{\code{A}_1}(x,y)\we{\code{A}_2}(z,t)
    \nonumber\\
    & \stackrel{(i)}{=} &\frac{1}{\card{\dual{\code{A}}_1}}\we{\dual{\code{A}}_1}(x+y,x-y)\frac{1}{\card{\dual{\code{A}}_2}}\we{\dual{\code{A}}_2}(z+t,z-t)
    \nonumber\\
    & \stackrel{(ii)}{=} &\frac{1}{\card{\dual{\code{A}}_1}}\we{\code{A}_2}(x+y,x-y)\frac{1}{\card{\dual{\code{A}}_2}}\we{\code{A}_1}(z+t,z-t)
    \nonumber\\
    & \stackrel{\eqref{eq:wes-jwe_C1-C2}}{=} &\frac{1}{\card{\dual{\code{A}}_2}\card{\dual{\code{A}}_1}}\jwe{\code{A}_2}{\code{A}_1}\bigl(xz+xt+yz+yt, xz-xt+yz-yt,
    \nonumber\\
    && \hspace*{3.50cm}\> xz+xt-yz-yt, xz-xt-yz+yt \bigr) \nonumber
    \nonumber\\
    & \stackrel{\eqref{eq:swap_jwe_C1-C2}}{=} &\frac{1}{\card{\dual{\code{A}}_1}\card{\dual{\code{A}}_2}}\jwe{\code{A}_1}{\code{A}_2}\bigl(xz+xt+yz+yt,xz+xt-yz-yt,
    \nonumber\\
    && \hspace*{3.50cm}\> xz-xt+yz-yt, xz-xt-yz+yt \bigr) 
    \IEEEeqnarraynumspace\label{eq:expansion-jwe}
  \end{IEEEeqnarray}
  where $(i)$ follows by the MacWilliams identity and $(ii)$ holds because $W_{\code{A}_2}(x,y) = W_{\dual{\code{A}}_1}(x,y)$ and $W_{\code{A}_1}(z,t) = W_{\dual{\code{A}}_2}(z,t)$. 

  Since $\dual{\code{C}}=\dual{\code{A}_2}+2\dual{\code{A}_1}$ by Lemma~\ref{lem:dual_C1-2C2}, the goal in this proof is to prove $\swe{\code{C}}(a,b,c) = \swe{\code{A}_1+2\code{A}_2}(a,b,c) = \swe{\dual{\code{A}_2}+2\dual{\code{A}_1}}(a,b,c) = \swe{\dual{\code{C}}}(a,b,c)$ for $\code{C}=\code{A}_1+2\code{A}_2$ to be formally self-dual. As $\we{\code{A}_1}(x,y)$ and $\we{\code{A}_2}(z,t)$, respectively, count the numbers of $0s$ and $1s$ in $\code{A}_1$ and $\code{A}_2$, for the particular case of $\code{C}=\code{A}_1+2\code{A}_2$, we should consider the following counting variables
  \begin{IEEEeqnarray*}{c}
    a = xz, b=yz=yt, \textnormal{ and } c=xt.    
  \end{IEEEeqnarray*}
  Thus, $\swe{\code{C}}(a,b,c) = \swe{\code{C}}(xz,xt,yz) = \swe{\code{A}_1+2\code{A}_2}(xz,xt,yz) = \jwe{\code{A}_1}{\code{A}_2}(xz,yz,xt,xt)$ and it is enough to demonstrate that
  \begin{IEEEeqnarray*}{c}
    \jwe{\code{A}_1}{\code{A}_2}(xz,yz,xt,xt) = \jwe{\dual{\code{A}_2}}{\dual{\code{A}_1}}(xz,yz,xt,xt).
  \end{IEEEeqnarray*}

  Indeed, from~\eqref{eq:expansion-jwe},
  \begin{IEEEeqnarray}{rCl}
    \IEEEeqnarraymulticol{3}{l}{%
      \jwe{\code{A}_1}{\code{A}_2}(xz,yz,xt,xt)}\nonumber\\*\quad%
    & = &\frac{1}{\card{\dual{\code{A}}_1}\card{\dual{\code{A}}_2}}\jwe{\code{A}_1}{\code{A}_2}\bigl(xz+yz+2xt,xz+yz-2xt, xz-yz, xz-yz \bigr) \nonumber\\
    & \stackrel{\eqref{eq:swap_jwe_C1-C2}}{=} &\frac{1}{\card{\dual{\code{A}}_1}\card{\dual{\code{A}}_2}}\jwe{\code{A}_2}{\code{A}_1}\bigl(xz+yz+2xt, xz-yz, xz+yz-2xt,  xz-yz \bigr) \nonumber \\
    & \stackrel{\eqref{eq:mac-williams-jwe}}{=}
    &\left(\frac{1}{\card{\dual{\code{A}}_1}\card{\dual{\code{A}}_2}}\right)^2\jwe{\dual{\code{A}}_2}{\dual{\code{A}}_1}\bigl(4xz, 4yz, 4xt, 4xt \bigr) \nonumber \\
    & \stackrel{(iii)}{=}
    & \jwe{\dual{\code{A}}_2}{\dual{\code{A}}_1}\bigl(xz, yz, xt, xt \bigr),
     \IEEEeqnarraynumspace
  \label{eq:fsd-from-jwe}
  \end{IEEEeqnarray}
where $(iii)$ follows from the hypothesis, provided that $W_{\code{A}_1}(x,y) = W_{\code{A}_2^\perp}(x,y)$ and $W_{\code{A}_2}(x,y) = W_{\code{A}_1^\perp}(x,y)$, then $\card{\code{A}_1} = \card{\dual{\code{A}}_2}$, $\card{\code{A}_2} = \card{\dual{\code{A}}_1}$ and $\card{\code{A}_1}\card{\code{A}_2}\card{\code{A}_1^\perp}\card{\code{A}_2^\perp}=4^n$. Therefore,
\begin{IEEEeqnarray*}{c}
\swe{\code{A}_1+2\code{A}_2}(xz,xt,yz) = \swe{\dual{\code{A}_2}+2\dual{\code{A}_1}}(xz,xt,yz)
\end{IEEEeqnarray*}
and the proof is complete.
\end{IEEEproof}

We remark that the result of Theorem~\ref{thm:FSD-Z4codes_A1plus2A2} holds only if $\code{A}_1$ and $\code{A}_2$ have even length, due to the restriction on their weight enumerators.

\begin{example}
  \label{ex:codes_dim12}
  Consider $\code{A}_1$ as the $[12,2,8]$ binary code and $\code{A}_2$ as the $[12,10,2]$ binary code, generated respectively by
  \begin{IEEEeqnarray*}{c}
    \mat{G}^{\code{A}_1}=
    \left(\begin{smallmatrix}
      1 & 0 & 1 & 1 & 1 & 0 & 0 & 0 & 1 & 1 & 1 & 1
      \\
      0 & 1 & 0 & 0 & 0 & 1 & 1 & 1 & 1 & 1 & 1 & 1
    \end{smallmatrix}\right),\,
    \mat{G}^{\code{A}_2}=
    \left(\begin{smallmatrix}
      1 & 0 & 1 & 1 & 1 & 0 & 0 & 0 & 1 & 1 & 1 & 1
      \\
      0 & 1 & 0 & 0 & 0 & 1 & 1 & 1 & 1 & 1 & 1 & 1
      \\
      0 & 0 & 0 & 0 & 0 & 0 & 0 & 0 & 1 & 1 & 1 & 1
      \\
      0 & 0 & 0 & 0 & 0 & 1 & 1 & 1 & 1 & 1 & 1 & 0
      \\
      0 & 1 & 1 & 1 & 1 & 0 & 0 & 0 & 0 & 1 & 1 & 1
      \\
      0 & 0 & 0 & 1 & 1 & 0 & 0 & 0 & 1 & 1 & 1 & 1
      \\
      0 & 0 & 0 & 0 & 0 & 0 & 1 & 1 & 1 & 1 & 1 & 1
      \\
      0 & 0 & 0 & 0 & 0 & 0 & 0 & 1 & 1 & 0 & 0 & 0
      \\
      0 & 0 & 0 & 0 & 0 & 0 & 0 & 0 & 1 & 1 & 0 & 0
      \\
      0 & 0 & 1 & 0 & 1 & 0 & 0 & 0 & 1 & 1 & 1 & 1
    \end{smallmatrix}\right).\IEEEeqnarraynumspace
  \end{IEEEeqnarray*}
  
  Observe that the first two rows of $\mat{G}^{\code{A}_2}$ correspond to the generators of $\code{A}_1$ and the third is the element-wise product between them. Therefore, we have a guarantee that $\code{A}_1 \subseteq \code{A}_2$ and this chain is closed under the element-wise product. Hence, $\code{C}=\code{A}_1+2\code{A}_2$ is a $\Integers_4$-linear code.
  
  The two codes $\code{A}_1$ and $\code{A}_2$ satisfy the conditions of Theorem~\ref{thm:FSD-Z4codes_A1plus2A2}, i.e.,  $W_{\code{A}_1}(x,y) = W_{\code{A}_2^\perp}(x,y)$ and $W_{\code{A}_2}(x,y) = W_{\code{A}_1^\perp}(x,y)$, but $\code{A}_2 \neq \code{A}_1^\perp$. The swe of $\code{C}$ is 
  \begin{IEEEeqnarray*}{rCl}
    \swe{\code{C}}(a,b,c)& = & a^{12}+1152 a^2 b^8 c^2+768 a^3 b^8 c +192 a^4 b^8 +18 a^{10} c^2+64 a^9 c^3 
    \nonumber\\
    && +\>111 a^8 b^4+192 a^7 c^5 +252 a^6 c^6 + 192 a^5 c^7 +111 a^4 c^8+64 a^3 c^9 
    \nonumber\\
    && +\>18 a^2 c^{10}+768 a b^8 c^3+192 b^8 c^4+c^{12},
  \end{IEEEeqnarray*}
  which satisfy the MacWilliams identity~\eqref{eq:swe-MacWilliams-identity_FSD-codes_Z4}, hence it is formally self-dual in $\Integers_4$. Moreover, $d_{\textnormal{Lee}}(\code{C})=4$, which is not optimal for this length, but coincides with the best Lee distance of self-dual codes~\cite[Table IV]{YooLeeKim17_1}.\hfill\exampleend
\end{example}
This construction will be of particular interest when $\code{A}_1$ and $\code{A}_2$ are chosen to be Reed-Muller codes and consequently generate Barnes-Wall lattices (see Section~\ref{sec:secrecy-gain_FSD-Z4-codes_A1plus2A2}).

\subsection{Double Circulant Construction and its Odd Extension}
\label{sec:double-circulant-its-odd-extension}

\subsubsection{Double Circulant Code (DCC)}
\label{sec:double-circulant-code}

The \emph{double circulant code (DCC)} is an important class of \emph{even-length} isodual codes (see, e.g.,~\cite[Ch.~9.8]{HuffmanPless03_1} or~\cite{BachocGulliverHarada00_1}), which consists of two subclasses of codes, namely the \emph{pure double circulant code (PDCC)} and the \emph{bordered double circulant code (BDCC)}. A PDCC and a BDCC have the generator matrices of the form
\begin{IEEEeqnarray}{c}
  \mat{G}^{\code{C}_\textnormal{pdc}}=
  \begin{pmatrix}
    \mat{I}_{\eta} & \mat{B}_{\eta}^{\textnormal{pc}}
  \end{pmatrix}\eqdef
  \begin{pmatrix}
    \mat{I}_{\eta} & \mat{R}_\eta
  \end{pmatrix}\quad\textnormal{and}\quad
  \mat{G}^{\code{C}_\textnormal{bdc}}=
  \begin{pmatrix}
    \mat{I}_{\eta} & \mat{B}_{\eta}^{\textnormal{bc}}
  \end{pmatrix}
  \eqdef
  \begin{pNiceMatrix}
    \Block[c]{4-4}<\Large>{\mat{I}_{\eta}} & & & & \alpha & \beta & \Cdots & \beta
    \\
    & & & & \gamma & \Block[c]{3-3}<\Large>{\mat{R}_{\eta-1}}  & & 
    \\
    & & & & \Vdots &  &  &  
    \\
    & & & & \gamma &  &  &  
  \end{pNiceMatrix},
  \label{eq:double-circulant-matrices}
\end{IEEEeqnarray}
respectively, where $\alpha, \beta, \gamma\in\Integers_4$, and
\begin{IEEEeqnarray}{c}
  \mat{R}_{\eta}\eqdef
  \begin{pNiceMatrix}
    r_1    & r_2 & r_3 & \cdots & r_\eta
    \\
    r_\eta & r_1 & r_2 & \cdots & r_{\eta-1}
    \\
    \vdots &  \vdots & \vdots  & \ddots &  \vdots
    \\
    r_2 & r_3 & r_4 & \cdots & r_{1}      
  \end{pNiceMatrix}
  \label{eq:def_circulant-matrix}
\end{IEEEeqnarray}
represents a \emph{circulant square matrix} of size $\eta\in\Naturals$, $r_i\in\Integers_4$, $i\in [1:\eta]$.

\subsubsection{Odd Extension of DCC}
\label{sec:odd-extension_DCC}

It is known that there exist self-dual codes of odd lengths over $\Integers_4$~\cite[Ch.~12.5]{HuffmanPless03_1}. In this work, we propose the odd-length code $\code{C}_\textnormal{oext}$ with a generator matrix of the form
\begin{IEEEeqnarray}{c}
  \mat{G}^{\code{C}_{\textnormal{oext}}}\eqdef
  \begin{pNiceMatrix}
    \Block[c]{3-3}<\Large>{\mat{I}_{\eta}} & & & a_1    & \Block{3-4}<\Large>{\mat{B}_{\eta}} & & & 
    \\
    &               &                          & \Vdots & & & & 
    \\
    &               &                          & a_\eta & & & & 
    \\
    0 & \Cdots & 0                             & 2      & 2c_1 & 2c_{2} & \Cdots & 2c_\eta
  \end{pNiceMatrix},
  \label{eq:def_odd-extension-G}
\end{IEEEeqnarray}
where $\mat{B}_{\eta}=\mat{B}_{\eta}^{\textnormal{pc}}$ or $\mat{B}_{\eta}^{\textnormal{bc}}$. We call such code an \textit{odd extension} code from a DCC. This code construction is inspired by~\eqref{eq:generator_cs} with $k_1=\eta$ and $k_2=1$, where $\mat{A}$ and $\mat{C}$ are chosen to be $\mat{A}=\trans{\vect{a}}=\trans{(a_1,\cdots, a_\eta)}$ and $\mat{C}=\vect{c}=(c_1, c_2, \cdots, c_\eta)$, respectively, $a_i,c_i\in\Integers_2$, $i\in[\eta]$. Note that this odd extension construction can also be defined using a general choice of $\mat{B}_{\eta}$. However, we will mostly consider the case where $\mat{B}_{\eta}=\mat{B}_{\eta}^{\textnormal{pc}}$ or $\mat{B}_{\eta}^{\textnormal{bc}}$, as in~\eqref{eq:double-circulant-matrices}.
    
As observed in \cite[p.~378]{HuffmanPless03_1}, pure double circulant codes are always isodual, and bordered double circulant codes are isodual if $\beta=\gamma=0$ or both $\beta$ and $\gamma$ are nonzero. An extensive list of optimal codes over $\Integers_4$ with respect to the minimum Lee weight is double circulant, see~\cite{BachocGulliverHarada00_1, GulliverHarada01_1}. The following result concerns the non-existence of pure double circulant self-dual codes.

\begin{proposition}[{\cite[Th.~5.1]{DoughertyGulliverHarada99_1}}]
  \label{prop:non-self-dual_PDCC_Z4}
  There exists no pure double circulant self-dual code over $\Integers_4$.
\end{proposition}

One can also obtain formally self-dual codes, particularly isodual codes, from the double circulant construction~\cite[Lemma~2.4]{BachocGulliverHarada00_1}. For bordered double circulant codes, the following result holds.
\begin{theorem}
  \label{thm:self-duality_BDCC} 
  If a BDCC $\code{C}_{\textnormal{bdc}}$ is self-dual, then the conditions i)--iv) all hold.
  \begin{enumerate}[nosep,label=\roman*)]
  \item $\alpha^2 + (\eta-1) \beta^2 \equiv 3 \pmod 4$,
  \item $\alpha\gamma+\beta \sum_{i=1}^{\eta-1} r_{i}  \equiv 0 \pmod 4$,
  \item $\gamma^2 + \sum_{i=1}^{\eta-1} r_i^2 \equiv 3 \pmod 4$,
  \item $(\eta -2)\gamma^2 + 2\sum_{i=1}^{\eta-1} \sum_{j=i+1}^{\eta-1} r_i r_j \equiv 0 \pmod 4$,
  \end{enumerate}
  where $\vect{r}=(r_1,\ldots,r_{\eta-1})$ is the first row of $\mat{R}_{\eta-1}$ defined in~\eqref{eq:def_circulant-matrix}.
\end{theorem}
\begin{IEEEproof} 
  If a BDCC $\code{C}_{\textnormal{bdc}}$ is self-dual, then $\mat{G}^{\code{C}_{\textnormal{bdc}}}\trans{(\mat{G}^{\code{C}_{\textnormal{bdc}}})} = \bigl(\mat{I}_{\eta}\,\,\,\mat{B}^{\textnormal{bc}}_\eta\bigr)\begin{psmallmatrix}\mat{I}_{\eta}\\ \trans{(\mat{B}^{\textnormal{bc}}_\eta)} \end{psmallmatrix}=\mat{O}_{\eta}$, and we have $\mat{B}^{\textnormal{bc}}\trans{\bigl(\mat{B}^{\textnormal{bc}}\bigr)} = -\mat{I}_{\eta}$. From~\eqref{eq:double-circulant-matrices}, this gives
  \begin{IEEEeqnarray*}{c}
    \begin{pNiceMatrix}
      \alpha & \beta & \Cdots & \beta
      \\
      \gamma & \Block[c]{3-3}<\Large>{\mat{R}_{\eta-1}}  & & 
      \\
      \Vdots &  &  &  
      \\
      \gamma &  &  &  
    \end{pNiceMatrix}
    \begin{pNiceMatrix}
      \alpha & \gamma & \Cdots & \gamma
      \\
      \beta & \Block[c]{3-3}<\Large>{\trans{\mat{R}_{\eta-1}}}  & & 
      \\
      \Vdots &  &  &  
      \\
      \beta &  &  &  
    \end{pNiceMatrix}
    =-\mat{I}_\eta.
  \end{IEEEeqnarray*}
  By comparing the matrix entries of the above equality, one can conclude that
  \begin{IEEEeqnarray*}{c}
    \begin{cases}
      \alpha^2+(\eta - 1)\beta^2 = -1
      \\
      \alpha\gamma + \beta \sum_{i=1}^{\eta -1} r_{i} = 0
      \\
      \gamma^2 + \sum_{i=1}^{\eta -1} r_i^2 = -1
      \\
      (\eta -2)\gamma^2 + 2\sum_{i=1}^{\eta-1} \sum_{j=i+1}^{\eta-1} r_i r_j = 0.
    \end{cases}
    \label{eq:conditions_self-duality_BDCC}
  \end{IEEEeqnarray*}
  Hence, from the above conditions i)--iv) arise.
\end{IEEEproof}
  
\begin{remark}
  From conditions iii) and iv), we particularly observe that
  \begin{IEEEeqnarray}{rCl}
     (r_1 + \cdots + r_{\eta -1})^2& = &\sum_{i=1}^{\eta-1} r_i^2 + 2 \sum_{i=1}^{\eta-1} \sum_{j=i+1}^{\eta-1} r_i r_j
    \nonumber \\
    & = & -1 - \gamma^2 - (\eta-2)\gamma^2 = -1 + (1-\eta)\gamma^2.
    \label{eq:squares_r-sum}
  \end{IEEEeqnarray}
  Since $x^2 \equiv 0 \textnormal{ or }1 \pmod 4,\,\forall\,x\in\Integers_4$, \eqref{eq:squares_r-sum} implies that $-1 + (1-\eta)\gamma^2$ can only be $0 \pmod 4$ or $1 \pmod 4$. Hence, we can conclude that the condition iv) of Theorem~\ref{thm:self-duality_BDCC} also leads to
  \begin{IEEEeqnarray*}{c}
    \begin{cases}
      \gamma^2 \equiv 1 \pmod 4 \textnormal{ and }\eta\equiv 0 \pmod 4 & \textnormal{if } (\sum_{i=1}^{\eta-1}r_i)^2 \equiv 0 \pmod 4,
      \\
      \gamma^2 \equiv 1 \pmod 4 \textnormal{ and }\eta \equiv 3 \pmod 4 & \textnormal{if } (\sum_{i=1}^{\eta-1}r_i)^2 \equiv 1 \pmod 4.
    \end{cases}    
  \end{IEEEeqnarray*}
\end{remark}


\begin{example}
  \label{ex:double_circulant_fsd}
  Consider a BDCC $\code{C}_{\textnormal{bdc}}$ of length $4$ with $\alpha=0, \gamma=2$ and $\beta=2$, and its dual code $\code{C}_{\textnormal{bdc}}^\perp$, generated respectively by
  \begin{IEEEeqnarray*}{c}
  \mat{G}^{\code{C}_{\textnormal{bdc}}}=
    \begin{pNiceMatrix}
      1 & 0 & 0 & 2
      \\
      0 & 1 & 2 & 1
    \end{pNiceMatrix},\quad
    \mat{G}^{\code{C}_{\textnormal{bdc}}^\perp}=
    \begin{pNiceMatrix}
      0 & 2 & 1 & 0
      \\
      2 & 3 & 0 & 1
     \end{pNiceMatrix}.
  \end{IEEEeqnarray*}
  
  Comparing to~\eqref{eq:Generator_Matrix} of the standard form, $k_1=2$ and $k_2=0$ in this example. Observe that $\mat{G}^{\code{C}_{\textnormal{bdc}}}\trans{(\mat{G}^{\code{C}_{\textnormal{bdc}}})} \neq \mat{O_{2}}$, and indeed, $\mat{G}^{\code{C}_{\textnormal{bdc}}}$ does not meet the conditions in Theorem~\ref{thm:self-duality_BDCC}. Therefore the code is not self-dual. However, $\swe{\code{C}_{{\textnormal{bdc}}}}(a,b,c) =a^4+a^3 c+4 a^2 b c+a^2 c^2+2 a b^2 c+a c^3+4 b^3 c+2 b^2 c^2$ satisfies the MacWilliams identity~\eqref{eq:swe-MacWilliams-identity_Z4}, meaning that the code is formally self-dual in $\Integers_4$. 
  Moreover, one can notice that the code $\code{C}_{\textnormal{bdc}}$ is indeed isodual, since $\beta$ and $\gamma$ are nonzero, and in addition to that $\mat{G}^{\code{C}_{\textnormal{bdc}}} = \mat{G}^{\code{C}_{\textnormal{bdc}}^\perp}\mat{Q}$, where
  \begin{IEEEeqnarray*}{c}
    \mat{Q}=
    \begin{pNiceMatrix}
      0 & 0 & 1 & 0 \\
      0 & 0 & 0 & -1 \\
      1 & 0 & 0 & 0 \\
      0 & 1 & 0 & 0 \\
    \end{pNiceMatrix}.
  \end{IEEEeqnarray*}
  \hfill\exampleend
\end{example}

\begin{example}
\label{ex:FSD-not-SD-codes_n12-14-16}
  The BDCCs of length $12$, $14$, and $16$, presented in Table~\ref{tab:long-table_FSD-Z4-codes-swes-SGs} of Appendix~\ref{sec:all-FSD-Z4-codes-swes-SGs}, fail conditions $\textnormal{i)}, \textnormal{iii)}$ and $\textnormal{ii)}$ of Theorem~\ref{thm:self-duality_BDCC}, respectively. Hence, they are all formally-self dual but not self-dual.
  \hfill\exampleend
\end{example}

We also give conditions for an odd extension code to be self-dual.
\begin{proposition}
  \label{prop:odd-extension-sel-dual} 
  Consider an odd extension code $\code{C}_{\textnormal{oext}}$ generated by $\mat{G}^{\code{C}_{\textnormal{oext}}}$ as in~\eqref{eq:def_odd-extension-G}. Then, $\code{C}_{\textnormal{oext}}$ is self-dual if and only if the following conditions hold
  \begin{enumerate}[nosep,label=\roman*)]
  \item $\trans{\vect{a}}\vect{a}+\mat{B}\trans{\mat{B}}\equiv 3\mat{I}_{\eta} \pmod 4$,
  \item $2\vect{a} + 2\vect{c}\trans{\mat{B}}\equiv \vect{0} \pmod 4$.
  \end{enumerate}
\end{proposition}
\begin{IEEEproof}
  Since $\code{C}_\textnormal{oext}$ is self-dual if and only if $\mat{G}^{\code{C}_\textnormal{oext}}\trans{\bigl(\mat{G}^\code{C}_{\textnormal{oext}}\bigr)}=\mat{O}_{\eta+1}$, we have
  \begin{IEEEeqnarray*}{c}
    \begin{pNiceMatrix}
      \mat{I}_{\eta} &\trans{\vect{a}} & \mat{B}_{\eta}
      \\
      \vect{0}       & 2               & 2\vect{c}
    \end{pNiceMatrix}
    \begin{pNiceMatrix}
      \mat{I}_{\eta} &\trans{\vect{0}} 
      \\
      \vect{a}       & 2               
      \\
      \trans{\mat{B}_{\eta}} & 2\trans{\vect{c}}
    \end{pNiceMatrix}
    =\mat{O}_{\eta+1}
  \end{IEEEeqnarray*}
  by using $\mat{G}^{\code{C}_\textnormal{oext}}$ as in~\eqref{eq:def_odd-extension-G}. Thus, this gives that $\mat{I}+\trans{\vect{a}}\vect{a}+\mat{B}\trans{\mat{B}}=\mat{O}_{\eta}$ and $2\vect{a} + 2\vect{c}\trans{\mat{B}}\equiv\vect{0} \pmod 4$, which leads to conditions i) and ii) stated in the proposition.
\end{IEEEproof}

\begin{example}
  \label{ex:n13k6_oextCode}
  Consider a $[13,2^{13}]$ formally self-dual code $\code{C}_{\textnormal{oext}}$ over $\Integers_4$ generated as in~\eqref{eq:def_odd-extension-G}, where 
  \begin{IEEEeqnarray*}{c}
    \mat{B}^{\textnormal{pc}}=
    \begin{pNiceMatrix}
      0 & 2 & 1 & 2 & 2 & 2 \\
      2 & 0 & 2 & 1 & 2 & 2 \\
      2 & 2 & 0 & 2 & 1 & 2 \\
      2 & 2 & 2 & 0 & 2 & 1 \\
      1 & 2 & 2 & 2 & 0 & 2 \\
      2 & 1 & 2 & 2 & 2 & 0
    \end{pNiceMatrix}
  \end{IEEEeqnarray*}
  is a pure double circulant matrix, $\vect{c}=(0,0,0,0,1,1)$ and $\vect{a}=(0,0,1,1,0,0)$. Since
  \begin{IEEEeqnarray*}{c}
    \trans{\vect{a}}\vect{a}+\mat{B}\trans{\mat{B}}= 
    \begin{pNiceMatrix}
      1 & 0 & 2 & 0 & 2 & 0
      \\
      0 & 1 & 0 & 2 & 0 & 2
      \\
      2 & 0 & 2 & 1 & 2 & 0
      \\
      0 & 2 & 1 & 2 & 0 & 2
      \\
      2 & 0 & 2 & 0 & 1 & 0
      \\
      0 & 2 & 0 & 2 & 0 & 1 
    \end{pNiceMatrix} \neq -\mat{I}_{6},
  \end{IEEEeqnarray*}
  this implies that $\code{C}_{\textnormal{oext}}$ is not self-dual. \hfill\exampleend
\end{example}

\begin{example}
  \label{ex:n9k14k21_Coext}
  One can verify that the code $\code{C}_\textnormal{oext}$ with generator matrix 
  \begin{IEEEeqnarray*}{c}
    \mat{G}^{\code{C}_{\textnormal{oext}}} =
    \begin{pNiceMatrix}
      1 & 0 & 0 & 0 & 0 & 2 & 1 & 1 & 1
      \\
      0 & 1 & 0 & 0 & 0 & 1 & 1 & 2 & 3
      \\
      0 & 0 & 1 & 0 & 0 & 1 & 3 & 1 & 2
      \\
      0 & 0 & 0 & 1 & 0 & 1 & 2 & 3 & 1
      \\
      0 & 0 & 0 & 0 & 2 & 0 & 0 & 0 & 0
    \end{pNiceMatrix}
  \end{IEEEeqnarray*}
  satisfies the conditions of Proposition~\ref{prop:odd-extension-sel-dual}, and therefore it is self-dual.\hfill\exampleend
\end{example}

\section{Theta Series of $2$-Level Construction C and Construction $\textnormal{A}_4$ Lattices}
\label{sec:theta-series_2-level-ConstrC-ConstructionAfour-lattices}

\subsection{$2$-Level Construction C Lattices}
\label{sec:2-level-ConstructionC-lattices}

From the fact that a $2$-level Construction C is geometrically uniform for any choice of binary linear codes $\code{A}_1$ and $\code{A}_2$, we can state the following result.
\begin{theorem}
  \label{thm:theta-series_2-level-constructionC}
  Consider a $2$-level Construction C packing given by $\PConstrC{\code{A}_1}{\code{A}_2} = \frac{1}{2}(\code{A}_1 + 2\code{A}_2 + 4\mathbb{Z}^n),$ where $\code{A}_1, \code{A}_2$ are binary linear codes. The theta series of $\PConstrC{\code{A}_1}{\code{A}_2}$ is
  \begin{IEEEeqnarray*}{c}
    \Theta_{\PConstrC{\code{A}_1}{{\code{A}_2}}(z)}
    = \sum_{\vect{a}_1 \in \code{A}_1} \sum_{\vect{a}_2 \in \code{A}_2} \vartheta_3^{d_{0,0}}(4z) \left(\frac{\vartheta_{2}(z)}{2}\right)^{{d_{1,0}} + {d_{1,1}}}  \vartheta_2^{{d_{0,1}}}(4z).
  \end{IEEEeqnarray*}
\end{theorem}

For the proof of this theorem, we will use Proposition~\ref{prop:thm_thetanonl}, which considers the expression of the theta series of a periodic packing~\cite{OdlyzkoSloane80_1}.
\begin{proposition}[\cite{OdlyzkoSloane80_1}]
  \label{prop:thm_thetanonl} 
  Given a periodic packing $\Gamma = \bigcup_{k=1}^{M} (\Lambda + \vect{u}_k)$, where $\Lambda \subset \Reals^n$  is a lattice and ${\bm u}_1, \dots, {\bm u}_M \in \Reals^n$ are the  $M$ coset representatives, then
  \begin{IEEEeqnarray}{c}
    \Theta_\Gamma(z) = \Theta_\Lambda(z) + \dfrac{2}{M} \displaystyle\sum_{k< \ell}\displaystyle\sum_{{\bm \lambda} \in \Lambda} q^{\|{\bm \lambda}+{\bm u}_k-{\bm u}_\ell \|^2}.
  \end{IEEEeqnarray}
  For a geometrically uniform packing $\Gamma$, where the set of distances is preserved for every point, then it reduces to
  \begin{IEEEeqnarray}{c}
    \label{eq:theta_equidistance}
    \Theta_\Gamma(z) = \displaystyle\sum_{k=1}^{M}\displaystyle\sum_{{\bm \lambda} \in \Lambda} q^{\|{\bm \lambda}+{\bm u}_k-{\bm u}_1 \|^2}.
  \end{IEEEeqnarray}
\end{proposition}
\begin{IEEEproof}[Proof of Theorem~\ref{thm:theta-series_2-level-constructionC}]
  Packings obtained from Construction C are periodic and in particular, a $2$-level Construction C is geometrically uniform, so we can apply Proposition~\ref{prop:thm_thetanonl}, more specifically, \eqref{eq:theta_equidistance}.
  
  In~\eqref{eq:theta_equidistance}, we identify $\vect{\lambda}\in 4\Integers^n$, $M=\ecard{\code{A}_1}\ecard{\code{A}_2}$, and $\vect{u}_1=(0,\dots,0)$, since $\code{A}_1, \code{A}_2$ are linear codes and thus contain the zero codeword. Notice that in our context, ${\bm u}_k \in \code{A}_1+ 2\code{A}_2$ and initially, let us fix $k$ and set ${\bm u} ={\bm u}_k$ to simplify.
  
  As ${\bm u} \in \code{A}_1+2\code{A}_2$, there exist $\vect{a}_{1} \in \code{A}_1$ and $\vect{a}_{2} \in \code{A}_2$ such that ${\bm u} = \vect{a}_{1} + 2\vect{a}_{2}$.
  The coordinates of ${\bm u}$ can be $0,1, 2,$ or $3$ and their recurrences are given respectively by $d_{0,0}(\vect{a}_{1}, \vect{a}_{2}),$ $d_{1,0}(\vect{a}_{1}, \vect{a}_{2}), d_{0,1}(\vect{a}_{1}, \vect{a}_{2})$, and $d_{1,1}(\vect{a}_{1}, \vect{a}_{2})$, as in~\eqref{eq:def_d00}-\eqref{eq:def_d11}.
  
  By fixing the $i$-th coordinate of ${\bm u},$ we have as possible exponents of $q$ in~\eqref{eq:theta_equidistance}  
  \begin{IEEEeqnarray}{c}
    4z_i + {\bm u}_{{i}} =
    \begin{cases}
      4z_i,                   & \textnormal{if } \vect{u}_{i}=0,
      \\
      4(z_i + \frac{1}{4}),  & \textnormal{if } \vect{u}_{i}=1,
      \\
      4(z_i + \frac{1}{2}),  & \textnormal{if } \vect{u}_{i}=2,
      \\
      4(z_i + \frac{3}{4}),  & \textnormal{if } \vect{u}_{i}=3.
    \end{cases}\label{eq:couting_4z}
  \end{IEEEeqnarray}
  
  The corresponding theta series associated to each one of the previous cases are
  \begin{IEEEeqnarray*}{rCl}
    \Theta_{4\mathbb{Z}}(z) = \vartheta_{3}(16z),\, \Theta_{4\big(\mathbb{Z}+ \frac{1}{2}\big)}(z) = \vartheta_{2}(16z),
    \nonumber \\
    \Theta_{4\big(\mathbb{Z}+\tfrac{1}{4}\big)}(z) = \Theta_{4\big(\mathbb{Z}+ \frac{3}{4}\big)}(z) = \frac{\vartheta_{2}(4z)}{2}, 
  \end{IEEEeqnarray*}
  
  By incorporating these  results into the fixed $n$-dimensional vector ${\bm u},$ we have that
  \begin{IEEEeqnarray*}{c}
    \sum_{{\bm z} \in \mathbb{Z}^n} q^{\|4{\bm z}+{\bm u} \|^2} = \vartheta_3^{d_{0,0}}(16z)  \left(\frac{\vartheta_{2}(4z)}{2}\right)^{{d_{1,0}}+ {d_{1,1}}} \vartheta_2^{d_{0,1}}(16z).
  \end{IEEEeqnarray*}
  
  Finally, running through all $k$ vectors ${\bm u}_k$ and considering the scaled version $\PConstrC{\code{A}_1}{\code{A}_2} = \frac{1}{2}(\code{A}_1 + 2\code{A}_2 + 4\mathbb{Z}^n),$ we get
  \begin{IEEEeqnarray}{rCl}
    \Theta_{\Gamma_\textnormal{C}}(z) &  = &   \displaystyle\sum_{k=1}^{M}\displaystyle\sum_{{\bm z} \in \mathbb{Z}^n} q^{\enorm{\frac{1}{2}(4{\bm z}+{\bm u}_k)}^2} \nonumber \\
    & = & \sum_{\vect{a}_{1_k} \in \code{A}_1} \sum_{\vect{a}_{2_k} \in \code{A}_2} \vartheta_3^{d_{0,0}}(4z) \left(\frac{\vartheta_{2}(z)}{2}\right)^{{d_{1,0}} + {d_{1,1}}}  \vartheta_2^{{d_{0,1}}}(4z),
    \label{eq:theta-series_GammaC}\IEEEeqnarraynumspace
  \end{IEEEeqnarray}
  where ${\bm u}_k = \vect{a}_{1_k} + 2\vect{a}_{2_k},$ for $\vect{a}_{1_k} \in \code{A}_1$ and $\vect{a}_{2_k} \in \code{A}_2$.
\end{IEEEproof}


Theorem~\ref{thm:theta-series_2-level-constructionC} is general and can also be applied to nonlattice packings. Next, we relate the theta series of $\PConstrC{\code{A}_1}{\code{A}_2}$ to the joint weight enumerator.
\begin{corollary}
  \label{coro:theta-series_2-level-ConstructionC}
  The theta series of a $2$-level Construction C packing, in terms of the jwe of two codes, is
  \begin{IEEEeqnarray*}{c}
    \Theta_{\PConstrC{\code{A}_1}{\code{A}_2}}(z)=\textnormal{jwe}_{\code{A}_1,\code{A}_2}\bigl(\vartheta_3(4z),\vartheta_2(4z), \nicefrac{\vartheta_2(z)}{2},\nicefrac{\vartheta_2(z)}{2} \bigr).\label{eq:Theta-ft_2-level-ConstructionC}
  \end{IEEEeqnarray*}
\end{corollary}
Remark that Corollary~\ref{coro:theta-series_2-level-ConstructionC} also holds for the lattice case $\ConstrC{\code{A}_1}{\code{A}_2}$.

\subsection{Construction $\textnormal{A}_4$ Lattices}
\label{sec:ConstrAfour-lattices}

We now define a few extra notions of weight enumerators and derive an expression of the theta series of the lattice generated via Construction $\textnormal{A}_4$, given the swe of the $\Integers_4$-linear code $\code{C}$. 

If we consider the $\Integers_4$-linear code $\code{C}$, the theta series of a Construction $\textnormal{A}_4$ lattice can be expressed as follows.
\begin{corollary}
    \label{coro:theta-series_ConstrAfour}
  Let $\code{C}$ be a $\Integers_4$-linear code with $\swe{\code{C}}(a,b,c)$, then the theta series of $\ConstrAfour{\code{C}}$ is 
  \begin{IEEEeqnarray*}{c}
    \Theta_{\eConstrAfour{\code{C}}}(z) = \swe{\code{C}}(\vartheta_3(4z), \nicefrac{\vartheta_2(z)}{2}, \vartheta_2(4z)).
  \end{IEEEeqnarray*}
\end{corollary}
\begin{IEEEproof}
  If the $\Integers_4$-linear code $\code{C}$ is such that $\code{C}=\code{A}_1+2\code{A}_2$, the result comes immediately from Corollary~\ref{coro:theta-series_2-level-ConstructionC}, since $\swe{\code{C}}(a,b,c)=\textnormal{jwe}_{\code{A}_1,\code{A}_2}(a,c,b,b)$. For a general $\Integers_4$-linear code $\code{C}$, the same proof of Theorem~\ref{thm:theta-series_2-level-constructionC} can be applied, since $\ConstrAfour{\code{C}} = \frac{1}{2}\bigl(\code{C}+4\Integers^n\bigl)$ is also a periodic packing and the coordinates of an element in $\ConstrAfour{\code{C}}$ are also described as in \eqref{eq:couting_4z}. The only difference is that the exponents in \eqref{eq:theta-series_GammaC} are replaced by $n_0(\vect{w}), n_1(\vect{w})+n_3(\vect{w})$, and $n_2(\vect{w})$ respectively, where $\vect{w}\in\code{C}$, and one can see that the result follows.
\end{IEEEproof}

We also highlight the following results.
\begin{corollary}
  \label{cor:FSD_Z4-FUM-lattices}
  If $\code{C}$ is a formally self-dual $\Integers_4$-linear code, then $\ConstrAfour{\code{C}}$ is formally unimodular. Moreover, if a $\Integers_4$-linear code $\code{C}$ is of Type I, then $\ConstrAfour{\code{C}}$ is of Type I.
\end{corollary}
\begin{IEEEproof}
  The first statement is a direct consequence of Corollary~\ref{coro:theta-series_ConstrAfour}. Then, we only need to prove the Type I property for the formally unimodular lattice $\ConstrAfour{\code{C}}$. Since $\code{C}$ is of Type I, we know that the $\Ewt{\vect{c}}\in 4\Integers$ for any $\vect{c}\in\code{C}$. Since by definition, any two vectors $\vect{x}$, $\vect{x}'\in\ConstrAfour{\code{C}}$ can be represented by
  \begin{IEEEeqnarray*}{c}
    \vect{x}=\frac{1}{2}(\vect{c}+4\vect{z}),\quad\vect{x}'=\frac{1}{2}(\vect{c}'+4\vect{z}').
  \end{IEEEeqnarray*}
  
  Hence, we have
  \begin{IEEEeqnarray}{c}
    \inner{\vect{x}}{\vect{x}'}=\frac{1}{4}\inner{\vect{c}}{\vect{c}'}+\inner{\vect{c}}{\vect{z}'}+\inner{\vect{c}'}{\vect{z}}+
    4\inner{\vect{z}}{\vect{z}'}\label{eq:inner-product_two-lattice-vectors}.
  \end{IEEEeqnarray}

  Observe that
  \begin{IEEEeqnarray*}{c}
    \inner{2}{2}\equiv 0 \pmod 4,\quad\inner{1}{1}=\inner{1}{3}=\inner{3}{3}\equiv 1 \pmod 4,\quad\inner{1}{2}=\inner{2}{3} \equiv 2 \pmod 4,
  \end{IEEEeqnarray*}
  we can claim that $\inner{\vect{c}}{\vect{c}'}\in4\Integers$ if $\Ewt{\vect{c}},\Ewt{\vect{c}'}\in4\Integers$. Therefore, from~\eqref{eq:inner-product_two-lattice-vectors}, it implies that $\inner{\vect{x}}{\vect{x}'}\in\Integers$, which completes the proof.  
\end{IEEEproof}

\section{Secrecy Gain of Formally Unimodular Lattices}
\label{sec:secrecy-gain_FU-lattices}

\subsection{The Secrecy Function of a Lattice}
\label{sec:secrecy-function_lattice}
    
Lattices are primarily used in Gaussian wiretap channels with an analogous \textit{coset encoding}~\cite{Wyner75_1} idea, where two lattices are considered: $\Lambda_b$ to ensure reliability and $\Lambda_e\subset\Lambda_b$ to guarantee security. Minimizing the success probability of correctly guessing the transmitted message received by the eavesdropper is equivalent to minimizing the theta function of $\Lambda_\textnormal{e}$. This study leads to the definitions of secrecy function and secrecy gain for a lattice~\cite{OggierSoleBelfiore16_1}.

\begin{definition}[Secrecy function and secrecy gain~{\cite[Defs.~1 and~2]{OggierSoleBelfiore16_1}}]
  \label{def:secrecy_function}
  Let $\Lambda$ be a lattice with volume $\vol{\Lambda}=\nu^n$. The secrecy function of $\Lambda$ is defined by
  \begin{IEEEeqnarray*}{c}
    \Xi_{\Lambda}(\tau)\eqdef\frac{\Theta_{\nu\Integers^n}(i\tau)}{\Theta_{\Lambda}(i\tau)},
    \label{eq:def_secrecy-function}
  \end{IEEEeqnarray*} 
  for $\tau\eqdef -i z>0$. The \emph{(strong) secrecy gain} of a lattice is given by $\xi_{\Lambda}\eqdef\sup_{\tau>0}\Xi_{\Lambda}(\tau)$.
\end{definition}

It was shown in~\cite{OggierSoleBelfiore16_1} that the higher the secrecy gain of a lattice, the more security is provided by the lattice wiretap code. Hence, the objective here is to design a  lattice $\Lambda$ that achieves a high secrecy gain.

Under the design criterion of the secrecy function, we summarize the following three important observations for the formally unimodular lattices~\cite{BollaufLinYtrehus22_1}.
\begin{enumerate}
\item The secrecy function of a formally unimodular lattice $\Lambda$ has exactly the same \emph{symmetry point} at $\tau=1$ as a unimodular or an isodual lattice, i.e., $\Xi_\Lambda(\tau) = \Xi_\Lambda\bigl(\frac{1}{\tau}\bigr)$.
\item Similar to Belfiore and Sol{\'{e}}'s conjecture from~\cite{BelfioreSole10_1}, it is also conjectured that the secrecy function of a formally unimodular lattice $\Lambda$ achieves its maximum at $\tau=1$, i.e., $\xi_{\Lambda}=\Xi_{\Lambda}(1)$.
\item It was demonstrated that formally unimodular lattices can outperform the secrecy gain of unimodular lattices. 
In particular, the unimodular and formally unimodular lattices constructed via Construction A are compared, and it indicates that formally unimodular lattices obtained from formally self-dual codes via Construction A always achieve better secrecy gains than the Construction A unimodular lattices obtained from self-dual codes (see~\cite[Tab.~I]{BollaufLinYtrehus22_1} for details).
\end{enumerate}

Using these observations, we next explore the secrecy gain of formally unimodular lattices obtained by Construction $\textnormal{A}_4$ from formally self-dual codes over $\Integers_4$.

\subsection{Secrecy Gain of Construction $\textnormal{A}_4$ Lattices obtained from Formally Self-Dual Codes over $\Integers_4$} 
\label{sec:secrecy-gain_ConstrAfour-lattices_fsd-codes_Z4}
    
In this subsection, we derive a closed-form expression of the theta series of a Construction $\textnormal{A}_4$ lattice obtained from a formally self-dual code $\code{C}$ over $\Integers_4$ and provide a new universal approach to derive the strong secrecy gain of the corresponding lattice. The result is stated in the following main theorem.
\begin{theorem}
  \label{thm:inv_secrecy-function_SymmetrizedWeightEnumerator}
  Let $\code{C}$ be a formally self-dual code over $\Integers_4$. Then
  \begin{IEEEeqnarray*}{c}
    \inv{\Bigl[\Xi_{\eConstrAfour{\code{C}}}(\tau)\Bigr]}=\frac{\swe{\code{C}}\bigl(1+t, \sqrt[4]{1-t^4}, 1-t\bigr)}{2^{n}},\label{eq:Xi-ft_ConstructionA_FSDcodes}\IEEEeqnarraynumspace
  \end{IEEEeqnarray*}
  where $0<t(\tau)\eqdef\nicefrac{\vartheta_4(i\tau)}{\vartheta_3(i\tau)} < 1$. Moreover, define $h_{\code{C}}(t)\eqdef\swe{\code{C}}\bigl(1+t, \sqrt[4]{1-t^4}, 1-t\bigr)$ for $0< t < 1$. Then, maximizing the secrecy function $\Xi_{\eConstrAfour{\code{C}}}(\tau)$ is equivalent to determining the minimum of $h_{\code{C}}(t)$ on $t\in(0,1)$.
\end{theorem}
\begin{IEEEproof}
From \eqref{eq:swe-MacWilliams-identity_FSD-codes_Z4} and the following identities from \cite[Eq.~(23), Ch.~4]{ConwaySloane99_1}, and \cite[Eq.~(31), Ch.~4]{ConwaySloane99_1}, respectively,
\begin{IEEEeqnarray}{rCl}
  \vartheta_3(z)+\vartheta_4(z)& = &2\vartheta_3(4z), \vartheta_3(z)-\vartheta_4(z)=2\vartheta_2(4z),\quad\label{eq:useful-identities-1}
  \\[1mm]
  \vartheta^4_2(z)+\vartheta^4_4(z)& = &\vartheta^4_3(z),\quad\label{eq:useful-identities-2}
\end{IEEEeqnarray}
we obtain
\begin{IEEEeqnarray}{rCl}
  \Theta_{\eConstrAfour{\code{C}}}(z)& = &\swe{\code{C}}(\vartheta_3(4z), \nicefrac{\vartheta_2(z)}{2}, \vartheta_2(4z))
  \nonumber\\
  & \stackrel{\eqref{eq:swe-MacWilliams-identity_FSD-codes_Z4}}{=} &\inv{\bigcard{\dual{\code{C}}}}\swe{\code{C}}\Big(\vartheta_3(4z)+2\frac{\vartheta_2(z)}{2}+\vartheta_2(4z), \vartheta_3(4z)-\vartheta_2(4z), \vartheta_3(4z)-2\frac{\vartheta_2(z)}{2}+\vartheta_2(4z) \Big) \nonumber \\
  & \stackrel{\eqref{eq:useful-identities-1}}{=} &\inv{\bigcard{\dual{\code{C}}}}\swe{\code{C}}\Big(\vartheta_3(z)+\vartheta_2(z), \vartheta_4(z), \vartheta_3(z)-\vartheta_2(z)\Big) \nonumber\\
  & \stackrel{\eqref{eq:swe-MacWilliams-identity_FSD-codes_Z4}}{=} &\bigcard{\dual{\code{C}}}^{-2}\swe{\code{C}}\Big(2\vartheta_3(z)+2\vartheta_4(z), 2\vartheta_2(z), 2\vartheta_3(z)-2\vartheta_4(z)\Big)  \nonumber\\
  & \stackrel{\eqref{eq:useful-identities-2}}{=} &\bigcard{\dual{\code{C}}}^{-2}\cdot 2^n\swe{\code{C}}\Bigl(\vartheta_3(z)+\vartheta_4(z), \sqrt[4]{\vartheta^4_3(z)-\vartheta^4_4(z)}, \vartheta_3(z)-\vartheta_4(z)\Bigr). \nonumber\\
  & = &2^{-n}\cdot\swe{\code{C}}\Bigl(\vartheta_3(z)+\vartheta_4(z), \sqrt[4]{\vartheta^4_3(z)-\vartheta^4_4(z)}, \vartheta_3(z)-\vartheta_4(z)\Bigr),\label{eq:theta-series_FSD-codes_Z4}\IEEEeqnarraynumspace  
\end{IEEEeqnarray}
where \eqref{eq:theta-series_FSD-codes_Z4} holds since if $\code{C}$ is formally self-dual, $\card{\dual{\code{C}}}=4^{\nicefrac{n}{2}}$.

From Definition~\ref{def:secrecy_function}, the inverse of the secrecy function of $\ConstrAfour{\code{C}}$ with volume $1$ becomes
  \begin{IEEEeqnarray*}{rCl}
    \inv{\Bigl[\Xi_{\eConstrAfour{\code{C}}}(\tau)\Bigr]}& = &\frac{\Theta_{\eConstrAfour{\code{C}}}(z)}{\Theta_{\Integers^n}(z)}
    \stackrel{(b)}{=}\frac{1}{2^{n}}\frac{\swe{\code{C}}\Bigl(\vartheta_3(z)+\vartheta_4(z), \sqrt[4]{\vartheta^4_3(z)-\vartheta^4_4(z)}, \vartheta_3(z)-\vartheta_4(z)\Bigr)}{\vartheta^n_3(z)}
    \\[1mm]
    & = &\frac{1}{2^{n}}\swe{\code{C}}\left(1+\frac{\vartheta_4(z)}{\vartheta_3(z)},\sqrt[4]{1-\frac{\vartheta^4_4(z)}{\vartheta^4_3(z)}}, 1-\frac{\vartheta_4(z)}{\vartheta_3(z)}\right)
    \\[1mm]
    & = &\frac{\swe{\code{C}}\bigl(1+t(\tau),\sqrt[4]{1-t^4(\tau)},1-t(\tau)\bigr)}{2^{n}},
  \end{IEEEeqnarray*}
  where $(b)$ holds because of $\Theta_{\Integers^n}(z)=\vartheta^n_3(z)$ and~\eqref{eq:theta-series_FSD-codes_Z4}. Lastly, using a similar argument as~\cite[Lemma 34 and Remark 35]{BollaufLinYtrehus23_3}, the second part of the theorem follows.
\end{IEEEproof}
\begin{example}
  \label{ex:FSDcode_dim8}
  Consider the formally self-dual code $\code{C}_8$ from~\cite[pp.~83--84]{BetsumiyaHarada03_1}, one can obtain
  \begin{IEEEeqnarray*}{rCl}
    \swe{\code{C}_{8}}(a,b,c)& = & c^8 + 64b^8 + 12a b^2 c^5 + 64a b^6 c +16 a^2 c^6\nonumber\\
    &&\, +\> 40a^3 b^2 c^3 + 30a^4 c^4 + 12a^5 b^2 c + 16a^6 c^2 + a^8.
  \end{IEEEeqnarray*} 
  Then, we have that $h_{\code{C}_8}(t) = 64 \bigl(2 t^8+t^6-\left(\sqrt{1-t^4}+2\right) t^4 -\left(\sqrt{1-t^4}-1\right) t^2+2 \left(\sqrt{1-t^4}+1\right)\bigr)$ and $h'_{\code{C}_8}(t)=0$ for $t=\nicefrac{1}{\sqrt[4]{2}}$, which can be numerically verified to be the global minimizer.\hfill\exampleend
\end{example}

\begin{example}
  \label{ex:optimal_codes}
  Gulliver and Harada presented in~\cite{GulliverHarada01_1} optimal formally self-dual codes over $\Integers_4$ in dimensions $6,8,10$ and $14$, together with their swe's. Each $h_{\code{C}_i}(t)$, $i=6,8,10,14$ achieves its minimum at $t=\nicefrac{1}{\sqrt[4]{2}}$. Therefore, we have $\xi_{\eConstrAfour{\code{C}_{6}}} \approx 1.172$, $\xi_{\eConstrAfour{\code{C}_{8}}} \approx 1.333$, $\xi_{\eConstrAfour{\code{C}_{10}}} \approx 1.379$, and $\xi_{\eConstrAfour{\code{C}_{14}}} \approx 1.871$, which coincide or are very close to best secrecy gains from~\cite[Tab.~I]{BollaufLinYtrehus22_1}.\hfill\exampleend
\end{example}

\begin{example}
  \label{ex:n22_FSD-code_Z4}
  In this example, we consider the swe of the isodual code $\code{D}_{4,22}$ presented in~\cite[p.~230, Prop.~4.2]{BachocGulliverHarada00_1}. Using Theorem~\ref{thm:inv_secrecy-function_SymmetrizedWeightEnumerator}, we get $\xi_{\eConstrAfour{\code{D}_{4,22}}}\approx 3.403$. We remark that the best-known secrecy gain for a formally unimodular lattice in this dimension is $3.34$, which is presented in~\cite[Tab.~I]{BollaufLinYtrehus22_1}. Therefore this result outperforms the current secrecy gain of lattices in dimension $22$.\hfill\exampleend
\end{example}

Next, we recall a general expression of $\swe{\code{C}}(a,b,c)$ from \emph{invariant theory} if $\code{C}$ is a Type I formally self-dual code over $\Integers_4$.
\begin{proposition}[{\cite[Eq.~(8.2.6)]{NebeRainsSloane06_1}}]
  \label{prop:swe_TypeI-FSDcodes_Z4}
  If $\code{C}$ is an $[n,2^{n}]$ Type I formally self-dual code over $\Integers_4$, then we have
  \begin{IEEEeqnarray}{c}
    \swe{\code{C}}(a,b,c)=\sum_{\substack{r,s\in\Naturals\cup\{0\}\\0\leq 4r+8s\leq n}}\alpha_{r,s}A^{n-4r-8s}\bigl(B^4+C^4\bigr)^r\bigl(B^4C^4\bigr)^s,
    \label{eq:Gleason-theorem_TypeI-FSDcodes_Z4}
  \end{IEEEeqnarray}
  where $A=a+c, B=2b, C=a-c$, and $\alpha_{r,s}\in\Rationals$.
\end{proposition}

If $\code{C}$ is a Type I formally self-dual code over $\Integers_4$, then $h_{\code{C}}(t)$ can be expressed as follows.
\begin{lemma}
  \label{lem:hC_TypeI-FSDcodes_Z4}
  If $\code{C}$ is an $[n,2^{n}]$ Type I formally self-dual code over $\Integers_4$, then
  \begin{IEEEeqnarray}{c}
    h_{\code{C}}(t)=2^{n}\sum_{s=0}^{\lfloor\frac{n}{8}\rfloor}\beta_{s}(-t^8+t^4)^{s},    
    \label{eq:hC_TypeI-FSDcodes_Z4}
  \end{IEEEeqnarray}
  where $\beta_s\eqdef\sum_{r=0}^{\lfloor\frac{n-8s}{4}\rfloor}\alpha_{r,s}$ for $s\in [0:\lfloor\nicefrac{n}{8}\rfloor]$
\end{lemma}
\begin{IEEEproof}
  Let $a=1+t, b= \sqrt[4]{1-t^4}, c=1-t$. This gives
  \begin{IEEEeqnarray*}{rCl}
    A& = &a+c=1+t+1-t=2,
    \\
    B^4+C^4& = &(2b)^4+(a-c)^4=16(1-t^4)+(2t)^4=2^4,
    \\
    B^4C^4& = &(2b)^4(a-c)^4=16(1-t^4)(2t)^4=2^8t^4(1-t^4).
  \end{IEEEeqnarray*}
  Hence, from Theorem \ref{thm:inv_secrecy-function_SymmetrizedWeightEnumerator} and Proposition~\ref{prop:swe_TypeI-FSDcodes_Z4}, we have
  \begin{IEEEeqnarray*}{c}
    h_{\code{C}}(t)=\sum_{\substack{r,s\in  \Naturals\cup\{0\} \\0\leq 4r+8s\leq n}}\alpha_{r,s} 2^{n-4r-8s}\cdot (2^4)^r\cdot (2^8t^4(1-t^4))^s=2^{n}\sum_{s=0}^{\lfloor\frac{n}{8}\rfloor}\beta_{s}(-t^8+t^4)^{s},    
  \end{IEEEeqnarray*}
  where $\beta_s\eqdef\sum_{r=0}^{\lfloor\frac{n-8s}{4}\rfloor}\alpha_{r,s}$ for $s\in [0:\lfloor\nicefrac{n}{8}\rfloor]$.
\end{IEEEproof}

In the following, we provide a sufficient condition for a Construction $\textnormal{A}_4$ lattice obtained from a Type I formally self-dual code over $\Integers_4$ to achieve its (strong) secrecy gain at $\tau=1$, or, equivalently, $t=\nicefrac{1}{\sqrt[4]{2}}$. 
\begin{theorem}
  \label{thm:strong-secrecy-gain_TypeI-FSDcodes_Z4}
  Consider an $[n,2^{n}]$ Type I formally self-dual code $\code{C}$ over $\Integers_4$. Let $u(t)\eqdef-t^8+t^4$. If the coefficients $\beta_s$ of $h_{\code{C}}(t)$ expressed in terms of~\eqref{eq:hC_TypeI-FSDcodes_Z4} satisfy
  \begin{IEEEeqnarray}{c}
    \sum_{s=1}^{\lfloor\frac{n}{8}\rfloor} s \beta_s u(t)^{s-1}
    < 0,
    \label{eq:condition_beta_TypeI-FSDcodes_Z4}
  \end{IEEEeqnarray}
   on $t\in (0,1)$, then the secrecy gain of $\ConstrAfour{\code{C}}$ is achieved at $\tau=1$, or, equivalently, $t=\nicefrac{1}{\sqrt[4]{2}}$. Moreover,
  \begin{IEEEeqnarray}{c}
    \xi_{\eConstrAfour{\code{C}}}=\frac{1}{\sum_{s=0}^{\lfloor\frac{n}{8}\rfloor}\beta_s(\frac{1}{4})^s}.
    \label{eq:secrecy-gain_beta_TypeI-FSDcodes_Z4}
  \end{IEEEeqnarray}
\end{theorem}
\begin{IEEEproof}
  It is enough to show that the function $h_{\code{C}}(t)$ as in \eqref{eq:hC_TypeI-FSDcodes_Z4} defined for $0 < t < 1$ achieves its minimum at $t=\nicefrac{1}{\sqrt[4]{2}}$. 

  
  The derivative of $h_{\code{C}}(t)$ satisfies
    \begin{IEEEeqnarray*}{c}
      \frac{\dd h_{\code{C}}(t)}{\dd t} = 2^{n} u'(t) \sum_{s=1}^{\lfloor\frac{n}{8}\rfloor} s \beta_s u(t)^{s-1}
    \end{IEEEeqnarray*}
    and $u'(t)=-8t^7+4t^3=-4t^3(2t^4-1)$. As the hypothesis holds, the behavior of the derivative is dominated
by $u'(t)$. Since
  \begin{IEEEeqnarray*}{c}
    u'(t)
    \begin{cases}
      >0 & \textnormal{if }0< t<\frac{1}{\sqrt[4]{2}},
      \\
      =0 & \textnormal{if }t=\frac{1}{\sqrt[4]{2}},
      \\
      <0 & \textnormal{if }\frac{1}{\sqrt[4]{2}}< t < 1,
    \end{cases}
  \end{IEEEeqnarray*}
  it implies that $h_{\code{C}}(t)$ is decreasing in $t\in(0,\nicefrac{1}{\sqrt[4]{2}})$ and increasing in $t\in(\nicefrac{1}{\sqrt[4]{2}},1)$. Finally, the expression of \eqref{eq:secrecy-gain_beta_TypeI-FSDcodes_Z4} follows directly from \eqref{eq:Xi-ft_ConstructionA_FSDcodes} and \eqref{eq:hC_TypeI-FSDcodes_Z4} by plugging $t=\nicefrac{1}{\sqrt[4]{2}}$. This completes the proof.
\end{IEEEproof}

\begin{example}[The Octacode $\code{O}_8$ and the Gosset lattice $\lattice{E}_8$]
  \label{ex:secrecy-gain_E8}
  Recall that in Example~\ref{ex:E8_octacode}, the octacode $\code{O}_8$ has the swe given by~\eqref{eq:swe_octacode}.

  Now, by comparing the coefficients of $\swe{\code{O}_8}(a,b,c)$ with \eqref{eq:hC_TypeI-FSDcodes_Z4} in Lemma~\ref{lem:hC_TypeI-FSDcodes_Z4}, we find that $\beta_0=1$ and $\beta_{1}=-1$. Applying Theorem~\ref{thm:strong-secrecy-gain_TypeI-FSDcodes_Z4}, those coefficients satisfy \eqref{eq:condition_beta_TypeI-FSDcodes_Z4} ($\beta_1=-1<0$) and thus the secrecy gain is $\xi_{\eConstrAfour{\code{O}_8}}=\inv{\bigl(\sum_{s=0}^{\lfloor\nicefrac{n}{8}\rfloor}\beta_s(\nicefrac{1}{4})^s\bigr)}=\nicefrac{4}{3}$, which coincides with the best-known secrecy gain up to now for $n=8$. Note that we have $\xi_{\eConstrAfour{\code{O}_8}}=\nicefrac{4}{3}>\xi_{\eConstrAfour{\code{C}_8}}\approx 1.282$, where $\code{C}_8$ is the code presented in Example~\ref{ex:FSDcode_dim8}.\hfill\exampleend
\end{example}

\begin{example}[Dimension $26$ code from~\cite{Harada12_1}]
  \label{ex:secrecy-gain_26}
  Consider the $\Integers_4$-linear code $\code{C}_{26}$ with swe given in Appendix~\ref{sec:all-FSD-Z4-codes-swes-SGs}. We have that
  \begin{IEEEeqnarray*}{rCl}
     h_{\code{C}}(t)  =  \swe{\code{C}_{26}}(1+t, \sqrt[4]{1-t^4},1-t) & = & 40894464 t^{16}-81788928 t^{12}+258998272 t^8 \\
     & & -\>218103808 t^4+67108864.
  \end{IEEEeqnarray*}  
  
  Comparing this expression with \eqref{eq:hC_TypeI-FSDcodes_Z4}, we get that $\beta_0=1, \beta_1=-\nicefrac{13}{4}, \beta_2=\nicefrac{39}{64}$ and $\beta_3=0$. 
  From Theorem~\ref{thm:strong-secrecy-gain_TypeI-FSDcodes_Z4}, those coefficients satisfy \eqref{eq:condition_beta_TypeI-FSDcodes_Z4}, i.e., $\sum_{s=1}^{\lfloor\frac{n}{8}\rfloor} s \beta_s\Bigl(\frac{1}{4}\Bigr)^{s-1}u(t) <0$ for all $t \in (0,1)$ and thus the secrecy gain is $\xi_{\eConstrAfour{\code{C}_{26}}}=\inv{\bigl(\sum_{s=0}^{\lfloor\nicefrac{n}{8}\rfloor}\beta_s(\nicefrac{1}{4})^s\bigr)}=\nicefrac{1024}{231} \approx 4.433$. \hfill\exampleend
\end{example}

\subsection{An Upper Bound on the Secrecy Gain of Type I Formally Unimodular Lattices}
\label{sec:an-upper-bound_SG_TypeI-FUL}

In this subsection, a secrecy gain upper bound of Type I formally unimodular lattices is presented, either for even or odd dimensions. The derivation follows to~\cite[Section IV]{LinOggier12_1} similarly. However, our approach is based on the technique in our prior work that is sufficient to prove the Belfiore and Sol{\'{e}} conjecture for Construction~A formally unimodular packings obtained from formally self-dual codes~\cite{BollaufLinYtrehus23_3}. For the sake of completeness, we provide the corresponding proof.
\begin{lemma}
  \label{lem:upper-bound_SG_TypeI-FUL}
  For any $n$-dimensional Type I formally unimodular lattice $\Lambda$ that satisfies the Belfiore and Sol{\'{e}} conjecture with $2\leq n\leq 40$,\footnote{In fact, our upper bound results can be applied to dimension $n=168$ as \cite[Table~II]{LinOggier12_1}. However, as we only investigate formally unimodular Construction $\textnormal{A}_4$ lattices until dimensions $n=32$ in this work, we do not provide the upper bound for dimensions more than $n=40$.} the secrecy gain is bounded from above by
  \begin{IEEEeqnarray*}{rCl}
    \xi_{\Lambda}& \leq &\frac{1}{\vect{\omega}\inv{\mat{S}}\trans{\vect{e}_1}},
  \end{IEEEeqnarray*}
  where $\vect{\omega}=\bigl(1,\nicefrac{3}{4},\ldots,(\nicefrac{3}{4})^\ell)$, $\mat{S}$ is an $(\ell+1) \times (\ell+1)$ matrix whose $(s+1)$-th column contains the first $\ell+1$ coefficients of the power series of $\vartheta_3^{n-8s}(z)\Theta_{\lattice{E}_8}(z)^s$ for $s\in [0:\ell]$, $\vect{e}_{s+1}$ is the vector with a $1$ in the $(s+1)$-th coordinate and zeros elsewhere, and $\ell\eqdef\lfloor\nicefrac{n}{8}\rfloor$.
\end{lemma}
\begin{IEEEproof}
  By definition, a Type I formally unimodular lattice satisfies
  \begin{IEEEeqnarray*}{c}
    \Theta_{\Lambda}(z)=\biggl(\frac{i}{z}\biggr)^{\frac{n}{2}}\Theta_{\Lambda}\Bigl(-\frac{1}{z}\Bigr)
    \textnormal{ and }
    \Theta_{\Lambda}(z+2)=\Theta_{\Lambda}(z).\label{eq:TypeI-FUL_Jacobi-formula}
  \end{IEEEeqnarray*}
  Thus, Hecke's theorem~\cite[Th.~7, Ch.~7]{ConwaySloane99_1} implies that its theta series can be represented by    
  \begin{IEEEeqnarray}{c}
    \Theta_\Lambda(z)=\sum_{r=0}^{\ell}a_r\vartheta_3^{n-8r}(z)\Delta_8^r(z),\label{eq:Hecke-theorem_Delta8}
  \end{IEEEeqnarray}
  where $\Delta_8(z)=\frac{1}{16}\vartheta_2^4(z)\vartheta_4^4(z)$ and $a_r\in\Rationals$.
  
  Since
  \begin{IEEEeqnarray*}{rCl}
    \vartheta^4_2(z)\vartheta^4_4(z)& \stackrel{\eqref{eq:useful-identities-2}}{=} &\bigl[\vartheta_3^4(z)-\vartheta_4^4(z)\bigr]\vartheta_4^4(z)
    =\bigl(\vartheta^2_3(z)\bigr)^4-\bigl[\vartheta^8_3(z)-\vartheta^4_3(z)\vartheta^4_4(z)+\vartheta^8_4(z)\bigr]
    \\
    & = &\bigl(\vartheta^2_3(z)\bigr)^4-\Theta_{\lattice{E}_8}(z),
  \end{IEEEeqnarray*}
  one can show that \eqref{eq:Hecke-theorem_Delta8} can also be expressed by
  \begin{IEEEeqnarray}{c}
    \Theta_\Lambda(z)=\sum_{s=0}^{\ell}n_s\vartheta_3^{n-8s}(z)\Theta_{\lattice{E}_8}^s(z)=\sum_{s=0}^{\ell}n_s\vartheta_3^{n-8s}(z)\bigl[\vartheta^8_3(z)-\vartheta^4_3(z)\vartheta^4_4(z)+\vartheta^8_4(z)\bigr]^s,\label{eq:Hecke-theorem_GossetE8}
  \end{IEEEeqnarray}
  where $n_s\in\Rationals$.
  
  Now, applying the same approach as the proof of~\cite[Th.~36]{BollaufLinYtrehus23_3}, we obtain
  \begin{IEEEeqnarray*}{rCl}
    \inv{\bigl[\Xi_{\Lambda}(\tau)\bigr]}& = &\frac{\Theta_{\Lambda}(z)}{\Theta_{\Integers^n}(z)}=\frac{\sum_{s=0}^{\ell}n_s\vartheta_3^{n-8s}(z)\bigl[\vartheta^8_3(z)-\vartheta^4_3(z)\vartheta^4_4(z)+\vartheta^8_4(z)\bigr]^s}{\vartheta^n_3(z)}
    \nonumber\\
    & = &\sum_{s=0}^{\ell}n_s\frac{\vartheta_3^{n-8s}(z)\bigl[\vartheta^8_3(z)-\vartheta^4_3(z)\vartheta^4_4(z)+\vartheta^8_4(z)\bigr]^s}{\vartheta^{n-8s}_3(z)\vartheta^{8s}_3(z)}
    \nonumber\\
    & = &\sum_{s=0}^{\ell}n_s\left[1-\frac{\vartheta^4_4(z)}{\vartheta^4_3(z)}+\frac{\vartheta^8_4(z)}{\vartheta^8_3(z)}\right]^s
    =\sum_{s=0}^{\ell}n_s[1-t^4+t^8]^s.
  \end{IEEEeqnarray*}
  Hence, if the secrecy function $\Xi_{\Lambda}(\tau)$ achieves its maximum at $\tau=1$, i.e., $t=\nicefrac{1}{\sqrt[4]{2}}$, the (strong) secrecy gain is equal to
  \begin{IEEEeqnarray}{c}
    \xi_{\Lambda}=\Xi_{\Lambda}(1)=\frac{1}{\sum_{s=0}^\ell n_s\bigl(\nicefrac{3}{4}\bigr)^s}.\label{eq:secrecy-gain_TypeI-FUL}
  \end{IEEEeqnarray}
  
  Next, to determine the coefficients of $\{n_s\}_{s=0}^\ell$, we re-write $\Theta_{\Lambda}(z)$ as
  \begin{IEEEeqnarray*}{rCl}
    \Theta_{\Lambda}(z)& \eqdef &\sum_{m=0}^\infty N_m q^m=\sum_{s=0}^{\ell}n_s\vartheta_3^{n-8s}(z)\bigl[\vartheta^8_3(z)-\vartheta^4_3(z)\vartheta^4_4(z)+\vartheta^8_4(z)\bigr]^s
    \IEEEyesnumber\label{eq:equaling-coefficients_theta-sereis_TypeI-FUL}\\
    & = &n_0\vartheta^n_3(z)+n_1\vartheta^{n-8}_3(z)\Theta_{\lattice{E}_8}(z)+\ldots+n_\ell\vartheta^{n-8\ell}_3(z)\Theta_{\lattice{E}_8}^\ell(z)
    \\
    & = &n_0(1+S_{2,1}q+S_{3,1}q^2+\ldots+S_{\ell+1,1}\cdot q^\ell+\ldots)
    \nonumber\\
    && +\>n_1(1+S_{2,2}q+S_{3,2}q^2+\ldots+S_{\ell+1,2}\cdot q^\ell+\ldots)+\ldots
    \nonumber\\
    && +\>n_\ell(1+S_{2,\ell}q+S_{3,\ell}q^2+\ldots+S_{\ell+1,\ell+1}\cdot q^\ell+\ldots)
  \end{IEEEeqnarray*}
  Denote by $\vect{n}\eqdef(n_0,n_1\ldots,n_\ell)$ and $\vect{N}\eqdef(N_0,N_1,\ldots,N_\ell)$, it is not so hard to obtain the following system of linear equations:
  \begin{IEEEeqnarray*}{c}
    \begin{pmatrix}
      1 & 1 & \cdots & 1
      \\
      S_{2,1} & S_{2,2} & \cdots & S_{2,\ell+1}
      \\
      S_{3,1} & S_{3,2} & \cdots & S_{3,\ell+1}
      \\
      \vdots
      \\
      S_{\ell+1,1} & S_{\ell+1,2} & \cdots & S_{\ell+1,\ell+1}
    \end{pmatrix}
    \begin{pmatrix}
      n_0
      \\
      n_1
      \\
      n_2
      \\
      \vdots
      \\
      n_\ell
    \end{pmatrix}
    =
    \begin{pmatrix}
      N_0
      \\
      N_1
      \\
      N_2
      \\
      \vdots
      \\
      N_\ell
    \end{pmatrix}
    \Leftrightarrow \mat{S}\trans{\vect{n}}=\trans{\vect{N}}.
  \end{IEEEeqnarray*}

  As the coefficients between the left-hand-side and right-hand-side of~\eqref{eq:equaling-coefficients_theta-sereis_TypeI-FUL} should be matched, $\mat{S}$ should be invertible. Thus, we have
  \begin{IEEEeqnarray}{c}
    \trans{\vect{n}}=\inv{\mat{S}}\trans{\vect{N}}=\sum_{s=0}^\ell N_s\inv{\mat{S}}\trans{\vect{e}_{s+1}}.\label{eq:n_s-solutions}
  \end{IEEEeqnarray}
  Combining~\eqref{eq:secrecy-gain_TypeI-FUL} and~\eqref{eq:n_s-solutions}, this leads to the upper bound we intend to show, with $\vect{\omega}=\bigl(1,\nicefrac{3}{4},\ldots,(\nicefrac{3}{4})^\ell)$:
  \begin{IEEEeqnarray*}{rCl}
    \xi_{\Lambda}& = &\frac{1}{\sum_{s=0}^\ell n_s\bigl(\nicefrac{3}{4}\bigr)^s}
    =\frac{1}{\vect{\omega}\trans{\vect{n}}}=\frac{1}{\sum_{s=0}^\ell N_s\vect{\omega}\inv{\mat{S}}\trans{\vect{e}_{s+1}}}
    \\
    & \stackrel{(a)}{\leq} &\frac{1}{N_0\vect{\omega}\inv{\mat{S}}\trans{\vect{e}_{1}}}\leq
    \frac{1}{\vect{\omega}\inv{\mat{S}}\trans{\vect{e}_{1}}},
  \end{IEEEeqnarray*}
  where $(a)$ holds because numerically, those terms of $\{\vect{\omega}\inv{\mat{S}}\trans{\vect{e}_{s+1}}\}_{s=1}^{\ell}$, are all positive for dimensions $2\leq n\leq 40$, and the second inequality holds by simply choosing $\vect{N}=(1,0,\ldots,0)$.
\end{IEEEproof}

Exact values for the upper bound on the secrecy gain of Type I formally unimodular lattices, for dimensions $2\leq n\leq 32$ can be found in Table~\ref{tab:table_secrecy-gains_FU-lattices_z4_summary}.

\ifthenelse{\boolean{short_version}}{}{
\section{Flatness Factor of Formally Unimodular Lattices}
\label{sec:flatness-factor_FU-lattices}

This paper mainly focuses on the secrecy gain criterion, which is derived based on the probability analysis for the eavesdropper. Another design criterion for wiretap lattice codes, the \emph{flatness factor}, has been proposed to quantify the mutual information between Eve's received signal and the transmitted confidential message~\cite{LingLuzziBelfioreStehle14_1}. For comparison, we further discuss our results regarding the flatness factor of formally unimodular lattices. We start by reviewing the definition of the flatness factor.

\begin{definition}[Flatness factor~{\cite[Def.~5]{LingLuzziBelfioreStehle14_1}}]
  \label{def:def_flatness-factor}
  Let $\Lambda$ be a lattice with volume $\vol{\Lambda}$. The flatness factor of $\Lambda$ for a parameter $\tau>0$ is defined by
  \begin{IEEEeqnarray}{c}
    \eps_{\Lambda}(\tau)=\max_{\vect{x}\in\set{R}(\Lambda)}\abs{\frac{f_{\tau,\Lambda}(\vect{x})}{\frac{1}{\vol{\Lambda}}}-1},\label{eq:def_flatness-factor}
  \end{IEEEeqnarray} 
  where $\set{R}(\Lambda)$ is a fundamental region and
  \begin{IEEEeqnarray}{c}
    f_{\tau,\Lambda}(\vect{x})\eqdef \tau^{\nicefrac{n}{2}} \sum_{\vect{\lambda}\in\Lambda}\ope^{-\pi\tau\enorm{\vect{x}-\vect{\lambda}}^2},\qquad\vect{x}\in\Reals^n.    
    \label{eq:Guassian-mass_lattice}
  \end{IEEEeqnarray}
\end{definition}

From~\eqref{eq:def_flatness-factor}, it can be seen that the flatness factor indicates how close the function $f_{\tau,\Lambda}(\vect{x})$ is to the uniform distribution of $\set{R}(\Lambda)$. The smaller the flatness factor, the more the function $f_{\tau,\Lambda}(\vect{x})$ behaves like a uniform distribution. In~\cite[Th.~4]{LingLuzziBelfioreStehle14_1}, given a dimension $n$, it was proved that a smaller flatness factor leads to a smaller upper bound on the mutual information leakage to the eavesdropper, demonstrating the secrecy design criterion for wiretap lattice coding. Note that to address the flatness factor and its connections to the secrecy function and secrecy gain in our context, we discuss results in terms of the parameter $\tau>0$, instead of the variance parameter $\sigma^2=\frac{1}{2\pi\tau}$ compared to~\cite{LingLuzziBelfioreStehle14_1}.

\begin{lemma}[{\cite[Prop.~2]{LingLuzziBelfioreStehle14_1}}]
  \label{lem:closed-form-expression_flatnes-factor}
  Consider an $n$-dimensional lattice $\Lambda$. Then, the flatness factor is equal to
  \begin{IEEEeqnarray}{c}
    \eps_{\Lambda}(\tau)=\vol{\Lambda}\tau^{\nicefrac{n}{2}}\Theta_{\Lambda}(i\tau)-1.
    \label{eq:closed-form-expression_flatnes-factor}
  \end{IEEEeqnarray}
\end{lemma}  
\begin{remark}
  \begin{itemize}
  \item Consider two $n$-dimensional lattices $\Lambda_1$ and $\Lambda_2$ with $\vol{\Lambda_1} = \vol{\Lambda_2}$. If $\Xi_{\Lambda_1}(\tau) > \Xi_{\Lambda_2}(\tau)$, it follows from Definition~\ref{def:secrecy_function} that $\Theta_{\Lambda_1}(i\tau) < \Theta_{\Lambda_2}(i\tau)$ and from Lemma~\ref{lem:closed-form-expression_flatnes-factor}, $\epsilon_{\Lambda_1}(\tau) < \epsilon_{\Lambda_2}(\tau)$.
  \item From~\cite[Prop.~12]{BollaufLinYtrehus23_3}, \eqref{eq:closed-form-expression_flatnes-factor} becomes
    \begin{IEEEeqnarray}{c}
      \eps_{\Lambda}(\tau)=\tau^{\nicefrac{n}{2}}\Theta_{\Lambda}(i\tau)-1.
      \label{eq:FF-closed-form-expression_FU-lattices}
    \end{IEEEeqnarray}
    for any formally unimodular lattices $\Lambda$.
  \end{itemize}
\end{remark}

From Definition~\ref{def:secrecy_function} and~\eqref{eq:FF-closed-form-expression_FU-lattices}, it becomes more apparent how the objectives of the two main secrecy criteria, secrecy gain and flatness factor, are related. For secrecy gain, the goal is to determine the smallest $\max_{\tau>0}\frac{\Theta_{\Lambda}(i\tau)}{\vartheta^n_3(i\tau)}$ among all possible formally unimodular lattices of a given dimension $n$. Using our theta series analysis and based on Belfiore and Sol{\'{e}}'s conjecture, we have shown that $\argmax_{\tau>0}\frac{\Theta_{\Lambda}(i\tau)}{\vartheta^n_3(i\tau)}=1$ for those formally unimodular lattices we consider. However, it can be shown that $\eps_{\Lambda}(\tau)$ is a monotonically increasing function of $\tau>0$~\cite[Remark 3]{LingLuzziBelfioreStehle14_1}, and thus $\eps_{\Lambda}(\tau)$ can be made arbitrarily small by decreasing $\tau$. Therefore, the objective for the flatness factor should be to find the best formally unimodular lattice $\Lambda$ such that $\eps_{\Lambda}(\tau)$ is small enough for a given $\tau>0$.

By connecting our performance analysis of secrecy gain, we would like to consider the quantity of flatness factor for a given value of $\tau=1$, which is the point where the secrecy function is maximized. It is also worth mentioning that $\tau=1$ means that the operating amount of the Gaussian noise at Eve side is $\sigma=\nicefrac{1}{\sqrt{2\pi}}$. However, following the discussion in~\cite[Sec.~III]{LinLingBelfiore14_1}, it is expected that $\eps_{\Lambda}(1)$ of even unimodular lattices are unlikely to be smaller than $1$ when $n$ goes to infinity. Thus, one needs to consider $\tau\in(0,1)$ to make $\eps_{\Lambda}(\tau)$ small enough. In the following, we address the relation between the secrecy gain and flatness factor by introducing two secrecy-goodness concepts.

\begin{definition}[{\cite[Def.~47]{BollaufLinYtrehus23_3}}]
  \label{def:optimal-FSD-code}
  A formally self-dual code $\code{C}^\ast$ of length $n$ is called strongly secrecy-optimal if
  \vspace*{-5mm}
  \begin{IEEEeqnarray*}{c}
    \code{C}^\ast=\argmax_{\code{C}\colon\textnormal{formally self-dual}}\xi_{\eConstrAfour{\code{C}}}.
  \end{IEEEeqnarray*}  
\end{definition}

\begin{definition}[{\cite[Def.~49]{BollaufLinYtrehus23_3}}]
  \label{def:good-FSD-code}
  A formally self-dual code $\Integers_4$-linear $\code{C}^\diamond$ of length $n$ is weakly secrecy-optimal if for all $\tau>0$,
  \vspace*{-5mm}
  \begin{IEEEeqnarray*}{c}
    \code{C}^\diamond=\argmax_{\code{C}\colon\textnormal{formally self-dual}}\Xi_{\eConstrAfour{\code{C}}}(\tau).
  \end{IEEEeqnarray*}
\end{definition}

The following proposition can be proved straightforwardly from Definition~\ref{def:secrecy_function} and~\eqref{eq:FF-closed-form-expression_FU-lattices}.
\begin{proposition}
  \label{prop:optimal-flatness-factor_from_secrecy-function}
  Given a length $n\geq 2$, if $\code{C}^\diamond$ is weakly secrecy-optimal, then for all $\tau>0$,
  \begin{IEEEeqnarray*}{c}
    \eps_{\ConstrAfour{\code{C}^\diamond}}(\tau)\leq\eps_{\ConstrAfour{\code{C}}}(\tau),
    \label{eq:optimal-flatness-factor_from_secrecy-function}
  \end{IEEEeqnarray*}
  for any formally self-dual code $\code{C}$ of length $n$ over $\Integers_4$.
\end{proposition}

\begin{figure*}[t!]
  \begin{minipage}[t]{7cm} 
    \input{best_secrecy_gain_pure_double_circulant_n12_v1.tex}
  \end{minipage}
  \hspace{1.00cm} 
  \begin{minipage}[t]{7cm} 
    \centering
    \input{best_secrecy_gain_pure_double_circulant_n18_v1.tex}
  \end{minipage}
  \caption{Secrecy functions $\Xi_{\eConstrAfour{\code{C}}}(\tau)=\inv{[h_{\code{C}}(t(\tau))]}$ in $t\in (0,1)$ based on PDCC searches for $n=12$ and $18$. Observe that $\xi_{\eConstrAfour{\code{C}^\ast}}>\xi_{\eConstrAfour{\code{C}}}$. Moreover, we have $\inv{[h_{\code{C}^\ast}(t)]}>\inv{[h_{\code{C}}(t)]}$ for all $t\in(0,1)$, which implies that $\Xi_{\eConstrAfour{\code{C}^\ast}}(\tau)>\Xi_{\eConstrA{\code{C}}}(\tau)$ for all $\tau>0$, and thus $\eps_{\eConstrAfour{\code{C}^\ast}}(\tau)<\eps_{\eConstrAfour{\code{C}}}(\tau)$ for all $\tau>0$.}
  \label{fig:best_secrecy_gain_PDCCs_n12-and-18}
\end{figure*}
  
Proposition~\ref{prop:optimal-flatness-factor_from_secrecy-function} implies that the best secrecy-good code in terms of flatness factor is determined by the weakly secrecy-optimality of codes. Furthermore, to demonstrate the connection to strongly secrecy-optimality, we have performed exhaustive PDCC and BDCC searches to find the best secrecy gain for all even lengths up to $n=20$ (where the algorithm will be described in Section~\ref{sec:secrecy-gain-comparisons-numerical-results}). Numerical results for lengths $n=12$ and $18$ are illustrated in Figure~\ref{fig:best_secrecy_gain_PDCCs_n12-and-18}, and one can see that strongly secrecy-optimality is equivalent to weakly secrecy-optimality for formally self-dual codes. I.e., $\argmax_{\code{C}}\xi_{\eConstrAfour{\code{C}}}\equiv\argmax_{\code{C}}\Xi_{\eConstrAfour{\code{C}}}(\tau)$ for any $\tau>0$. Hence, if the best code is determined regarding secrecy gain, then such code is also the best in terms of the flatness factor.
This indicates that our code design based on the secrecy gain is also useful regarding the flatness factor, as the larger the secrecy gain of a lattice, the smaller the flatness factor is at any $\tau>0$. 

According to~\cite[Cor.~3]{LingLuzziBelfioreStehle14_1}, it is necessary to have $\eps_{\Lambda}(\tau)$ grow less than $\nicefrac{1}{n}$ to achieve a smaller upper bound on the mutual information leakage to the eavesdropper. Hence, we have provided some numerical values of $\tau^{(n)}_{\eConstrAfour{\code{C}}}\eqdef\max\{\tau\in (0,1)\colon\eps_{\eConstrAfour{\code{C}}}(\tau)\leq\nicefrac{1}{n}\}$ for those formally self-dual $\Integers_4$-linear codes $\code{C}$ achieve large secrecy gains in the last column of Table~\ref{tab:table_secrecy-gains_FU-lattices_z4_summary}.
}

\section{Secrecy Criteria Comparisons and Numerical Results}
\label{sec:secrecy-gain-comparisons-numerical-results}

\subsection{Formally Self-Dual $\Integers_4$-Linear Codes $\code{C}=\code{A}_1+2\code{A}_2$}
\label{sec:secrecy-gain_FSD-Z4-codes_A1plus2A2}

The formally self-dual code $\code{C}_{12}$ in Example~\ref{ex:codes_dim12} has secrecy gain $\xi_{\eConstrAfour{\code{C}_{12}}} \approx 1.6$, which coincides with the performance of self-dual codes. However, it is slightly worse than the best record until now, $1.657$. On the other hand, we can notice that optimizing $d_{\textnormal{Lee}}$ of formally self-dual codes does not imply higher secrecy gain: for length-$12$ codes, $\xi_{\eConstrAfour{\code{C}_{12}}}\approx 1.6$ is achieved by $\code{C}_{12}$ with $d_\textnormal{Lee}=4$, while a Lee-optimal code with $d_{\textnormal{Lee}} = 6$~\cite[Tab.~IV]{YooLeeKim17_1} only has secrecy gain $\xi_{\eConstrAfour{\code{C}_{12}}}\approx 1.456 < 1.6$.

Potential candidates for good secrecy gain performance in this class of construction are obtained from Reed-Muller codes, as we describe next.
\begin{definition}[{Reed-Muller codes~\cite[Ch.~13]{MacWilliamsSloane77_1}}]
  \label{def:reed-muller-codes}
  For a given $v\in\Naturals$, the $r$-th order binary Reed-Muller code $\code{R}(r,v)$ is a linear $[n=2^v,k=\sum_{i=0}^r\binom{v}{i}]$ code for $r\in [0:v]$, constructed as the vector space spanned by the set of all $v$-variable Boolean monomials of degree at most $r$.
\end{definition}

Reed-Muller codes have interesting properties, such as being nested. In order to get $\Integers_4$-linear codes from pairs of Reed-Muller binary codes, we still need to guarantee that the chain is closed under the element-wise product, which is true for the chains described next.

A result connecting the construction of $\Integers_4$-linear codes and Reed-Muller chains is the following.
\begin{proposition}[{\cite[Ex.~12.8]{Wan97_1}}]
  \label{prop:reedmuller_unimodular}
  The $\Integers_4$-linear code ${\code{C}}_{2^m} \eqdef \code{R}(1,m)+2\code{R}(m-2,m)$ induces a unimodular lattice $\eConstrAfour{{\code{C}}_{2^m}} = \tfrac{1}{2} \left( {\code{C}}_{2^m} + 4\mathbb{Z}^{2^m} \right)$. 
\end{proposition}

\begin{example}
  \label{ex:BWs_dims16-32}
  Proposition~\ref{prop:reedmuller_unimodular} gives an even unimodular lattice in dimension $16$ obtained via $\Lambda_{\textnormal{A}_4}({\code{C}}_{16})$ from ${\code{C}}_{16} =  \code{R}(1,4)+2\code{R}(2,4)$, which is isomorphic to the lattice $\lattice{E}_8 \times \lattice{E}_8$. We have that, $\xi_{\eConstrAfour{{\code{C}}_{16}}} \approx 1.778 < 2.141$, see Table~\ref{tab:long-table_FSD-Z4-codes-swes-SGs} in Appendix~\ref{sec:all-FSD-Z4-codes-swes-SGs}. If one considers
  \begin{IEEEeqnarray}{c}
    {\code{C}}_{32}  =  \code{R}(1,5)+2\code{R}(3,5),
    \label{eq:c32}
  \end{IEEEeqnarray}
  then $\lattice{BW}_{32}=\sqrt{2}\eConstrAfour{{\code{C}}_{32}}=\tfrac{\sqrt{2}}{2} \left({\code{C}}_{32}+4\mathbb{Z}^{32}\right)$ is a unimodular lattice in dimension $32$.
  
  The code ${\code{C}}_{32}$ as in~\eqref{eq:c32} is self-dual. Hence, Theorem~\ref{thm:inv_secrecy-function_SymmetrizedWeightEnumerator} can be applied and we get $\xi_{\eConstrAfour{{\code{C}}_{32}}} \approx 7.11$, which is the best-known secrecy gain up to now for such dimension~\cite[p.~5698]{OggierSoleBelfiore16_1}.\hfill\exampleend
\end{example}

\begin{remark}
  Note that Theorem~\ref{thm:FSD-Z4codes_A1plus2A2} gives a sufficient but not necessary condition for formal self-duality. Relaxing these conditions can, among other consequences, remove restrictions on $\code{A}_1,\code{A}_2$ and may allow improved constructions.
\end{remark}

\subsection{DCC and Its Odd Extension}
\label{sec:DCC-odd-extension}

The best even-length DCCs and odd extension codes in terms of secrecy gain presented in this work are obtained based on numerical searches. We first present a simple result showing that using particular choices of $\vect{a}$ and $\vect{c}$, the swe of $\code{C}_{\textnormal{oext}}$ can be equal to the one of $\code{C}$, generated by $G^{\code{C}}=(\mat{I}\,\,\,\mat{B})$, and as a consequence, their secrecy gains coincide.
\begin{proposition} 
  \label{prop:swe_odd-extension_equal_swe_a+c}
  Let $\code{C}$ be a $[2\eta,M]$ code over $\Integers_4$ with generator matrix $\mat{G}^{\code{C}} = (\mat{I}\,\,\,\mat{B})$, $\eta\in\Naturals$. Consider a $[2\eta+1, 2M]$ odd extension code $\code{C}_{\textnormal{oext}}$ with the generator matrix
  \begin{IEEEeqnarray*}{c}
    \mat{G}^{\code{C}_{\textnormal{oext}}} =
    \begin{pNiceMatrix}
      {\mat{I}_{\eta}} & \trans{\vect{0}} & \mat{B}_{\eta}
      \\
      \vect{0}         & 2                & \vect{0}
    \end{pNiceMatrix}.
  \end{IEEEeqnarray*}
  Then, $\swe{\code{C}_{\textnormal{oext}}}(a,b,c)=\swe{\code{C}}(a,b,c)\cdot(a+c)$ and $\Xi_{\eConstrAfour{\code{C}}}(\tau) = \Xi_{\eConstrAfour{\code{C}_{\textnormal{oext}}}}(\tau)$.
\end{proposition}
\begin{IEEEproof}
   A codeword $\vect{v}_{\textnormal{oext}}\in\code{C}_\textnormal{oext}$ can be expressed as
  \begin{IEEEeqnarray*}{c}
    \vect{v}_{\textnormal{oext}}=(u_1,\ldots,u_{\eta},u_{\eta+1})\mat{G}_{\code{C}_{\textnormal{oext}}}=\biggl(u_1,\ldots,u_{\eta}, 2 u_{\eta+1},\sum_{i=1}^{\eta} b_{i,1}u_i,\cdots,\sum_{i=1}^{\eta} b_{i,\eta}u_i\biggr),
  \end{IEEEeqnarray*}
  where $(u_1,\ldots,u_{\eta},u_{\eta+1})\in\Integers_4^{\eta}\times\Integers_2$ is a message vector.
  
  For convenience, denote by $\vect{v}_\textnormal{oext}\eqdef[\vect{u},2u_{\eta+1},\vect{w}]$ and $\vect{v}\eqdef\vect{u}\mat{G}^{\code{C}}=[\vect{u},\vect{w}]$. Then, we have
  \begin{IEEEeqnarray*}{rCl}
    \IEEEeqnarraymulticol{3}{l}{%
      \swe{\code{C}_\textnormal{oext}}(a,b,c)
    }\nonumber\\*\quad%
    & = &\sum_{[\vect{u}, 2u_{\eta+1},\vect{w}]\in\code{C}_\textnormal{oext}}a^{\na{0}{\vect{v}_\textnormal{oext}}}b^{\na{1}{\vect{v}_\textnormal{oext}}+\na{3}{\vect{v}_\textnormal{oext}}}c^{\na{2}{\vect{v}_\textnormal{oext}}}
    \\
    & \stackrel{(a)}{=} &\sum_{[\vect{u},0,\vect{w}]\in\code{C}_\textnormal{oext}}a^{\na{0}{\vect{v}_\textnormal{oext}}}b^{\na{1}{\vect{v}_\textnormal{oext}}+\na{3}{\vect{v}_\textnormal{oext}}}c^{\na{2}{\vect{v}_\textnormal{oext}}}+\sum_{[\vect{u},2,\vect{w}]\in\code{C}_\textnormal{oext}}a^{\na{0}{\vect{v}_\textnormal{oext}}}b^{\na{1}{\vect{v}_\textnormal{oext}}+\na{3}{\vect{v}_\textnormal{oext}}}c^{\na{2}{\vect{v}_\textnormal{oext}}}
    \\
    & = &\sum_{[\vect{u},\vect{w}]\in\code{C}}a^{\na{0}{\vect{v}}+1}b^{\na{1}{\vect{v}}+\na{3}{\vect{v}}}c^{\na{2}{\vect{v}}}+\sum_{[\vect{u},\vect{w}]\in\code{C}}a^{\na{0}{\vect{v}}}b^{\na{1}{\vect{v}}+\na{3}{\vect{v}}}c^{\na{2}{\vect{v}}+1}
    \\
    & = &\swe{\code{C}}(a,b,c)\cdot(a+c),\label{eq:swe_odd-extension_equal_swe_a+c}\IEEEyesnumber
  \end{IEEEeqnarray*}
  where $(a)$ holds since $u_{\eta+1}\in\Integers_2$.

  Next, using Theorem~\ref{thm:inv_secrecy-function_SymmetrizedWeightEnumerator} and~\eqref{eq:swe_odd-extension_equal_swe_a+c}, we have
  \begin{IEEEeqnarray*}{rCl}
    \Xi_{\eConstrAfour{\code{C}_{\textnormal{oext}}}}(\tau)
    & = & \frac{2^{2\eta + 1}}{\swe{\code{C}_{\textnormal{oext}}}\bigl(1+t, \sqrt[4]{1-t^4}, 1-t\bigr)} =  \frac{2^{2\eta+1}}{\swe{\code{C}}\bigl(1+t, \sqrt[4]{1-t^4}, 1-t\bigr) \cdot (1+t+1-t)}
    \\
    & = & \frac{2^{2\eta}}{\swe{\code{C}}\bigl(1+t, \sqrt[4]{1-t^4}, 1-t\bigr)} 
    =  \Xi_{\eConstrAfour{\code{C}}}(\tau).
  \end{IEEEeqnarray*}
\end{IEEEproof}

In Table~\ref{tab:long-table_FSD-Z4-codes-swes-SGs} of Appendix~\ref{sec:all-FSD-Z4-codes-swes-SGs}, the codes with parameters $[5,2^5, 2]$, $[7,2^7,2]$, and $[9,2^9,2]$ are obtained according to Proposition~\ref{prop:swe_odd-extension_equal_swe_a+c} from the 
corresponding even length codes $[4, 2^4, 2],$ $[6,2^6,4]$, and $[8,2^8,6]$. The next example points out that the swe's between $\code{C}_\textnormal{oext}$ and $\code{C}$ can be different.
\begin{example}
  \label{ex:distinct_swe_n12to13} 
    Consider a $[12,2^{12}]$ code $\code{C}_{\textnormal{pdc}}$ constructed as in~\eqref{eq:double-circulant-matrices}, with
  the same $\mat{B}^{\textnormal{pc}}$ in Example~\ref{ex:n13k6_oextCode}. Observe that $\xi_{\eConstrAfour{\code{C}_{\textnormal{pdc}}}} \approx 1.657$. If we now consider the same $[13,2^{13}]$ code $\code{C}_{\textnormal{oext}}$ as in Example~\ref{ex:n13k6_oextCode}, 
  we would get $\swe{\code{C}_{\textnormal{oext}}}(a,b,c) \neq \swe{\code{C}_{\textnormal{pdc}}}(a,b,c)\cdot (a+c)$ (the respective swes can be found in Table~\ref{tab:long-table_FSD-Z4-codes-swes-SGs}). However, one can see that $\xi_{\eConstrAfour{\code{C}_{\textnormal{oext}}}} \approx 1.704 > \xi_{\eConstrAfour{\code{C}_{\textnormal{pdc}}}}$.  \hfill\exampleend 
\end{example}
 
Proposition~\ref{prop:swe_odd-extension_equal_swe_a+c} points out that every secrecy gain in a certain even dimension $2\eta$ can also be achieved by an odd extension code in dimension $2\eta+1$ with certain choices of $\vect{a}$ and $\vect{c}$, meaning that the secrecy gain of lattices constructed from $\code{C}_{\textnormal{oext}}$ can be at least as good as the ones from lattices constructed from codes $\code{C}$ in the precedent even dimension. On the other hand, Example~\ref{ex:distinct_swe_n12to13} shows that there are cases when the swe's differ, which could lead to improvements. To this end, we perform an exhaustive odd extension code search from DCCs in terms of secrecy gain, which can be briefly summarized by the following algorithm.

\begin{description}
\item[Step 1:] We fix a DCC $\code{C}$ that can be generated according to \eqref{eq:double-circulant-matrices} (the code can be either PDCC or BDCC). We select two initial vectors $\vect{a}=\vect{c}=\vect{0}$ and construct the corresponding odd extension code $\code{C}_{\textnormal{oext}}$ with a generator matrix as in~\eqref{eq:def_odd-extension-G}. We compute the secrecy gain $\xi_{\eConstrAfour{\code{C}_\textnormal{oext}}}$ according to Theorem~\ref{thm:inv_secrecy-function_SymmetrizedWeightEnumerator}.
\item[Step 2:] We select another two nonzero vectors $\vect{a},\vect{c}$ and construct the new odd extension code $\code{C}'_{\textnormal{oext}}$. For this new odd extension code we verify its formally-self duality via the MacWilliams identity~\eqref{eq:swe-MacWilliams-identity_FSD-codes_Z4} and compute the corresponding secrecy gain $\xi_{\eConstrAfour{\code{C}'_\textnormal{oext}}}$ and if $\xi_{\eConstrAfour{\code{C}'_\textnormal{oext}}}>\xi_{\eConstrAfour{\code{C}_\textnormal{oext}}}$, we replace $\code{C}_\textnormal{oext}$ by $\code{C}'_\textnormal{oext}$. Otherwise, we keep the current $\code{C}_\textnormal{oext}$.
\item[Step 3:] We repeat \textbf{Step 2} until all the possible vectors of $\vect{a},\vect{c}\in\Integers_2^\eta$ are selected.
\end{description}

Note that the search of the best even-length DCCs with respect to secrecy gain is done similarly in the above algorithm, where we initially set $\xi_{\eConstrAfour{\code{C}}}=1$ as $1$ is the trivial secrecy gain that can be achieved by an uncoded lattice, i.e., $\Integers^n$. If we find a better formally-self dual code $\code{C}'$ such that $\xi_{\eConstrAfour{\code{C}'}}>\xi_{\eConstrAfour{\code{C}}}$, we update the latest best-found $\xi_{\eConstrAfour{\code{C}}}$. For PDCC, we run over all possible choices of vectors $\vect{r}\in\Integers_4^\eta$ for $\mat{B}^{\textnormal{pc}}_{\eta}$ in~\eqref{eq:double-circulant-matrices}, while for BDCCs, an exhaustive search over all possible elements of $\alpha,\beta,\gamma\in\Integers_4$, and vectors $\vect{r}\in\Integers_4^{\eta-1}$ for $\mat{B}^{\textnormal{bc}}_{\eta}$ in~\eqref{eq:double-circulant-matrices}, is performed.

As a result, the Construction $\textnormal{A}_4$ lattices obtained from DCCs and odd extension codes can achieve a better secrecy gain than the ones known in the literature. We summarize and discuss the best-found codes in Section~\ref{sec:summary}.

\subsection{Gray Map to Binary Codes}
\label{sec:gray-maps}

It is well known~\cite{HammonsKumarCalderbankSloaneSole94_1} that codes of length $n$ over $\Integers_4$ can be mapped into binary codes of length $2n$ through the so called Gray map $\psi\colon\Integers_4\rightarrow\Field_2\times\Field_2$, which acts coordinate-wisely as
\begin{IEEEeqnarray*}{c}
  0 \mapsto (0,0),\quad 1 \mapsto (0,1),\quad 2 \mapsto (1,1),\quad 3 \mapsto (1,0).
\end{IEEEeqnarray*}
We denote $\code{C}_{\textnormal{g}}=\psi(\code{C}) \subseteq \Field_2^n$, where $\code{C} \subseteq \Integers_4^n$ is a linear code. The code $\code{C}_{\textnormal{g}}$ is not necessarily linear. However, due to the distance-preserving properties, it can be deduced from~\cite[Th.~1]{Dougherty12_1} that if $\code{C}$ is formally self-dual over $\Integers_4$ with respect to the swe, then $\code{C}_{\textnormal{g}}$ is formally self-dual with respect to the Hamming weight enumerator.

The secrecy gain analysis of formally unimodular packings derived from the Gray map was originally discussed in~\cite{BollaufLinYtrehus23_3}, where improvements were demonstrated with respect to lattices. Hence, it seems like a natural direction to investigate whether the Gray map of the secrecy-good codes over $\Integers_4$ constructed in this paper yields improvements with respect to what was previously achieved in the binary case.

Recall from~\cite[eq.~(2.15)]{Wan97_1} that the weight enumerator of the code $\code{C}_{\textnormal{g}} \subseteq \Field_2^{2n}$ can be obtained by the respective swe of $\code{C} \subseteq \Integers_4^n$ according to
\begin{IEEEeqnarray*}{c}
  W_{\code{C}_{\textnormal{g}}}(x,y)=\sum_{\vect{c}\in\code{C}_{\textnormal{g}}} x^{n-\Hwt{\vect{c}}}y^{\Hwt{\vect{c}}} = \swe{\code{C}}(x^2,xy,y^2).
  \label{eq:relation_weight-enumerators}
\end{IEEEeqnarray*}
By performing this operation and calculating the secrecy gain of the Construction A lattice obtained from $\code{C}_{\textnormal{g}}$ as in~\cite[Th.~2]{BollaufLinYtrehus22_1}, we got the following improvements compared to previous literature:
\begin{itemize}
\item $n=44:$ $\code{C} \subseteq \Integers_4^{22}$ is the $[22,2^{22},10]$ code $\code{D}_{4,22}$ presented in~\cite[p.~230, Prop.~4.2]{BachocGulliverHarada00_1} (see also Ex.~\ref{ex:n22_FSD-code_Z4}) and $\code{C}_{\textnormal{g}}=\psi(\code{C})$ is a $(44,2^{22},10)$ nonlinear code. The secrecy gain of the Construction A nonlattice packing $\GammaA{\code{C}_{\textnormal{g}}}$ is $\xi_{\eGammaA{\code{C}_{\textnormal{g}}}} \approx 16.957$, which is larger than what was previously found in~\cite{BollaufLinYtrehus23_3, PerssonBollaufLinYtrehus23_1}. 
  
\item $n=48:$ $\code{C} \subseteq \Integers_4^{24}$ is the $[24,2^{24},12]$ code from~\textnormal{\cite[p.~494]{HuffmanPless03_1}} and $\code{C}_{\textnormal{g}}=\psi(\code{C})$ is a $(48,2^{24},12)$ nonlinear code. The secrecy gain of $\eGammaA{\code{C}_{\textnormal{g}}}$ is $\xi_{\eGammaA{\code{C}_{\textnormal{g}}}} \approx 23.257$, which outperforms the results presented for this dimension in~\cite{BollaufLinYtrehus23_3, PerssonBollaufLinYtrehus23_1}. 
\end{itemize}

The respective weight enumerators are in Appendix~\ref{sec:weight-enumerators-codes-gray-44-48}. The nonlinearity of the codes presented above was verified computationally.

\subsection{Summary}
\label{sec:summary}

\begin{table}[t!]
  \centering
  \caption{Comparison of (strong) secrecy gains of Construction $\textnormal{A}_4$ lattices for (some) dimensions $4\leq n \leq 32$. 
  }
  \label{tab:table_secrecy-gains_FU-lattices_z4_summary}
  \vskip -2.0ex
  \Scale[0.85]{\begin{IEEEeqnarraybox}[
    \IEEEeqnarraystrutmode
    \IEEEeqnarraystrutsizeadd{3.5pt}{3.0pt}]{V/c/V/c/V/c/V/c/V/c/V/c/V}
    \IEEEeqnarrayrulerow\\
    & [n, M, d_{\textnormal{Lee}}] 
    && \textnormal{Reference~/~Type}
    && \xi_{{\Lambda_{\textnormal{A}_4}(\code{C})}}
    && \textnormal{Best-known}~\textnormal{\cite{BollaufLinYtrehus22_1, LinOggier13_1, PerssonBollaufLinYtrehus23_1}}
    && \textnormal{Upper bound (Type I)}
    && {\tau^{(n)}_{\eConstrAfour{\code{C}}}} 
    &\\
    \hline\hline
    & [4,2^4,2] && \textnormal{bdc, Ex.~\ref{ex:double_circulant_fsd}} && \mathbf{1.052} && 1 && 1 && {0.939} &
    \\*\IEEEeqnarrayrulerow \\
    & [5,2^5,2] && \textnormal{oext} && \mathbf{1.052} && - && 1 && {0.835} &
    \\*\IEEEeqnarrayrulerow \\
    & [6,2^6,4]  && \textnormal{\cite[p.~125]{GulliverHarada01_1}} && \mathbf{1.172}  && \mathbf{1.172} && 1 && {0.853} &
    \\*\IEEEeqnarrayrulerow \\
    & [7,2^7,2] && \textnormal{oext} && {\mathbf{1.172}} && -  && 1 && {0.788} &
    \\*\IEEEeqnarrayrulerow \\
    & [8,2^8,6]  && \textnormal{\cite[p.~505]{HuffmanPless03_1}, Ex.~\ref{ex:E8_octacode} / II} && \mathbf{1.333}  && \mathbf{1.333} && 1.333 && {0.831} &
    \\*\IEEEeqnarrayrulerow \\
    & [8,2^8,6]  && \textnormal{\cite[p.~84]{BetsumiyaHarada03_1}, Ex.~\ref{ex:FSDcode_dim8}} && 1.282  && \mathbf{1.333} && 1.333 && {0.801} &
    \\*\IEEEeqnarrayrulerow \\
    & [9,2^9,4]  && \textnormal{oext} && {\mathbf{1.333}}  && - && 1.391 && {0.776} &
    \\*\IEEEeqnarrayrulerow \\
    & [10,2^{10},6] && \textnormal{\cite[p.~127]{GulliverHarada01_1}}  && \mathbf{1.478} && \mathbf{1.478} && 1.455 && {0.802} & 
    \\*\IEEEeqnarrayrulerow \\
    & [11,2^{11},4] && \textnormal{opdc} && {\mathbf{1.512}} && - && 1.524 && {0.773} &
    \\*\IEEEeqnarrayrulerow \\
    & [12,2^{12},6] &&  \textnormal{pdc, Ex.~\ref{ex:distinct_swe_n12to13}}  && 1.657 && 1.657 && 1.6 && {0.787} &
    \\*\IEEEeqnarrayrulerow \\
    & [12,2^{12},4] &&  \textnormal{Ex.~\ref{ex:codes_dim12}~/~I}  && \mathbf{1.6} && 1.657 && 1.6 && {0.767} &
    \\*\IEEEeqnarrayrulerow \\  
    & [13,2^{13},4] && \textnormal{opdc, Exs.~\ref{ex:n13k6_oextCode}, \ref{ex:distinct_swe_n12to13}}  &&  \mathbf{1.704} && -  && 1.684  && {0.764} &
    \\*\IEEEeqnarrayrulerow \\
    & [14,2^{14},7]  && \textnormal{bdc} &&  \mathbf{1.876}  && 1.875 && 1.778  && {0.780} & \\
    \IEEEeqnarrayrulerow \\
    & [15,2^{15},6]  && \textnormal{obdc} && \mathbf{1.972} && 1.882 && 1.882 && {0.771} & \\
    \IEEEeqnarrayrulerow \\
    & [16,2^{16},8] && \textnormal{bdc}  && 2.147 &&  \mathbf{2.207} && 2.246 && {0.780} &  \\
    \IEEEeqnarrayrulerow \\
    & [16,2^{16},8] && \textnormal{Ex.~\ref{ex:BWs_dims16-32}~/~II}  && 1.778 && \mathbf{2.207} && 2.246 && {0.701} &  \\
    \IEEEeqnarrayrulerow \\
    & [17,2^{17},4] && \textnormal{opdc}  && \mathbf{2.203} && 2.133 && 2.387 && {0.757} & \\
    \IEEEeqnarrayrulerow \\
     & [18,2^{18},7] &&  \textnormal{pdc}  && 2.458 && \mathbf{2.485} && 2.541 && {0.779} & \\
    \IEEEeqnarrayrulerow  \\
    & [19,2^{19},4] &&  \textnormal{obdc}  && \mathbf{2.641} && 2.462 && 2.709  && {0.780} & \\
    \IEEEeqnarrayrulerow  \\
    & [20,2^{20},4] && \textnormal{pdc}  &&  \mathbf{2.868} &&  \mathbf{2.868} &&  2.893 && {0.784} & \\
    \IEEEeqnarrayrulerow \\
    & [21,2^{21},6] && \textnormal{\cite[App.~A]{PlessSoleQian97_1}}  && \mathbf{2.909} && \mathbf{2.909} && 3.094 && {0.759} & \\
    \IEEEeqnarrayrulerow \\
    & [22,2^{22},10] &&  \textnormal{\cite[p.~230]{BachocGulliverHarada00_1}, Ex.~\ref{ex:n22_FSD-code_Z4}}  && \mathbf{3.403} && 3.335 && 3.314 && {0.800} & \\
    \IEEEeqnarrayrulerow \\
    & [23,2^{23},10] &&  \textnormal{\cite[App.~A]{PlessSoleQian97_1}}  && \mathbf{3.556} && \mathbf{3.556} && 3.556 && {0.789} & \\
    \IEEEeqnarrayrulerow \\
     & [24,2^{24},12] &&  \textnormal{\cite[p.~494]{HuffmanPless03_1}}\textnormal{~/~II}  && \mathbf{4.063} && \mathbf{4.063} && 4.063 && {0.816} & \\
    \IEEEeqnarrayrulerow \\
     & [26,2^{26},6] && \textnormal{\cite[pp.~535--536]{Harada12_1}, Ex.~\ref{ex:secrecy-gain_26}}\textnormal{~/~I}  && \mathbf{4.433} && 4.356 && 4.68 && {0.792} & \\
    \IEEEeqnarrayrulerow \\
    & [31,2^{31},6] && \textnormal{\cite[App.~A]{PlessSoleQian97_1}}  && \mathbf{6.564} && - && 6.774 && {0.804} & \\
    \IEEEeqnarrayrulerow \\
    & [32,2^{32},8] && \textnormal{\cite[App.~A]{PlessSoleQian97_1}, Ex.~\ref{ex:BWs_dims16-32}}  && \mathbf{7.111} && \mathbf{7.111} && 7.583 && {0.809} & \\
    \IEEEeqnarrayrulerow
  \end{IEEEeqnarraybox}}
\vspace{-5.0mm}
\end{table}

Table~\ref{tab:table_secrecy-gains_FU-lattices_z4_summary} (and Table~\ref{tab:long-table_FSD-Z4-codes-swes-SGs}, which is an extended version that contains the swe's) summarizes the best results in terms of secrecy gain 
in even and odd dimensions. Boldfaced values for the secrecy gain correspond to the best-found code/lattice with the respective length/dimension. The values 
in the last column of Table~\ref{tab:table_secrecy-gains_FU-lattices_z4_summary} correspond to the  numerical values of $\tau^{(n)}_{\eConstrAfour{\code{C}}}$ defined in Section~\ref{sec:flatness-factor_FU-lattices}. 

Note that we are comparing the best-known lattices and nonlattice packings constructed from binary codes via Construction A~\cite{BollaufLinYtrehus23_3, PerssonBollaufLinYtrehus23_1}. Interestingly, we found some even-dimensional formally unimodular lattices outperforming the upper bounds on the secrecy gain of Type I formally unimodular lattices, e.g., $n=14, 22$. We also observe that the Construction $\textnormal{A}_4$ lattices obtained from DCCs generally achieve better secrecy gain than the ones obtained from nested binary codes. We remark that since not every formally unimodular lattice is Type I (and neither Type II), it is possible to obtain a good secrecy gain of a formally unimodular lattice that exceeds the upper bound on the secrecy gain of Type I formally unimodular lattices.

In addition to improvements on the secrecy gain provided by the even-length formally unimodular lattices, the odd extension codes proposed in this paper also outperform the best secrecy gains previously achieved in such odd dimensions by unimodular lattices~\cite[Tab.~II]{LinOggier13_1}. Some particular cases also even exceed the upper bounds on the secrecy gain of Type I formally unimodular lattices, such as $n=7, 13$, and $15$, with secrecy gains given respectively by $\xi_{\eConstrAfour{\code{C}_{7}}} \approx 1.172$, $\xi_{\eConstrAfour{\code{C}_{13}}} \approx 1.704$, and $\xi_{\eConstrAfour{\code{C}_{15}}} \approx 1.972$.

We use the following acronyms for Table~\ref{tab:table_secrecy-gains_FU-lattices_z4_summary}: 
\begin{itemize}
\item pdc: a PDCC with a generator matrix $\mat{G}^{\code{C}_{\textnormal{pdc}}}$ as in~\eqref{eq:double-circulant-matrices};
\item bdc: a BDCC with a generator matrix $\mat{G}^{\code{C}_{\textnormal{bdc}}}$ as in~\eqref{eq:double-circulant-matrices};
\item oext: an odd extension code obtained according to Proposition~\ref{prop:swe_odd-extension_equal_swe_a+c}, from the respective precedent even dimension;
\item opdc: an odd extension code where its $\mat{B}$ as in~\eqref{eq:def_odd-extension-G} is $\mat{B}^{\textnormal{pc}}$;
\item obdc: an odd extension code where its $\mat{B}$ as in~\eqref{eq:def_odd-extension-G} is $\mat{B}^{\textnormal{bc}}$.
\end{itemize}

\ifthenelse{\boolean{short_version}}{
\section{Conclusion}
\label{sec:conclusion}
}{
\section{Conclusion and Discussion}
\label{sec:conclusion-discussion}
}

This work investigated the secrecy gain of Construction $\textnormal{A}_4$ lattices obtained from formally self-dual $\Integers_4$-linear codes. To this end, several novel code constructions of formally self-dual $\Integers_4$-linear codes of both even and odd lengths were presented. In contrast to binary codes, there are odd-length formally self-dual codes over the integers modulo $4$. Thus, we can address odd-dimensional formally unimodular lattices, which are not extensively explored in the past literature. The theta series of Construction $\textnormal{A}_4$ lattices obtained from formally self-dual $\Integers_4$-linear codes were derived, along with a universal approach to determine their secrecy gains. We found that it is possible to obtain a better secrecy gain from Construction $\textnormal{A}_4$ formally unimodular lattices than that from Construction $\textnormal{A}$ formally unimodular lattices. 
Extensive code searches were performed to support these observations.

Moreover, as we briefly discussed in Section~\ref{sec:flatness-factor_FU-lattices}, our analysis techniques for the ratio of the theta series of the studied lattice to $\vartheta_3^n(i\tau)$ (i.e., $\inv{[\Xi_{\Lambda}(\tau)]}$) can be directly applied to the flatness factor $\eps_{\Lambda}(\tau)$ at the point $\tau=1$. However, to make the upper bound on the information leakage to the eavesdropper small, it is necessary to have $\eps_{\Lambda}(\tau)$ grow less than $\nicefrac{1}{n}$, as stated in~\cite[Cor.~3]{LingLuzziBelfioreStehle14_1}, and it is expected to consider $\eps_{\Lambda}(\tau)$ for $\tau<1$ of unimodular lattices (see the indications from \cite[Sec.~III]{LinLingBelfiore14_1}). By exhaustive PDCC and BDCC searches for all even lengths up to $20$, we demonstrated that our code design based on the secrecy gain also translates into determining the best code in terms of the flatness factor $\eps_{\Lambda}(\tau)$ at any given $\tau>0$. To limit the scope of this paper, we postpone the investigation of theoretical relations between $\Xi_{\Lambda}(\tau)$ and $\eps_{\Lambda}(\tau)$ to the future work.

On the other hand, while our main objective has been to study the secrecy gain of lattices obtained from  $\Integers_4$-linear codes, it is natural to ask whether one can use linear codes over $\Integers_m$ with $m\neq 2, 4$ to further improve the secrecy gain performance. This is indeed possible, but note that our performance analysis mainly relies on the fact that there are concise relations between the swe's of $\Integers_4$-linear codes, the theta series of Construction $\textnormal{A}_4$ lattices obtained from the corresponding $\Integers_4$-linear codes, and the useful identities of the Jacobi theta functions $\vartheta_i(z), i=2,3,4$. For general $m=2\kappa$, $\kappa\in\Naturals$, expressing the theta series of Construction $\textnormal{A}_{2\kappa}$ lattices using the codes' swe's is possible~\cite{BachocGulliverHarada00_1}. But unlike the $\Integers_2$ and $\Integers_4$ cases, to the best of our knowledge, no closed-form identities like Jacobi theta functions that can be used for these theta series in the general $\Integers_{2\kappa}$ cases. Still, it can be an interesting future research topic to investigate the secrecy gain of codes over $\Integers_{2\kappa}$ with $\kappa\neq 1,2$.


\balance 

\bibliographystyle{IEEEtran}
\bibliography{defshort1,biblioHY}

\clearpage
\appendices
\ifthenelse{\boolean{short_version}}{
  \makeatletter\afterpage{\if@firstcolumn \else\afterpage{
      \onecolumn
      \section{$\Integers_4$-Linear Formally Self-Dual Codes and Their Symmetrized Weight Enumerators}
\label{sec:all-FSD-Z4-codes-swes-SGs}

{\footnotesize
\begin{longtable}{|p{1.90cm}|p{2.15cm}|p{8.00cm}|c|p{1.35cm}|}
  \caption{$\mathbb{Z}_4$-Linear Formally Self-Dual Codes, Their Symmetrized Weight Enumerators, and the Corresponding Secrecy  Criteria} 
  \label{tab:long-table_FSD-Z4-codes-swes-SGs}
  \\*\hline
  {$[n, M, d_{\textnormal{Lee}}, d_{\textnormal{E}}]$}
  & Reference / Type
  & {$\swe{\code{C}}(a,b,c)$}
  & {$\xi_{\Lambda_{\textnormal{A}_4}(\code{C})}$}
  & {$\xi_{\Lambda}$}
  \\*\hline\hline
  {$[4,2^4,2,4]$} & Ex.~\ref{ex:double_circulant_fsd}, bdc  & {$a^4+a^3 c+4 a^2 b c+a^2 c^2+2 a b^2 c+a c^3+4 b^3 c+2 b^2 c^2$} & {$\mathbf{1.052}$} & {1~\cite{LinOggier13_1}} 
   \\*\hline 
    {$[5,2^5,2,4]$} & {oext} & $a^5+2 a^4 c+4 a^3 b c+2 a^3 c^2+2 a^2 b^2 c+4 a^2 b c^2+2 a^2 c^3+4 a b^3 c+4 a b^2 c^2+a c^4+4 b^3 c^2+2 b^2 c^3$ & {$\mathbf{1.052}$} & {-} 
   \\*\hline 
   {$[6,2^6,4,6]$} & \cite[p.~125]{GulliverHarada01_1} &  $a^6+3 a^4 c^2+12 a^3 b^2 c+3 a^2 c^4+24 a b^4 c+12 a b^2 c^3+8 b^6+c^6$ & {$\mathbf{1.172}$} & {$1.172$~\cite{BollaufLinYtrehus23_3}} 
   \\*\hline  
   {$[7,2^7,2,4]$} & {oext} & $a^7+a^6 c+3 a^5 c^2+12 a^4 b^2 c+3 a^4 c^3+12 a^3 b^2 c^2+3 a^3 c^4+24 a^2 b^4 c+12 a^2 b^2 c^3+3 a^2 c^5+8 a b^6+24 a b^4 c^2+12 a b^2 c^4+a c^6+8 b^6 c+c^7$ & {$\mathbf{1.172}$} & {-} 
   \\*\hline  
  {$[8,2^8,6,8]$} & {\cite[p.~505]{HuffmanPless03_1}, Ex.~\ref{ex:E8_octacode} / II} &{$a^8+14 a^4 c^4+112 a^3 b^4 c+112 a b^4 c^3+16 b^8+c^8$}  & {$\mathbf{1.333}$} & {$1.333$~\cite{OggierSoleBelfiore16_1}} 
   \\*\hline   
  {$[8,2^8,4,6]$} & {\cite[p.~84]{BetsumiyaHarada03_1}, Ex.~\ref{ex:FSDcode_dim8}} &  $a^8+16 a^6 c^2+12 a^5 b^2 c+30 a^4 c^4+40 a^3 b^2 c^3+16 a^2 c^6+64 a b^6 c+12 a b^2 c^5+64 b^8+c^8$  & {$1.282$} & {$1.333$~\cite{OggierSoleBelfiore16_1}} 
   \\*\hline  
   {$[9,2^9,2,4]$} & {oext} & $a^9+a^8 c+14 a^5 c^4+112 a^4 b^4 c+14 a^4 c^5+112 a^3 b^4 c^2+112 a^2 b^4 c^3+16 a b^8+112 a b^4 c^4+a c^8+16 b^8 c+c^9$  & {$\mathbf{1.333}$} & {-} 
   \\*\hline   
   {$[10,2^{10},4,8]$} & {\cite[p.~229]{BachocGulliverHarada00_1}} & $a^{10}+5 a^8 c^2+40 a^6 b^2 c^2+10 a^6 c^4+40 a^5 b^4 c+80 a^4 b^2 c^4+10 a^4 c^6+160 a^3 b^6 c+240 a^3 b^4 c^3+80 a^2 b^8+40 a^2 b^2 c^6+5 a^2 c^8+160 a b^6 c^3+40 a b^4 c^5+32 b^{10}+80 b^8 c^2+c^{10}$ & {$\mathbf{1.478}$} & {$1.478$~\cite{BollaufLinYtrehus23_3}} 
   \\*\hline 
   {$[11,2^{11},4,6]$} &  {opdc} & $a^{11}+10 a^9 c^2+5 a^8 c^3+8 a^7 b^3 c+12 a^7 b^2 c^2+5 a^7 c^4+16 a^6 b^4 c+32 a^6 b^3 c^2+60 a^6 b^2 c^3+11 a^6 c^5+2 a^5 b^6+32 a^5 b^5 c+16 a^5 b^4 c^2+56 a^5 b^3 c^3+120 a^5 b^2 c^4+5 a^5 c^6+16 a^4 b^7+10 a^4 b^6 c+64 a^4 b^5 c^2+64 a^4 b^3 c^4+120 a^4 b^2 c^5+11 a^4 c^7+48 a^3 b^8+64 a^3 b^7 c+20 a^3 b^6 c^2+64 a^3 b^5 c^3+8 a^3 b^4 c^4+56 a^3 b^3 c^5+60 a^3 b^2 c^6+10 a^3 c^8+64 a^2 b^9+144 a^2 b^8 c+96 a^2 b^7 c^2+20 a^2 b^6 c^3+64 a^2 b^5 c^4+8 a^2 b^4 c^5+32 a^2 b^3 c^6+12 a^2 b^2 c^7+5 a^2 c^9+32 a b^{10}+128 a b^9 c+144 a b^8 c^2+64 a b^7 c^3+10 a b^6 c^4+32 a b^5 c^5+8 a b^4 c^6+8 a b^3 c^7+a c^{10}+32 b^{10} c+64 b^9 c^2+48 b^8 c^3+16 b^7 c^4+2 b^6 c^5+8 b^4 c^7$ &  {$\mathbf{1.512}$} &  {-} 
    \\*\hline 
    {$[12,2^{12},4,8]$} & {\cite[p.~229]{BachocGulliverHarada00_1}} & $a^{12}+6 a^{10} c^2+48 a^8 b^2 c^2+15 a^8 c^4+48 a^7 b^4 c+144 a^6 b^2 c^4+20 a^6 c^6+96 a^5 b^6 c+336 a^5 b^4 c^3+48 a^4 b^8+384 a^4 b^6 c^2+192 a^4 b^4 c^4+144 a^4 b^2 c^6+15 a^4 c^8+384 a^3 b^8 c+320 a^3 b^6 c^3+336 a^3 b^4 c^5+192 a^2 b^{10}+96 a^2 b^8 c^2+384 a^2 b^6 c^4+48 a^2 b^2 c^8+6 a^2 c^{10}+384 a b^8 c^3+96 a b^6 c^5+48 a b^4 c^7+64 b^{12}+192 b^{10} c^2+48 b^8 c^4+c^{12}$ & {$1.635$} & {$1.657$~\cite{BollaufLinYtrehus23_3}} 
     \\*\hline 
    {$[12,2^{12},4,8]$} &  {Ex.~\ref{ex:codes_dim12} / I}  & $a^{12}+18 a^{10} c^2+64 a^9 c^3+111 a^8 c^4+192 a^7 c^5+252 a^6 c^6+192 a^5 c^7+192 a^4 b^8+111 a^4 c^8+768 a^3 b^8 c+64 a^3 c^9+1152 a^2 b^8 c^2+18 a^2 c^{10}+768 a b^8 c^3+192 b^8 c^4+c^{12}$ &  {$1.6$} &  {$1.657$~\cite{BollaufLinYtrehus23_3}} 
    \\*\hline 
    {$[12,2^{12},6,8]$} & { Ex.~\ref{ex:distinct_swe_n12to13} / pdc}  & $a^{12}+15 a^8 c^4+24 a^7 b^4 c+144 a^6 b^4 c^2+32 a^6 c^6+384 a^5 b^6 c+168 a^5 b^4 c^3+72 a^4 b^8+288 a^4 b^4 c^4+15 a^4 c^8+192 a^3 b^8 c+1280 a^3 b^6 c^3+168 a^3 b^4 c^5+432 a^2 b^8 c^2+144 a^2 b^4 c^6+192 a b^8 c^3+384 a b^6 c^5+24 a b^4 c^7+64 b^{12}+72 b^8 c^4+c^{12}$ & {$\mathbf{1.657}$} & {$1.657$~\cite{BollaufLinYtrehus23_3}} 
     \\*\hline 
     {$[13,2^{13},4,7]$} & { Exs.~\ref{ex:n13k6_oextCode}, \ref{ex:distinct_swe_n12to13} / opdc}  & $a^{13}+6 a^{11} c^2+4 a^{10} c^3+15 a^9 c^4+20 a^8 b^4 c+16 a^8 b^2 c^3+16 a^8 c^5+48 a^7 b^5 c+96 a^7 b^4 c^2+32 a^7 b^3 c^3+128 a^7 b^2 c^4+20 a^7 c^6+32 a^6 b^7+128 a^6 b^6 c+128 a^6 b^5 c^2+112 a^6 b^4 c^3+64 a^6 b^3 c^4+112 a^6 b^2 c^5+24 a^6 c^7+160 a^5 b^7 c+96 a^5 b^6 c^2+208 a^5 b^5 c^3+208 a^5 b^4 c^4+64 a^5 b^3 c^5+128 a^5 b^2 c^6+15 a^5 c^8+32 a^4 b^9+80 a^4 b^8 c+352 a^4 b^7 c^2+256 a^4 b^6 c^3+256 a^4 b^5 c^4+232 a^4 b^4 c^5+64 a^4 b^3 c^6+112 a^4 b^2 c^7+16 a^4 c^9+128 a^3 b^{10}+256 a^3 b^9 c+384 a^3 b^8 c^2+448 a^3 b^7 c^3+320 a^3 b^6 c^4+208 a^3 b^5 c^5+128 a^3 b^4 c^6+32 a^3 b^3 c^7+6 a^3 c^{10}+64 a^2 b^{11}+128 a^2 b^{10} c+448 a^2 b^9 c^2+352 a^2 b^8 c^3+352 a^2 b^7 c^4+128 a^2 b^6 c^5+128 a^2 b^5 c^6+80 a^2 b^4 c^7+16 a^2 b^2 c^9+4 a^2 c^{11}+64 a b^{12}+128 a b^{11} c+128 a b^{10} c^2+256 a b^9 c^3+64 a b^8 c^4+160 a b^7 c^5+96 a b^6 c^6+48 a b^5 c^7+16 a b^4 c^8+a c^{12}+64 b^{12} c+64 b^{11} c^2+128 b^{10} c^3+32 b^9 c^4+16 b^8 c^5+32 b^7 c^6+4 b^4 c^9$ & {$\mathbf{1.704}$} & {-} 
      \\*\hline 
       {$[14,2^{14},7,8]$} & {bdc}  & $a^{14}+3 a^{10} c^4+24 a^9 c^5+48 a^8 b^5 c+30 a^8 b^4 c^2+36 a^8 c^6+240 a^7 b^6 c+240 a^7 b^5 c^2+48 a^7 b^4 c^3+16 a^7 c^7+28 a^6 b^8+224 a^6 b^7 c+384 a^6 b^6 c^2+528 a^6 b^5 c^3+48 a^6 b^4 c^4+11 a^6 c^8+96 a^5 b^9+120 a^5 b^8 c+912 a^5 b^6 c^3+720 a^5 b^5 c^4+96 a^5 b^4 c^5+24 a^5 c^9+96 a^4 b^{10}+480 a^4 b^9 c+324 a^4 b^8 c^2+1120 a^4 b^7 c^3+1536 a^4 b^6 c^4+720 a^4 b^5 c^5+108 a^4 b^4 c^6+12 a^4 c^{10}+384 a^3 b^{10} c+960 a^3 b^9 c^2+464 a^3 b^8 c^3+912 a^3 b^6 c^5+528 a^3 b^5 c^6+48 a^3 b^4 c^7+32 a^2 b^{12}+576 a^2 b^{10} c^2+960 a^2 b^9 c^3+324 a^2 b^8 c^4+672 a^2 b^7 c^5+384 a^2 b^6 c^6+240 a^2 b^5 c^7+a^2 c^{12}+64 a b^{12} c+384 a b^{10} c^3+480 a b^9 c^4+120 a b^8 c^5+240 a b^6 c^7+48 a b^5 c^8+32 b^{12} c+96 b^{10} c^4+96 b^9 c^5+28 b^8 c^6+32 b^7 c^7+6 b^4 c^{10}$ &  {$\mathbf{1.876}$} &  {$1.875$~\cite{BollaufLinYtrehus23_3}} 
      \\*\hline 
      {$[15,2^{15},6,8]$} & {obdc}  & $a^{15}+a^{12} c^3+3 a^{11} c^4+6 a^{10} b^4 c+36 a^{10} c^5+30 a^9 b^4 c^2+60 a^9 c^6+336 a^8 b^6 c+48 a^8 b^4 c^3+27 a^8 c^7+60 a^7 b^8+1104 a^7 b^6 c^2+96 a^7 b^4 c^4+27 a^7 c^8+372 a^6 b^8 c+2352 a^6 b^6 c^3+204 a^6 b^4 c^5+60 a^6 c^9+288 a^5 b^{10}+1116 a^5 b^8 c^2+3888 a^5 b^6 c^4+204 a^5 b^4 c^6+36 a^5 c^{10}+1440 a^4 b^{10} c+1908 a^4 b^8 c^3+3888 a^4 b^6 c^5+96 a^4 b^4 c^7+3 a^4 c^{11}+32 a^3 b^{12}+2880 a^3 b^{10} c^2+1908 a^3 b^8 c^4+2352 a^3 b^6 c^6+48 a^3 b^4 c^8+a^3 c^{12}+96 a^2 b^{12} c+2880 a^2 b^{10} c^3+1116 a^2 b^8 c^5+1104 a^2 b^6 c^7+30 a^2 b^4 c^9+96 a b^{12} c^2+1440 a b^{10} c^4+372 a b^8 c^6+336 a b^6 c^8+6 a b^4 c^{10}+32 b^{12} c^3+288 b^{10} c^5+60 b^8 c^7+c^{15}$ & {$\mathbf{1.972}$} & {$1.882$~\cite{LinOggier13_1}} 
      \\*\hline 
   {$[16,2^{16},8,8]^*$} & {$\widebar{\code{C}}_{16}$, {Ex.~\ref{ex:BWs_dims16-32} / II}}  & $a^{16}+140 a^{12} c^4+448 a^{10} c^6+480 a^8 b^8+870 a^8 c^8+13440 a^6 b^8 c^2+448 a^6 c^{10}+33600 a^4 b^8 c^4+140 a^4 c^{12}+13440 a^2 b^8 c^6+2048 b^{16}+480 b^8 c^8+c^{16}$ & {$1.778$} & {$2.207$~\cite{BollaufLinYtrehus23_3}} 
    \\*\hline 
     {$[16,2^{16},8,8]$} &  {bdc} & $a^{16}+14 a^{12} c^4+112 a^{10} b^4 c^2+56 a^{10} c^6+224 a^9 b^6 c+16 a^8 b^8+672 a^8 b^6 c^2+672 a^8 b^4 c^4+114 a^8 c^8+912 a^7 b^8 c+1344 a^7 b^6 c^3+896 a^7 b^4 c^5+448 a^6 b^{10}+3584 a^6 b^8 c^2+2912 a^6 b^6 c^4+224 a^6 b^4 c^6+56 a^6 c^{10}+896 a^5 b^{10} c+6384 a^5 b^8 c^3+4032 a^5 b^6 c^5+896 a^5 b^4 c^7+224 a^4 b^{12}+3136 a^4 b^{10} c^2+7392 a^4 b^8 c^4+2912 a^4 b^6 c^6+672 a^4 b^4 c^8+14 a^4 c^{12}+896 a^3 b^{12} c+5376 a^3 b^{10} c^3+6384 a^3 b^8 c^5+1344 a^3 b^6 c^7+1344 a^2 b^{12} c^2+3136 a^2 b^{10} c^4+3584 a^2 b^8 c^6+672 a^2 b^6 c^8+112 a^2 b^4 c^{10}+896 a b^{12} c^3+896 a b^{10} c^5+912 a b^8 c^7+224 a b^6 c^9+256 b^{16}+224 b^{12} c^4+448 b^{10} c^6+16 b^8 c^8+ c^{16}$ &  {$2.147$} &  {$2.207$~\cite{BollaufLinYtrehus23_3}} 
    \\*\hline 
    {$[17,2^{17},4,7]$} & {opdc} & $a^{17}+8  a^{15}c^2+4 a^{14} c^3 +28 a^{13}c^4 +48 a^{13}b^2 c^2 +4 a^{13} b^3 c +24 a^{12} c^5 +24 a^{12} b^2 c^3 +16 a^{12} b^3 c^2 +56 a^{11} c^6 +240 a^{11} b^2 c^4 +40 a^{11} b^3 c^3 +104  a^{11} b^4 c^2+24 a^{11} b^5 c +8  a^{10} b^7+60 a^{10} c^7 +216a^{10}  b^2 c^5 +80 a^{10} b^3 c^4 +256 a^{10}b^4 c^3 +112 a^{10} b^5 c^2 +72 a^{10} b^6 c +70 c^8 a^9+480a^9 b^2 c^6 +124 a^9b^3 c^5 +808 a^9b^4 c^4 +376 a^9 b^5 c^3 +112 a^9 b^6 c^2 +96 a^9 b^7 c +48 a^8b^9 +80 a^8c^9 +528 a^8 b^2 c^7 +160  a^8 b^3 c^6+1216 a^8 b^4 c^5 +768 a^8 b^5 c^4 +1192a^8 b^6 c^3 +616 a^8b^7 c^2 +144 a^8 b^8 c +64 a^7 b^{10} +56 a^7 c^{10} +480 a^7 b^2 c^8 +176 a^7 b^3 c^7 +1680 a^7 b^4 c^6 +1136 a^7 b^5 c^5 +2176 a^7 b^6 c^4 +1792 a^7 b^7 c^3 +1088 a^7 b^8 c^2 +640 a^7 b^9 c +320 a^6 b^{11} +60 a^6 c^{11} +528 a^6 b^2 c^9 +160 a^6 b^3 c^8 +1728 a^6 b^4 c^7 +1312 a^6 b^5 c^6 +3024 a^6 b^6 c^5 +3216 a^6 b^7 c^4 +2304 a^6 b^8 c^3 +1984 a^6 b^9 c^2 +864 a^6 b^{10} c +512 a^5 b^{12} +28 a^5 c^{12} +240 a^5 b^2 c^{10} +124 a^5 b^3 c^9 +1232 a^5 b^4 c^8 +1136 a^5 b^5 c^7 +3296 a^5 b^6 c^6 +3904  a^5 b^7 c^5 +4160 a^5 b^8 c^4 +4480 a^5 b^9 c^3 +1984 a^5 b^{10} c^2 +1472 a^5 b^{11} c +576 a^4 b^{13} +24 a^4 c^{13} +216 a^4 b^2 c^{11} +80 a^4 b^3 c^{10} +832 a^4 b^4 c^9 +768 a^4 b^5 c^8 +2128 a^4 b^6 c^7 +3216 a^4 b^7 c^6 +4192 a^4 b^8 c^5 +6176 a^4 b^9 c^4 +3744 a^4 b^{10} c^3 +3520 a^4 b^{11} c^2 +1280 a^4 b^{12} c +768 a^3 b^{14} +8 a^3 c^{14} +48 a^3 b^2 c^{12} +40 a^3 b^3 c^{11} +264 a^3 b^4 c^{10} +376 a^3 b^5 c^9 +1024 a^3 b^6 c^8 +1792 a^3 b^7 c^7 +2240 a^3 b^8 c^6 +4480 a^3 b^9 c^5 +3776 a^3 b^{10} c^4 +4736 a^3 b^{11} c^3 +2304 a^3b^{12} c^2 +1536 a^3b^{13} c +256 a^2 b^{15} +4 a^2 c^{15} +24 a^2 b^2 c^{13} +16  a^2 b^3 c^{12}+64 a^2 b^4 c^{11} +112 a^2 b^5 c^{10} +232 a^2 b^6 c^9 +616 a^2 b^7 c^8 +1024 a^2 b^8 c^7 +1984 a^2 b^9 c^6 +1952 a^2 b^{10} c^5 +3520 a^2 b^{11} c^4 +2304 a^2 b^{12} c^3 +1920 a^2 b^{13} c^2 +768 a^2 b^{14} c +256 a b^{16} +ac^{16} +4 ab^3 c^{13}+8 ab^4 c^{12} +24a b^5 c^{11} +48 ab^6 c^{10} +96 ab^7 c^9 +192 ab^8 c^8 +640a b^9 c^7 +832 ab^{10} c^6 +1472 ab^{11} c^5 +1280 ab^{12} c^4 +1536 ab^{13} c^3 +768a b^{14} c^2 +512 ab^{15} c +8 b^6 c^{11}+8 b^7 c^{10}+16 b^8 c^9+48 b^9 c^8+96 b^{10} c^7+320 b^{11} c^6+512 b^{12} c^5+576 b^{13} c^4+768 b^{14} c^3+256 b^{15} c^2+256 b^{16} c$ & {$\mathbf{2.203}$} & {$2.133$~\cite{LinOggier13_1}} 
     \\*\hline 
    {$[18,2^{18},7,9]$} & {pdc} & $a^{18}+18 c^5 a^{13}+45 c^6 a^{12}+18 b^5 c a^{12}+72 c^7 a^{11}+108 b^5 c^2 a^{11}+90 b^6 c a^{11}+81 c^8 a^{10}+252 b^5 c^3 a^{10}+342 b^6 c^2 a^{10}+288 b^7 c a^{10}+20 b^9 a^9+76 c^9 a^9+846 b^5 c^4 a^9+1206 b^6 c^3 a^9+864 b^7 c^2 a^9+504 b^8 c a^9+198 b^{10} a^8+90 c^{10} a^8+1512 b^5 c^5 a^8+2952 b^6 c^4 a^8+2880 b^7 c^3 a^8+1854 b^8 c^2 a^8+864 b^9 c a^8+324 b^{11} a^7+72 c^{11} a^7+1836 b^5 c^6 a^7+4284 b^6 c^5 a^7+6300 b^7 c^4 a^7+4608 b^8 c^3 a^7+2628 b^9 c^2 a^7+1368 b^{10} c a^7+240 b^{12} a^6+30 c^{12} a^6+1980 b^5 c^7 a^6+5076 b^6 c^6 a^6+8172 b^7 c^5 a^6+8640 b^8 c^4 a^6+6228 b^9 c^3 a^6+5112 b^{10} c^2 a^6+1908 b^{11} c a^6+288 b^{13} a^5+18 c^{13} a^5+1440 b^5 c^8 a^5+4788 b^6 c^7 a^5+8028 b^7 c^6 a^5+10368 b^8 c^5 a^5+9612 b^9 c^4 a^5+10152 b^{10} c^3 a^5+6084 b^{11} c^2 a^5+1440 b^{12} c a^5+288 b^{14} a^4+9 c^{14} a^4+774 b^5 c^9 a^4+2664 b^6 c^8 a^4+6156 b^7 c^7 a^4+8316 b^8 c^6 a^4+9972 b^9 c^5 a^4+12420 b^{10} c^4 a^4+10260 b^{11} c^3 a^4+3600 b^{12} c^2 a^4+1440 b^{13} c a^4+360 b^5 c^{10} a^3+1098 b^6 c^9 a^3+3060 b^7 c^8 a^3+4896 b^8 c^7 a^3+5964 b^9 c^6 a^3+10152 b^{10} c^5 a^3+9900 b^{11} c^4 a^3+4800 b^{12} c^3 a^3+2880 b^{13} c^2 a^3+1152 b^{14} c a^3+72 b^5 c^{11} a^2+486 b^6 c^{10} a^2+900 b^7 c^9 a^2+1872 b^8 c^8 a^2+2844 b^9 c^7 a^2+5112 b^{10} c^6 a^2+6012 b^{11} c^5 a^2+3600 b^{12} c^4 a^2+2880 b^{13} c^3 a^2+1728 b^{14} c^2 a^2+18 b^5 c^{12} a+54 b^6 c^{11} a+180 b^7 c^{10} a+360 b^8 c^9 a+720 b^9 c^8 a+1368 b^{10} c^7 a+2124 b^{11} c^6 a+1440 b^{12} c^5 a+1440 b^{13} c^4 a+1152 b^{14} c^3 a+36 b^7 c^{11}+54 b^8 c^{10}+60 b^9 c^9+198 b^{10} c^8+252 b^{11} c^7+240 b^{12} c^6+288 b^{13} c^5+288 b^{14} c^4$  & {$\mathbf{2.458}$} &  {$2.286$~\cite{LinOggier13_1}} 
     \\*\hline 
     {$[19,2^{19},4,10]$} & {obdc} & $a^{19} + a^{17}c^2 + 16a^{16}c^3 + 28a^{15}c^4 + 64a^{13}b^4*c^2 + 64a^{13}b^3c^3 + 28a^{13}c^6 + 416a^{12}b^4c^3 + 64a^{12}b^3c^4 + 112a^{12}c^7 + 256a^{11}b^7c + 1280a^{11}b^4c^4 + 832a^{11}b^3c^5 + 70a^{11}c^8 + 464a^{10}b^8c + 1152a^{10}b^7c^2 + 2720a^{10}b^4c^5 + 832a^{10}b^3c^6 + 128a^{10}c^9 + 256a^9b^{10} + 1696a^9b^8c^2 + 2688a^9b^7c^3 + 4352a^9b^4c^6 + 1664a^9b^3c^7 + 326a^9c^{10} + 768a^8b^{11} + 2304a^8b^{10}c + 5392a^8b^8c^3 + 7168a^8b^7c^4 + 5440a^8b^4c^7 + 1664a^8b^3c^8 + 240a^8c^{11} + 384a^7b^{12} + 5632a^7b^{11}c + 9216a^7b^{10}c^2 + 12288a^7b^8c^4 + 10752a^7b^7c^5 + 5504a^7b^4c^8 + 1152a^7b^3c^9 + 28a^7c^{12} + 1792a^6b^{12}c + 20992a^6b^{11}c^2 + 21504a^6b^{10}c^3 + 16800a^6b^8c^5 + 12544a^6b^7c^6 + 4416a^6b^4c^9 + 1152a^6b^3c^10 + 4224a^5b^{12}c^2 + 40960a^5b^{11}c^3 + 32256a^5b^{10}c^4 + 15680a^5b^8c^6 + 11520a^5b^7c^7 + 2752a^5b^4c^{10} + 320a^5b^3c^{11} + 28a^5c^{14} + 7168a^4b^{12}c^3 + 51712a^4b^{11}c^4 + 32256a^4b^{10}c^5 + 11168a^4b^8c^7 + 7168a^4b^7c^8 + 1312a^4b^4c^{11} + 320a^4b^3c^{12} + 16a^4c^{15} + 128a^3b^{16} + 3072a^3b^{15}c + 8320a^3b^{12}c^4 + 41472a^3b^{11}c^5 + 21504a^3b^{10}c^6 + 5632a^3b^8c^8 + 3328a^3b^7c^9 + 384a^3b^4c^{12} + 64a^3b^3c^{13} + a^3c^{16} + 384a^2b^{16}c + 3072a^2b^{15}c^2 + 5376a^2b^{12}c^5 + 19968a^2b^{11}c^6 + 9216a^2b^{10}c^7 + 1936a^2b^8c^9 + 640a^2b^7c^{10} + 32a^2b^4c^{13} + 64a^2b^3c^{14} + 384ab^{16}c^2 + 1024ab^{15}c^3 + 1408ab^{12}c^6 + 6144ab^{11}c^7 + 2304ab^{10}c^8 + 544ab^8c^{10} + 128ab^7c^{11} + ac^{18} + 128b^{16}c^3 + 1024b^{15}c^4 + 768b^{11}c^8 + 256b^{10}c^9 + 80b^8c^{11}$  & {$\mathbf{2.641}$} &  {$2.462$~\cite{LinOggier13_1}} 
     \\*\hline 
     {$[20,2^{20},4,10]$}  & {pdc} & $a^{20}+10 a^{18} c^2+45 a^{16} c^4+160 a^{15} b^2 c^3+120 a^{14} c^6+320 a^{13} b^4 c^3+960 a^{13} b^2 c^5+960 a^{12} b^6 c^2+1600 a^{12} b^4 c^4+210 a^{12} c^8+960 a^{11} b^8 c+2880 a^{11} b^4 c^5+2400 a^{11} b^2 c^7+2880 a^{10} b^8 c^2+13760 a^{10} b^6 c^4+6400 a^{10} b^4 c^6+252 a^{10} c^{10}+5760 a^9 b^{10} c+9920 a^9 b^8 c^3+7040 a^9 b^4 c^7+3200 a^9 b^2 c^9+640 a^8 b^{12}+26880 a^8 b^8 c^4+46720 a^8 b^6 c^6+9600 a^8 b^4 c^8+210 a^8 c^{12}+8960 a^7 b^{12} c+61440 a^7 b^{10} c^3+40320 a^7 b^8 c^5+7040 a^7 b^4 c^9+2400 a^7 b^2 c^{11}+3840 a^6 b^{14}+25600 a^6 b^{12} c^2+53120 a^6 b^8 c^6+46720 a^6 b^6 c^8+6400 a^6 b^4 c^{10}+120 a^6 c^{14}+42240 a^5 b^{12} c^3+123648 a^5 b^{10} c^5+40320 a^5 b^8 c^7+2880 a^5 b^4 c^{11}+960 a^5 b^2 c^{13}+1280 a^4 b^{16}+57600 a^4 b^{14} c^2+60160 a^4 b^{12} c^4+26880 a^4 b^8 c^8+13760 a^4 b^6 c^{10}+1600 a^4 b^4 c^{12}+45 a^4 c^{16}+10240 a^3 b^{16} c+42240 a^3 b^{12} c^5+61440 a^3 b^{10} c^7+9920 a^3 b^8 c^9+320 a^3 b^4 c^{13}+160 a^3 b^2 c^{15}+23040 a^2 b^{16} c^2+57600 a^2 b^{14} c^4+25600 a^2 b^{12} c^6+2880 a^2 b^8 c^{10}+960 a^2 b^6 c^{12}+10 a^2 c^{18}+10240 a b^{18} c+10240 a b^{16} c^3+8960 a b^{12} c^7+5760 a b^{10} c^9+960 a b^8 c^{11}+1024 b^{20}+1280 b^{16} c^4+3840 b^{14} c^6+640 b^{12} c^8+c^{20}$ & {$\mathbf{2.868}$} &  {$2.868$~\cite{PerssonBollaufLinYtrehus23_1}} 
     \\*\hline 
    {$[21,2^{21},6,8]$} & {\cite[App.~A]{PlessSoleQian97_1}} & $a^{21}+28 a^{18} c^3+84 a^{17} c^4+273 a^{16} c^5+924 a^{15} c^6+1956 a^{14} c^7+2982 a^{13} c^8+4340 a^{12} c^9+5796 a^{11} c^{10}+5796 a^{10} c^{11}+4340 a^9 c^{12}+2982 a^8 c^{13}+1956 a^7 c^{14}+924 a^6 c^{15}+273 a^5 c^{16}+84 a^4 c^{17}+28 a^3 c^{18}+b^{12} (2688 a^9+24192 a^8 c+96768 a^7 c^2+225792 a^6 c^3+338688 a^5 c^4+338688 a^4 c^5+225792 a^3 c^6+96768 a^2 c^7+24192 a c^8+2688 c^9)+b^8 (84 a^{13}+1092 a^{12} c+6552 a^{11} c^2+24024 a^{10} c^3+60060 a^9 c^4+108108 a^8 c^5+144144 a^7 c^6+144144 a^6 c^7+108108 a^5 c^8+60060 a^4 c^9+24024 a^3 c^{10}+6552 a^2 c^{11}+1092 a c^{12}+84 c^{13})+c^{21}$ & {$\mathbf{2.909}$} &  {$2.909$~\cite{LinOggier13_1}} 
     \\*\hline 
     {$[22,2^{22},10,12]$} & {\cite[p.~230]{BachocGulliverHarada00_1}, Ex.~\ref{ex:n22_FSD-code_Z4}} & $a^{22}+176 a^{15} c^7+330 a^{14} c^8+616 a^{13} b^8 c+2464 a^{13} b^7 c^2+4004 a^{12} b^8 c^2+14784 a^{11} b^8 c^3+29568 a^{11} b^7 c^4+672 a^{11} c^{11}+1232 a^{10} b^{12}+14784 a^{10} b^{11} c+40656 a^{10} b^8 c^4+616 a^{10} c^{12}+12320 a^9 b^{12} c+83160 a^9 b^8 c^5+110880 a^9 b^7 c^6+55440 a^8 b^{12} c^2+221760 a^8 b^{11} c^3+124740 a^8 b^8 c^6+5632 a^7 b^{15}+147840 a^7 b^{12} c^3+140800 a^7 b^8 c^7+140800 a^7 b^7 c^8+176 a^7 c^{15}+2464 a^6 b^{16}+258720 a^6 b^{12} c^4+620928 a^6 b^{11} c^5+123200 a^6 b^8 c^8+77 a^6 c^{16}+14784 a^5 b^{16} c+118272 a^5 b^{15} c^2+310464 a^5 b^{12} c^5+83160 a^5 b^8 c^9+66528 a^5 b^7 c^{10}+36960 a^4 b^{16} c^2+258720 a^4 b^{12} c^6+443520 a^4 b^{11} c^7+41580 a^4 b^8 c^{10}+49280 a^3 b^{16} c^3+197120 a^3 b^{15} c^4+147840 a^3 b^{12} c^7+14784 a^3 b^8 c^{11}+9856 a^3 b^7 c^{12}+36960 a^2 b^{16} c^4+55440 a^2 b^{12} c^8+73920 a^2 b^{11} c^9+3696 a^2 b^8 c^{12}+14784 a b^{16} c^5+39424 a b^{15} c^6+12320 a b^{12} c^9+616 a b^8 c^{13}+352 a b^7 c^{14}+2464 b^{16} c^6+1232 b^{12} c^{10}+1344 b^{11} c^{11}+44 b^8 c^{14}$ &  {$\mathbf{3.403}$} &   {$3.335$~\cite{BollaufLinYtrehus23_3}} 
     \\*\hline 
     {$[23,2^{23},10,12]$} &  {\cite[App.~A]{PlessSoleQian97_1}} & $a^{23}+253 a^{16} c^7+506 a^{15} c^8+1288 a^{12} c^{11}+1288 a^{11} c^{12}+506 a^8 c^{15}+253 a^7 c^{16}+b^{16} (8096 a^7+56672 a^6 c+170016 a^5 c^2+283360 a^4 c^3+283360 a^3 c^4+170016 a^2 c^5+56672 a c^6+8096 c^7)+b^{12} (2576 a^{11}+28336 a^{10} c+141680 a^9 c^2+425040 a^8 c^3+850080 a^7 c^4+1190112 a^6 c^5+1190112 a^5 c^6+850080 a^4 c^7+425040 a^3 c^8+141680 a^2 c^9+28336 a c^{10}+2576 c^{11})+b^8 (1012 a^{14} c+7084 a^{13} c^2+28336 a^{12} c^3+85008 a^{11} c^4+191268 a^{10} c^5+318780 a^9 c^6+404800 a^8 c^7+404800 a^7 c^8+318780 a^6 c^9+191268 a^5 c^{10}+85008 a^4 c^{11}+28336 a^3 c^{12}+7084 a^2 c^{13}+1012 a c^{14})+c^{23}$ &  {$\mathbf{3.556}$} & {$3.556$~\cite{LinOggier13_1}} 
     \\*\hline 
     {$[24,2^{24},12,16]$} & {\cite[p.~494]{HuffmanPless03_1} / II}  & $a^{24}+2576 a^{12} c^{12}+61824 a^{11} b^{12} c+a b^{12} c^{11}+24288 a^8 b^{16}+b^{16} c^8+1214400 a^8 b^8 c^8+1700160 a^4 b^{16} c^4+759 a^{16} c^8+a^8 c^{16}+12144 a^{14} b^8 c^2+a^2 b^8 c^{14}+170016 a^{12} b^8 c^4+a^4 b^8 c^{12}+765072 a^{10} b^8 c^6+a^6 b^8 c^{10}+1133440 a^9 b^{12} c^3+a^3 b^{12} c^9+4080384 a^7 b^{12} c^5+a^5 b^{12} c^7+680064 a^6 b^{16} c^2+a^2 b^{16} c^6+4096 b^{24}+c^{24}$ & {$\mathbf{4.063}$} & {$4.063$~\cite{OggierSoleBelfiore16_1}} 
      \\*\hline 
      {$[26,2^{26},6,12]$} & {\cite{Harada12_1}, Ex.~\ref{ex:secrecy-gain_26} / I}
      & $a^{26}+30 a^{23} c^3+255 a^{22} c^4+1100 a^{21} c^5+3571 a^{20} c^6+9990 a^{19} c^7+24330 a^{18} c^8+49680 a^{17} c^9+83237 a^{16} c^{10}+119004 a^{15} c^{11}+2880 a^{14} b^{12}+150750 a^{14} c^{12}+40320 a^{13} b^{12} c+164680 a^{13} c^{13}+262080 a^{12} b^{12} c^2+150750 a^{12} c^{14}+1048320 a^{11} b^{12} c^3+119004 a^{11} c^{15}+17408 a^{10} b^{16}+2882880 a^{10} b^{12} c^4+83237 a^{10} c^{16}+174080 a^9 b^{16} c+5765760 a^9 b^{12} c^5+49680 a^9 c^{17}+783360 a^8 b^{16} c^2+8648640 a^8 b^{12} c^6+24330 a^8 c^{18}+2088960 a^7 b^{16} c^3+9884160 a^7 b^{12} c^7+9990 a^7 c^{19}+16384 a^6 b^{20}+3655680 a^6 b^{16} c^4+8648640 a^6 b^{12} c^8+3571 a^6 c^{20}+98304 a^5 b^{20} c+4386816 a^5 b^{16} c^5+5765760 a^5 b^{12} c^9+1100 a^5 c^{21}+245760 a^4 b^{20} c^2+3655680 a^4 b^{16} c^6+2882880 a^4 b^{12} c^{10}+255 a^4 c^{22}+327680 a^3 b^{20} c^3+2088960 a^3 b^{16} c^7+1048320 a^3 b^{12} c^{11}+30 a^3 c^{23}+245760 a^2 b^{20} c^4+783360 a^2 b^{16} c^8+262080 a^2 b^{12} c^{12}+98304 a b^{20} c^5+174080 a b^{16} c^9+40320 a b^{12} c^{13}+16384 b^{20} c^6+17408 b^{16} c^{10}+2880 b^{12} c^{14}+c^{26}$ & {$\mathbf{4.433}$} & {$4.356$~\cite{PerssonBollaufLinYtrehus23_1}} 
      \\*\hline 
       {$[31,2^{31},6,12]$} &  {\cite[App.~A]{PlessSoleQian97_1}} & $a^{31}+155 a^{28} c^3+1085 a^{27} c^4+5208 a^{26} c^5+22568 a^{25} c^6+82615 a^{24} c^7+247845 a^{23} c^8+628680 a^{22} c^9+1383096 a^{21} c^{10}+2648919 a^{20} c^{11}+4414865 a^{19} c^{12}+6440560 a^{18} c^{13}+8280720 a^{17} c^{14}+9398115 a^{16} c^{15}+9398115 a^{15} c^{16}+8280720 a^{14} c^{17}+6440560 a^{13} c^{18}+4414865 a^{12} c^{19}+2648919 a^{11} c^{20}+1383096 a^{10} c^{21}+628680 a^9 c^{22}+247845 a^8 c^{23}+82615 a^7 c^{24}+22568 a^6 c^{25}+5208 a^5 c^{26}+1085 a^4 c^{27}+155 a^3 c^{28}+b^{16} (63488 a^{15}+952320 a^{14} c+6666240 a^{13} c^2+28887040 a^{12} c^3+86661120 a^{11} c^4+190654464 a^{10} c^5+317757440 a^9 c^6+408545280 a^8 c^7+408545280 a^7 c^8+317757440 a^6 c^9+190654464 a^5 c^{10}+86661120 a^4 c^{11}+28887040 a^3 c^{12}+6666240 a^2 c^{13}+952320 a c^{14}+63488 c^{15})+c^{31}$ &  {$\mathbf{6.564}$} &  {-} 
      \\*\hline 
      {$[32,2^{32},8,16]$} &  {\cite[App.~A]{PlessSoleQian97_1}, Ex.~\ref{ex:BWs_dims16-32}} & $a^{32}+1240 a^{28} c^4+27776 a^{26} c^6+330460 a^{24} c^8+2011776 a^{22} c^{10}+7063784 a^{20} c^{12}+14721280 a^{18} c^{14}+18796230 a^{16} c^{16}+14721280 a^{14} c^{18}+7063784 a^{12} c^{20}+2011776 a^{10} c^{22}+330460 a^8 c^{24}+27776 a^6 c^{26}+1240 a^4 c^{28}+b^{16} (126976 a^{16}+15237120 a^{14} c^2+231096320 a^{12} c^4+1016823808 a^{10} c^6+1634181120 a^8 c^8+1016823808 a^6 c^{10}+231096320 a^4 c^{12}+15237120 a^2 c^{14}+126976 c^{16})+67108864 b^{32}+c^{32}$ &  {$\mathbf{7.111}$} & {7.111} 
      \\*\hline 
\end{longtable}}

    }
    \fi}\makeatother
}{}

\newpage 

\ifthenelse{\boolean{short_version}}{
  \makeatletter\afterpage{\if@firstcolumn \else\afterpage{
      \onecolumn
      \section{Weight Enumerators of Codes obtained via Gray map}
\label{sec:weight-enumerators-codes-gray-44-48}

\begin{longtable}{|c|c|p{10.0cm}|c|}
  \caption{Weight Enumerators of Codes obtained via Gray map}
  \\*\hline
  $\code{C}_{\textnormal{g}} = \psi(\code{C})$
  & $\code{C}$
  & $W_{\code{C}_{\textnormal{g}}}(x,y)$
  & $\xi_{\eGammaA{\code{C}_{\textnormal{g}}}}$
  \\*\hline\hline
   \multirow{8}{*}{$[44,2^{22}, 10]$}  &  \multirow{8}{*}{$[22,2^{22},10]$} & {$x^{44}+616 x^{34} y^{10}+2464 x^{33} y^{11}+5236 x^{32} y^{12}+14784 x^{31} y^{13}+27280 x^{30} y^{14} + 35200 x^{29} y^{15}+98890 x^{28} y^{16}+221760 x^{27} y^{17}+245784 x^{26} y^{18}+229152 x^{25} y^{19} + 420420 x^{24} y^{20}+ 620928 x^{23} y^{21} +  501216 x^{22} y^{22}+337920 x^{21} y^{23}+419496 x^{20} y^{24} + 443520 x^{19} y^{25}+245784 x^{18} y^{26}+105952 x^{17} y^{27} +99484 x^{16} y^{28}+73920 x^{15} y^{29} + 27280 x^{14} y^{30}+9856 x^{13} y^{31}+5005 x^{12} y^{32}+1344 x^{11} y^{33} + 616 x^{10} y^{34} + 352 x^9 y^{35}+44 x^8 y^{36}$} &  \multirow{8}{*}{$16.956$} 
  \\*\hline
  \multirow{4}{*}{$[48,2^{24}, 12]$}  &  \multirow{4}{*}{$[24,2^{24},12]$} & {$x^{48}+12144 x^{36} y^{12}+61824 x^{34} y^{14}+195063 x^{32} y^{16}+ 1133440 x^{30} y^{18}+ 1445136 x^{28} y^{20}+ 4080384 x^{26} y^{22}+2921232 x^{24} y^{24}+4080384 x^{22} y^{26}+ 1445136 x^{20} y^{28}+1133440 x^{18} y^{30}+ 195063 x^{16} y^{32}+61824 x^{14} y^{34}+12144 x^{12}$} &  \multirow{4}{*}{$23.257$} 
  \\*\hline
\end{longtable}

    }
    \fi}\makeatother
}{\section{Weight Enumerators of Codes obtained via Gray map}
\label{sec:weight-enumerators-codes-gray-44-48}

\begin{longtable}{|c|c|p{10.0cm}|c|}
  \caption{Weight Enumerators of Codes obtained via Gray map}
  \\*\hline
  $\code{C}_{\textnormal{g}} = \psi(\code{C})$
  & $\code{C}$
  & $W_{\code{C}_{\textnormal{g}}}(x,y)$
  & $\xi_{\eGammaA{\code{C}_{\textnormal{g}}}}$
  \\*\hline\hline
   \multirow{8}{*}{$[44,2^{22}, 10]$}  &  \multirow{8}{*}{$[22,2^{22},10]$} & {$x^{44}+616 x^{34} y^{10}+2464 x^{33} y^{11}+5236 x^{32} y^{12}+14784 x^{31} y^{13}+27280 x^{30} y^{14} + 35200 x^{29} y^{15}+98890 x^{28} y^{16}+221760 x^{27} y^{17}+245784 x^{26} y^{18}+229152 x^{25} y^{19} + 420420 x^{24} y^{20}+ 620928 x^{23} y^{21} +  501216 x^{22} y^{22}+337920 x^{21} y^{23}+419496 x^{20} y^{24} + 443520 x^{19} y^{25}+245784 x^{18} y^{26}+105952 x^{17} y^{27} +99484 x^{16} y^{28}+73920 x^{15} y^{29} + 27280 x^{14} y^{30}+9856 x^{13} y^{31}+5005 x^{12} y^{32}+1344 x^{11} y^{33} + 616 x^{10} y^{34} + 352 x^9 y^{35}+44 x^8 y^{36}$} &  \multirow{8}{*}{$16.956$} 
  \\*\hline
  \multirow{4}{*}{$[48,2^{24}, 12]$}  &  \multirow{4}{*}{$[24,2^{24},12]$} & {$x^{48}+12144 x^{36} y^{12}+61824 x^{34} y^{14}+195063 x^{32} y^{16}+ 1133440 x^{30} y^{18}+ 1445136 x^{28} y^{20}+ 4080384 x^{26} y^{22}+2921232 x^{24} y^{24}+4080384 x^{22} y^{26}+ 1445136 x^{20} y^{28}+1133440 x^{18} y^{30}+ 195063 x^{16} y^{32}+61824 x^{14} y^{34}+12144 x^{12}$} &  \multirow{4}{*}{$23.257$} 
  \\*\hline
\end{longtable}
}



\end{document}